\documentclass[a4paper,fleqn,usenatbib,useAMS]{mnras} 
\usepackage{float}
\usepackage{graphicx}	
\usepackage{amssymb}
\usepackage{amsmath}	
\usepackage[T1]{fontenc}
\usepackage{ae,aecompl}
\usepackage{subcaption}
\usepackage{mwe}
\usepackage{rotating}
\usepackage{multirow}

\newcommand{\az}[1]{\textcolor{black}{{#1}}}
\newcommand{\ka}[1]{\textcolor{black}{{#1}}}

\newcommand{\ngca}{NGC\,3310 }
\newcommand{\ngcb}{NGC\,2276 }
 
\newcommand{\ergs}{~erg s$^{-1}$}

\newcommand{\nh}{$\rm{N_{H}}$}
\newcommand{\chisq}{$\rm{\chi^{2}}$ }
\newcommand{\chandra}{\textit{Chandra} }

\title[NGC\,3310 and NGC\,2276]{Do sub-galactic regions follow the galaxy-wide X-ray scaling relations? The example of NGC\,3310 and NGC\,2276}
\author[K. Anastasopoulou et al.]{
K. Anastasopoulou,$^{1,2}$\thanks{E-mail: kanast@physics.uoc.gr} A. Zezas,$^{1,2,3}$ V. Gkiokas$^{1,4}$, K. Kovlakas$^{1,2}$ \\
$^{1}$Physics Department \& Institute of Theoretical \& Computational Physics, University of Crete, 71003 Heraklion, Crete, Greece\\
$^{2}$Foundation for Research and Technology-Hellas, 71110 Heraklion, Crete, Greece \\
$^{3}$Harvard-Smithsonian Center for Astrophysics, 60 Garden Street, Cambridge, MA 02138, USA\\
$^{4}$Institute of Astronomy \& Astrophysics, National Observatory of Athens, Palaia Penteli, 152 36 Athens, Greece}

\date{Accepted XXX. Received YYY; in original form ZZZ}

\pubyear{2016}

\begin{document}
\label{firstpage}
\pagerange{\pageref{firstpage}--\pageref{lastpage}}
\maketitle

\begin{abstract}
We present results from \textit{Chandra} observations of the X-ray starburst galaxies NGC\,3310 and NGC\,2276.
We detect 27 discrete sources in NGC\,3310, and 19 discrete sources in NGC\,2276 with luminosities above 1.0$\times \mathrm{10^{38}\ erg\ s^{-1}}$. The majority of the sources have photon indices of $1.7$-$2.0$, typical for X-ray binaries. Both galaxies have large numbers of ultra-luminous X-ray sources (ULXs;  sources with $\mathrm{L(0.3-10.0\ keV)>10^{39}\ erg\ s^{-1}}$), 14 for NGC\,3310 concentrated on the circumnuclear star-forming ring and north spiral arm and 11 for NGC\,2276 with the brighter ones on the west side of the galaxy which is compressed due to harassment by the intra-group medium it is moving into.
We find for both galaxies that the ULX-hosting areas are located above the general $\mathrm{L_{X}}$-SFR scaling relations while other areas either follow or fall below the scaling relations. This indicates that sub-galactic regions follow the galaxy-wide scaling relations but with much larger scatter resulting from the age (and possibly metallicity) of their local stellar populations in agreement with recent theoretical and observational results. Such differences in age could be the origin of the scatter we observe in the low SFR regime in the Lx-SFR scaling relations. 
\end{abstract}


\section{Introduction}

Ultra luminous X-ray sources are off-nuclear sources that have luminosities $\mathrm{L(0.3-10.0\,keV)>10^{39}\ erg\ s^{-1}}$. This generally exceeds the Eddington luminosity of a typical stellar-mass black hole and therefore it indicates very high accretion rates. The most generally accepted model to explain the nature of ULXs is that of super-Eddington accretion (with possibly mild beaming) onto a stellar-mass black hole or a neutron star (NS) X-ray binary (XRB) \citep[e.g.][and references therein]{king09,kaaret17}. Other models include accretion onto an intermediate-mass black hole (IMBH) although with very little observational evidence \citep[e.g.][]{kaaret17}.
In general ULXs are more abundant in low metallicity galaxies \citep[e.g.][]{prestwich13,douna15} and are found in large numbers in merging and star-forming galaxies \citep[e.g.][]{swartz11,konna16}.

Two star-forming galaxies which have been found to host large number of ULXs are NGC\,3310 and NGC\,2276 \citep[e.g.][]{wolter11,wolter15,lehmer15}. 
They are relatively nearby, at distances of 22\,Mpc for NGC 3310 and at 41\,Mpc for NGC 2276. 
These distances are based on the latest cosmology \citep[][$\mathrm{H_o=67.4\ km\ s^{-1}\ Mpc^{-1}}$, $\mathrm{\Omega_M=0.32}$, and  $\mathrm{\Omega_{\Lambda}=0.68}$]{planck18} using the Virgo infall corrected redshift. For both galaxies the available redshift-independent distances, are based on the Tully-Fischer relation, which given their disturbed morphology and signs of interaction (described in the next paragraph) are not reliable. 

Both galaxies exhibit unique morphologies. NGC\,3310, which is among the most luminous star-forming galaxies in the local Universe, shows a disturbed morphology, which is possibly the result of a recent merger \citep[$\sim$30 Myr ago;][]{elmegreen02,degrijs03a} with a low metallicity dwarf galaxy which triggered a circumnuclear star forming ring of about 20 arcsec. The galaxy also shows two distinct spiral arms, one on the north and one on the south. NGC\,2276, which is a member of the loose group NGC\,2300 displays also a uniquely disturbed morphology, where the west side of the galaxy is being compressed as it moves supersonically (900~$\mathrm{km\,\ s^{-1}}$) through the NGC\,2300 intra-group medium \citep[IGM;][]{rasmussen06}.

In the X-rays, NGC\,3310 showed evidence for the existence of an active galactic nucleus (AGN) based on the presence of an FeK$\alpha$ line in \textit{Chandra} spectra of the nucleus \citet{tzanavaris07}. However this is not supported by any other AGN indicator \citep[e.g. optical lines,][]{ho97}. \citet{lehmer15} combined simultaneous \textit{Chandra} and \textit{NuSTAR} observations of the galaxy and found an excess of X-ray emission per unit SFR in the 6-30\,keV band compared to other lower sSFR (specific SFR=SFR per stellar mass) star-forming galaxies. This was interpreted as the result of the over-abundance of ULXs in NGC\,3310 compared to typical galaxies. They argue that this excess of ULXs is most likely explained by the relatively low metallicity of the young stellar population in this galaxy.
\textit{HST} optical observations have identified hundreds of star clusters \citep[][]{elmegreen02} younger than 10\,Myr and with masses of 10$^{4}$-10$^{5}\mathrm{M_{\odot}}$ for the largest clumps, as well as, 17 candidate super star clusters, mainly in the innermost southern spiral arm. Furthermore \citet{degrijs03a,degrijs03b} using the same data, found that the age and metallicity distributions of the clusters in and outside the circumnuclear ring in NGC\,3310 are statistically indistinguishable, although there is a clear and significant excess of higher mass clusters in the ring compared to the non-ring cluster sample. \citet{miralles14a} using data from the PPAK Integral Field Spectroscopy (IFS) Nearby Galaxies Survey (PINGS) found a rather flat gas-phase abundance gradient for about a hundred HII regions located on the disk and the spiral arms. This indicates that the minor merger event had a substantial impact on metal mixing in the galaxy, resulting in uniform metallicity across the galaxy. \citet{miralles14b} studied the Wolf-Rayet population of NGC\,3310 by spatially resolving 18 star-forming knots with typical sizes of 200-300\,pc in the disc of the galaxy hosting a substantial population of Wolf-Rayet stars, which assuming metallicity-dependent luminosities results to an integrated number of more than 4000.

NGC\,2276 has been extensively studied in the X-rays. \textit{Chandra} observations have shown that it hosts 16 ULXs \citep{wolter11,wolter15}. Its diffuse X-ray morphology shows several similarities to its optical morphology with the shock-like feature along the west side and a faint tail to the east side of NGC\,2276 \citep{rasmussen06,wolter15} and has temperatures of $\mathrm{kT\sim 0.3-0.8\ keV}$ and luminosities of $\mathrm{L(0.3-2.0\ keV)=1.9}$-$\mathrm{18.0\times10^{39}\ erg\ s^{-1}}$ for the main body and the faint tail of the galaxy. \citet{wolter15} using hydrodynamic simulations found that, though the periapsis passage of NGC\,2276 and NGC\,2300 $\sim85$\,Myr ago helped to produce tidal arms and thicken the gaseous disk, these effects are marginal compared to the effects from ram-pressure and viscous stripping of the galaxy by the IGM.
In the optical, the galaxy contains numerous HII regions \citep[e.g.][]{hodge83,davis97} and supernov\ae\ \citep[e.g.][]{isk67,dimai05}. Furthermore \citet{mezcua15}
analysing quasi-simultaneous \textit{Chandra} X-ray observations and European VLBI Network radio observations, report an IMBH candidate (NGC2276-3c) of $5\times10^{4}\mathrm{M_{\odot}}$ associated with a \textit{Chandra} source with $\mathrm{L(0.3-10.0\,keV)=5.5\times 10^{39} erg\ s^{-1}}$.

These two galaxies due to their proximity, unique morphology and large number of X-ray sources, are excellent laboratories for studying the connection of XRBs and ULXs with galaxy parameters like the SFR and the stellar mass, even at sub-galactic scales. 
{\az{An investigation of the ULX population in the interacting pair NGC\,2207/IC\,2163 \cite{mineo13,mineo14} showed that at sub-galactic scales the number of ULXs and their luminosity scales with SFR in a similar way as in galaxy-wide scales. However, they do find tentative evidence for a dependence of the number (and possibly the luminosity) of ULXs on the FIR to UV luminosity ratio, a proxy for the age of the stellar populations and dust extinction.  However it is unclear if this trend is the result of local variations of the ULX population stemming from local variations of the star formation history, or it is the result of stochastic sampling of the X-ray binary luminosity function.  }}

In this paper we use the available high quality data for NGC\,3310 and NGC\,2276 in order to address two questions: (a) the validity of the general relations between X-ray binaries and star-forming activity \citep[e.g.][]{lehmer10,mineo12a} to galaxies with large numbers of ULXs, and (b) the validity of these galaxy-wide scaling relations to sub-galactic scales.
The structure of the paper is as follows: in section \ref{observationanddataanalysis} we describe the observation, the data analysis and present our results. We discuss our results in section \ref{discussion} and in section \ref{summary} we summarize our findings. All errors correspond to the 90\% confidence interval unless otherwise stated.

\section{Observation and Data Analysis}\label{observationanddataanalysis}

In this work we use observations obtained with the ACIS-S camera \citep{garmire} on board the \textit{Chandra} X-ray observatory \citep{weisskopf} for the galaxies NGC\,3310 and NGC\,2276. The observations for NGC\,3310 were performed on January 15th 2003 (OBSID 2939; 47.16$\,\textrm{ks}$) and October 22nd 2016 (OBSID 19891; 35.84$\,\textrm{ks}$). A much shorter observation (10$\,\textrm{ks}$; OBSID 16025) performed on the 11th of June 2014 is not included in our analysis since it will complicate the analysis without improving the statistics. The observations for NGC\,2276 were performed on June 23rd 2004 (OBSID 4968; 45.57$\,\textrm{ks}$) and May 14th 2013 (OBSID 15648; 24.74$\,\textrm{ks}$). {\az{Although detailed analysis of the first observation of NGC\,2276 has already been presented \citep{rasmussen06,wolter11,wolter15}, we re-analyse the data for consistency and to incorporate all available data sets.}

{\az{In our analysis we followed the same procedures described in \citet{konna16}}}.
We used the CIAO {\az{data analysis suite}} version 4.8 and CALDB version 4.7.0 for the analysis of the data. We first applied the latest calibration data (using the \textit{acis\_process\_events} tool) by reprocessing the Level-1 event files and  then by filtering for bad grades and status bits (using the \textit{dmcopy} tool) we created the Level-2 events files. We kept only grades=0, 2, 3, 4, 6 and status=0. The net exposure times after the removal of background flares are for \ngca: {\az{46.67$\,\textrm{ks}$ (OBSID 2939) and 35.84$\,\textrm{ks}$ (OBSID 19891) and for \ngcb: 44.29$\,\textrm{ks}$ (OBSID 4968) and 21.15$\,\textrm{ks}$ (OBSID 15648).}}

We used the \textit{fluximage} tool on the events-2 files with \textit{binsize=1.0} for each OBSID for the creation of the images and the exposure maps in the broad (0.5-7.0 \textrm{keV}), soft (0.5-1.2 \textrm{keV}), medium (1.2-2.0 \textrm{keV}), and hard (2.0-7.0 \textrm{keV}) bands. We also created images (using the \textit{merge\_obs} tool) for the co-added exposure based on the two observations for each galaxy. Additionally, we created sub-pixel resolution images (\textit{binsize=0.2}) in order to look at small-scale structures which helps to distinguish sources in crowded areas of the galaxies. In order to measure the relative variation of the effective area in different regions of each image, we normalized all the exposure maps to the exposure of a reference pixel at approximately the centre of the galaxy \citep[c.f.][]{zezas06}.

Finally we used the \textit{csmooth} CIAO tool with a Gaussian convolution kernel in order to {\az{create adaptively smoothed images}}. We applied a minimum signal-to-noise ratio of 3 and a maximum of 5. {\az{We also used the scales of the broad band in order to smooth the images in the soft, medium, and hard bands.} {\az{Figs. \ref{fig.smoothcolor_ngc3310} and \ref{fig.smoothcolor_ngc2276} show `true colour images'' for NGC\,3310 and NGC\,2276 respectively, where the soft, medium, and hard band adaptively smoothed images are show in red, green, and blue respectively.} In these figures we see a population of hard discrete sources revealed by their blue colours, and soft, diffuse, emission shown in red.

\begin{figure*}
	\centering
	\begin{minipage}{160mm} 
		\includegraphics[scale=0.43]{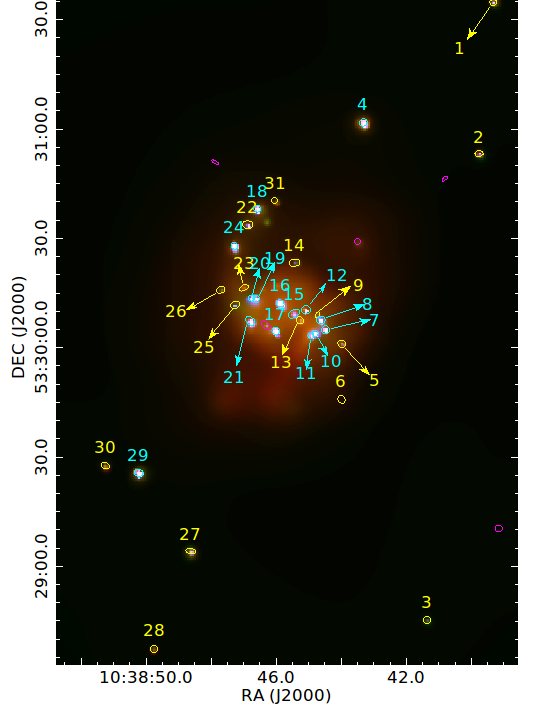}
	    \includegraphics[scale=0.40]{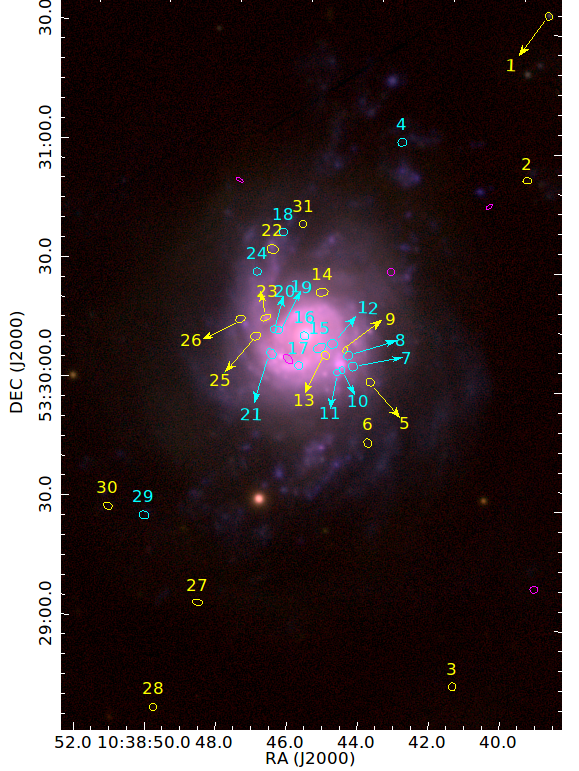} 
		\caption{
        {\az{Left: An adaptively smoothed true colour X-ray image of \ngca, with the soft (0.5-1.2 \textrm{keV}), medium (1.2-2.0 \textrm{keV}), and hard band (2.0-7.0 \textrm{keV}) shown in red, green, and blue respectively.  Right: A PanSTARRS colour image with the y, i, and g bands shown in red, green, and blue respectively. The 31 sources with $\mathrm{SNR>3.0}$ are also overlaid on the two images, with the numbers corresponding to the source-IDs in Table \ref{tab.propertiesngc3310}.  Cyan and yellow circles indicate sources with luminosities above and below $10^{39}$\ergs respectively (i.e. the ULX limit). The two images have the same scale.}}  
		}
		\label{fig.smoothcolor_ngc3310}
	\end{minipage}
\end{figure*}

\begin{figure*}
	\centering
		\begin{minipage}{160mm}
		\includegraphics[scale=0.30]{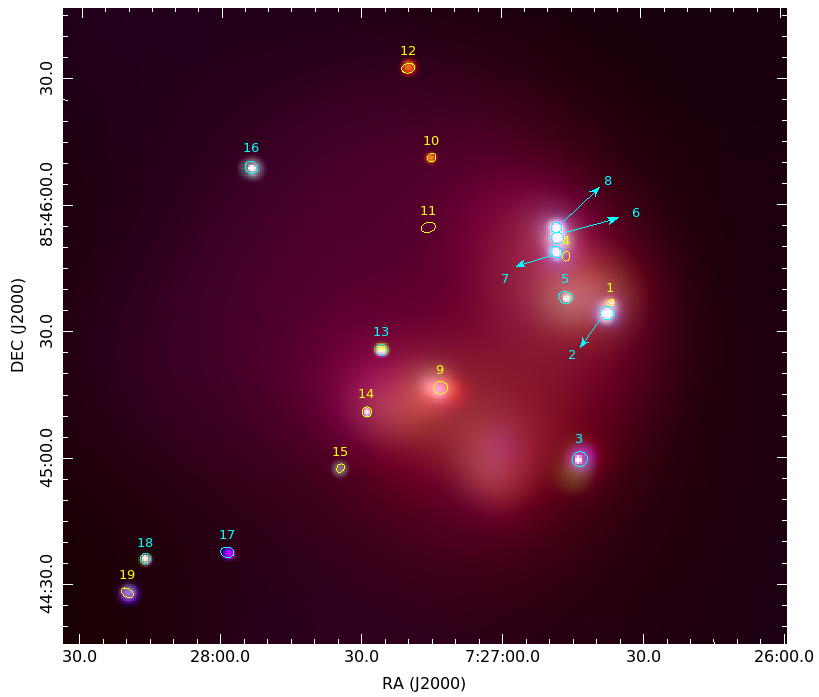}
		\includegraphics[scale=0.27]{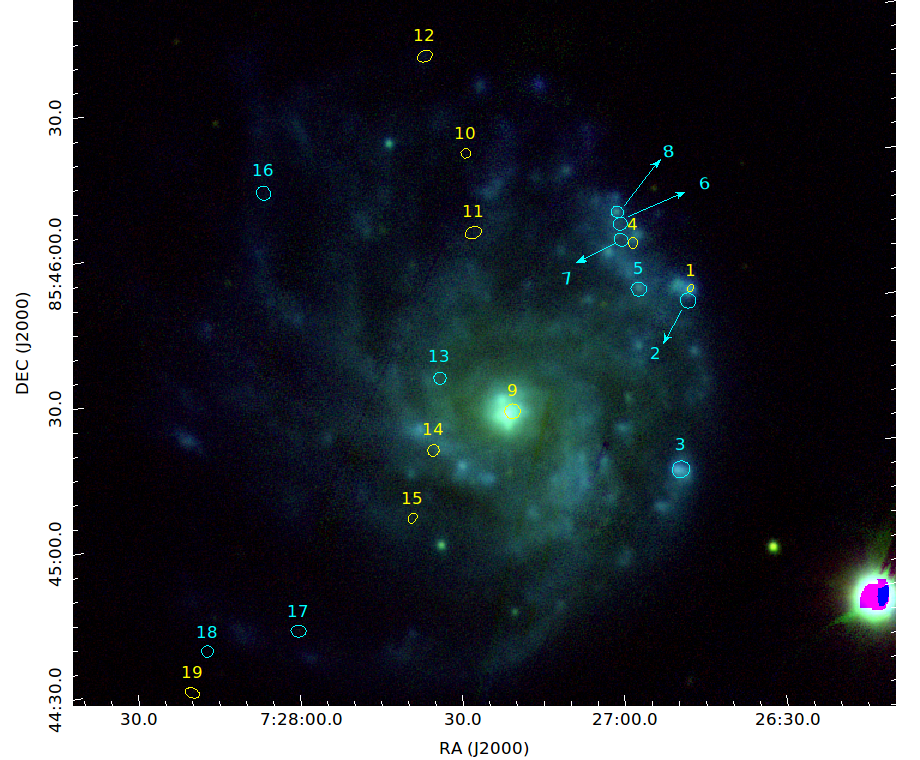}
		\caption{{\az{Left: An adaptively smoothed true colour X-ray image of \ngcb, with the soft (0.5-1.2 \textrm{keV}), medium (1.2-2.0 \textrm{keV}), and hard band (2.0-7.0 \textrm{keV}) shown in red, green, and blue respectively.  Right: A PanSTARRS colour image with the y, i, and g bands shown in red, green, and blue respectively. The 19 sources with $\mathrm{SNR>3.0}$ are also overlaid on the two images, with the numbers corresponding to the source-IDs in Table \ref{tab.propertiesngc2276}.  Cyan and yellow circles indicate sources with luminosities above and below $10^{39}$\ergs respectively (i.e. the ULX limit). The two images have the same scale.}}      
        }
		\label{fig.smoothcolor_ngc2276}
	\end{minipage}
\end{figure*}

\subsection{Source Detection and Photometry}\label{sourcedetection}
The source detection was performed, using the \textit{wavdetect} tool \citep{b2}, on the co-added exposure as well as the individual {\az{observations}}. We searched in the broad (0.5-7.0 \textrm{keV}), soft (0.5-1.2 \textrm{keV}), medium (1.2-2.0 \textrm{keV}), and hard-band (2.0-7.0 \textrm{keV}) images for sources on scales of 2, 4, 8, 16, and 32 pixels for the images with \textit{binsize=0.2} (pixel scale 0.0984 arcsec/pixel) and on scales of 2, 4, 8, and 16 pixels for the images with \textit{binsize=1.0} (with the native pixel scale of 0.492 arcsec). In the sub-pixel image of NGC\,3310 we found two additional sources, more specifically sources 10, 11 and 19, 20 appeared as a single source in the \textit{binsize=1.0} image. No additional sources were found in the sub-pixel image of NGC\,2276. The final source list was created from the combination of the source lists in each band {\az{in the individual and the co-added observation}}. We found in total 37 sources {\az{encompassed by} NCG\,3310 and 23 sources {\az{in}} NGC\,2276. {\az{The extent of \ngcb was defined as the }} D25 region {\az{reported in the 3rd Reference Catalogue}} \citep[RC3;][]{rc31,rc32}. However, {\az{since the D25 region of \ngca\ reported in the RC3 was smaller than the optical extent of the galaxy, we defined the outline of the galaxy on the based on optical DSS  images. This is parametrized as an ellipse of semimajor axis $1.545\arcmin$, semiminor axis $1.199\arcmin$, and position angle PA=$180 \deg$ centred on the nucleus of the galaxy (RA= 10:38:45.8, Dec=+53:30:12)}}.
 
 {\az{For each source we performed aperture photometry using an elliptical aperture encompassing at least 90\% of the energy of a point source at $\mathrm{1.49\ keV}$ (based on a PSF map calculated with \textit{mkpsfmap}), while taking care to avoid emission from any nearby sources. The background was measured from an annulus with inner radius $\sim1$ pixel larger than the source aperture and an outer radius ($\sim$ 4-20 arcsec), excluding any encompassed sources.}
This way we avoid contamination by the wings of the source PSF, and obtain good count statistics for reliable photometry.

\begin{table*}
	\centering
	\begin{minipage}{140mm}
		\caption{Broad-band ($\mathrm{0.5-7.0\ keV}$) photometry of the discrete sources in NGC\,3310.}
		\hskip-2.0cm
		\setlength{\tabcolsep}{4pt}
		\begin{tabular}[!htbp]{@{}lcccc|ccc|ccc|ccc@{}}
		\hline 
				Src  &  RA      & Dec  & $\mathrm{r_1}$  &   $\mathrm{r_2}$  &      Net counts$\pm$error& Bkg & S/N & Net counts$\pm$error &Bkg & S/N&Net counts$\pm$error&Bkg &S/N\\
			ID &h m s &$^{\circ}$ $'$ $''$   & $''$ & $''$&OBSID 2939 &&&OBSID 19891&&&co-added OBSIDs&&\\
   (1) &  (2)      & (3)   &       (4)& (5) & (6) &(7)  &   (8)  & (9)&(10)&(11)&(12)&(13)&(14)\\ 
   \hline 
   	  1 & 10:38:39.3 & +53:31:35 & 0.95 & 0.94  & $74.6\pm{9.9}$    & 0.3 &  8.6	& $ 16.8 \pm  5.2 $ &  0.2  &  3.2 &	           $ 90.5 \pm 10.6 $ & 0.5 & 8.6    \\     
   	  2 & 10:38:39.7 & +53:30:53 & 1.13 & 0.78  & $62.7\pm{9.2}$    & 0.2 &  7.8 	& $ 14.8 \pm  5.0 $ &  0.2  &  3.0 &	         $ 74.5 \pm 9.7 $ & 0.5 & 7.7     \\    
   	  3 & 10:38:41.3 & +53:28:45 & 0.98 & 1.00  & $21.6\pm{6.1}$    & 0.3 &  4.5 	& $ 9.8 \pm  4.3 $ &  0.2  &  2.3 &	         $ 29.4 \pm 6.5 $ & 0.6 & 4.5     \\     
   	  4 & 10:38:43.3 & +53:31:02 & 1.13 & 0.96  & $945.6\pm{31.8}$  & 1.3 &  30.7 	& $ 487.2 \pm  23.1 $ &  0.8  &  21.1 &          $ 1326.8 \pm 37.5 $ & 2.2 & 35.4  \\     
   	  5 & 10:38:43.9 & +53:30:01 & 1.1 & 0.94   & $29.7\pm{8.2}$    & 8.2 &  4.3 	& $ 18.5 \pm  5.8 $ &  3.5  &  3.2 &	           $ 43.4 \pm 8.5 $ & 11.6 & 5.1     \\    
   	  6 & 10:38:43.9 & +53:29:46 & 1.10 & 0.90  & $12.3\pm{5.5}$    & 1.6 &  3.1 	& $ -0.6 \pm  1.9 $ &  0.6  &  -0.3 &            $ 12.1 \pm 4.8 $ & 1.9 & 2.5     \\    
   	  7 & 10:38:44.5 & +53:30:05 & 1.13 & 1.14  & $306.5\pm{20.2}$  & 26.4 &  16.1 	& $ 86.7 \pm  11.4 $ &  15.3  &  7.6 &          $ 377.8 \pm 22.1 $ & 48.2 & 17.1   \\     
   	  8 & 10:38:44.6 & +53:30:07 & 1.04 & 1.17  & $280.5\pm{19.7}$  & 29.4 &  15.2	 &$ 87.5 \pm  11.3 $ &  13.5  &  7.8 &           $ 373.6 \pm 21.6 $ & 41.4 & 17.3   \\     
   	  9 & 10:38:44.7 & +53:30:09 & 0.85 &  0.56  & $24.3\pm{7.9}$   & 8.6 &  3.7 	 &$ 17.7 \pm  5.7 $ &  2.3  &  3.1 &	          $ 32.6 \pm 8.0 $ & 12.4 & 4.1     \\    
   	  10 & 10:38:44.8 & +53:30:04 & 0.72 & 0.89  & $475.3\pm{24.2}$  & 26.6 &  20.6 & $ 125.8 \pm  12.9 $ &  11.2  &  9.8 &          $ 530.0 \pm 25.0 $ & 36.0 & 21.2  \\     
   	  11 & 10:38:44.9 & +53:30:03 & 0.83 & 0.83  & $191.3\pm{16.9}$  & 25.6 &  12.2 & $ 82.0 \pm  10.7 $ &  9.0  &  7.6 &          $ 270.5 \pm 18.9 $ & 37.5 & 14.3  \\     
   	  12 & 10:38:45.0 & +53:30:10 & 1.28 & 1.17  & $154.3\pm{15.6}$  & 24.6 &  10.8 & $ 43.0 \pm  9.0 $ &  14.0  &  4.8 &          $ 190.0 \pm 16.5 $ & 36.0 & 11.5  \\      
   	  13 & 10:38:45.2 & +53:30:07 & 1.06 & 0.94  & $38.6\pm{11.5}$   & 31.3 &  3.8 	& $ 20.7 \pm  7.2 $ &  12.3  &  2.9 &          $ 46.5 \pm 11.7 $ & 47.5 & 4.0   \\     
   	  14 & 10:38:45.4 & +53:30:23 & 1.51 & 1.00  & $19.5\pm{7.8}$    & 10.4 &  3.0 	& $ 8.3 \pm  4.4 $ &  2.7  &  1.9 &	          $ 26.5 \pm 7.4 $ & 13.5 & 3.6    \\    
   	  15 & 10:38:45.4 & +53:30:09 & 1.68 & 1.05  & $93.4\pm{15.3}$   & 50.5 &  6.7	 &$ 53.7 \pm  10.8 $ &  22.3  &  5.0 &           $ 134.8 \pm 16.9 $ & 75.2 & 8.0   \\
	  16\footnote{Nucleus of the galaxy; variable} & 10:38:45.8 & +53:30:12 & 1.10 & 0.95  & $1384.2\pm{39.8}$ & 53.7 &  35.8  &$ 312.0 \pm  19.5 $ &  20.0  &  16.0 &           $ 1587.6 \pm 42.3 $ & 85.4 & 37.5  \\    
   	  17 & 10:38:46.0 & +53:30:04 & 1.04 & 1.02  & $1160.5\pm{36.2}$  & 34.4 &  33.0 &$ 351.4 \pm  20.2 $ &  12.6  &  17.4 &           $ 1438.4 \pm 40.0 $ & 59.6 & 36.0  \\     
   	  18 & 10:38:46.5 & +53:30:38 & 0.96 & 0.94  & $307.6\pm{18.9}$   & 4.3 &  17.2  &$ 436.5 \pm  22.1 $ &  6.5  &  19.8 &           $ 681.2 \pm 27.3 $ & 8.8 & 25.0   \\     
   	  19 & 10:38:46.6 & +53:30:13 & 0.96 & 0.89  & $205.7\pm{17.9}$   & 34.2 &  12.4 &$ 94.7 \pm  11.8 $ &  13.3  &  8.0 &           $ 288.6 \pm 20.2 $ & 51.4 & 14.3  \\     
   	  20 & 10:38:46.7 & +53:30:13 & 0.81 & 0.86  & $381.4\pm{21.4}$   & 14.5 &  18.8 &$ 122.2 \pm  12.6 $ &  7.8  &  9.7 &           $ 482.9 \pm 23.8 $ & 25.1 & 20.3  \\      
   	  21 & 10:38:46.8 & +53:30:07 & 1.40 & 1.05  & $143.8\pm{15.9}$   & 34.1 &  9.8  &$ 680.9 \pm  27.8 $ &  22.1  &  24.5 &           $ 780.7 \pm 30.3 $ & 53.3 & 25.7  \\     
   	  22 & 10:38:46.9 & +53:30:33 & 1.43 & 1.08  & $80.5\pm{11.0}$   & 6.5 &  8.3 	 &$ 41.5 \pm  7.9 $ &  4.5  &  5.3 &	            $ 118.3 \pm 12.6 $ & 13.7 & 9.4   \\      
   	  23 & 10:38:47.0 & +53:30:16 & 1.37 & 0.72  & $27.9\pm{8.5}$    & 11.0 &  3.9 	 &$ 4.3 \pm  4.8 $ &  7.7  &  0.9 &	           $ 29.2 \pm 8.2 $ & 17.8 & 3.6    \\      
   	  24 & 10:38:47.2 & +53:30:28 & 0.97 & 0.94  & $825.1\pm{30.0}$   & 6.9 &  28.4  &$ 365.7 \pm  20.2 $ &  3.3  &  18.1 &           $ 1137.3 \pm 34.9 $ & 12.7 & 32.5  \\      
   	  25 & 10:38:47.2 & +53:30:11 & 1.34 & 0.96  & $51.8\pm{10.4}$   & 15.1 &  5.7 	 &$ 20.5 \pm  7.0 $ &  11.5  &  3.0 &            $ 69.0 \pm 10.7 $ & 19.0 & 6.5   \\      
   	  26 & 10:38:47.7 & +53:30:16 & 1.21 & 0.91  & $19.1\pm{7.4}$    & 7.8 &  3.2 	 &$ 9.3 \pm  4.5 $ &  1.7  &  2.0 &	           $ 23.1 \pm 7.0 $ & 9.9 & 3.3     \\     
   	  27 & 10:38:48.6 & +53:29:04 & 1.34 & 0.77  & $89.6\pm{10.7}$   & 0.4 &  9.4 	 &$ 45.8 \pm  7.8 $ &  0.2  &  5.8 &	            $ 123.4 \pm 12.2 $ & 0.6 & 10.1   \\      
   	  28 & 10:38:49.7 & +53:28:37 & 1.00 & 1.00  & $14.7\pm{5.4}$    & 0.3 &  3.7 	 &$ 14.8 \pm  5.0 $ &  0.2  &  3.0 &	            $ 28.5 \pm 6.5 $ & 0.5 & 4.4     \\     
   	  29 & 10:38:50.2 & +53:29:25 & 1.25 & 1.01  & $675.0\pm{27.1}$   & 0.9 &  25.9  &$ 460.4 \pm  22.5 $ &  0.6  &  20.5 &           $ 1109.3 \pm 34.3 $ & 1.7 & 32.3  \\      
   	  30 & 10:38:51.2 & +53:29:27 & 1.14 & 0.86  & $31.5\pm{7.0}$    & 0.5 &  5.5 	 &$ 15.7 \pm  5.1 $ &  0.3  &  3.1 &	           $ 43.1 \pm 7.7 $ & 0.9 & 5.6     \\     
   	  31 & 10:38:46.0 & +53:30:40 & 0.94 & 0.94  & $11.0\pm{5.6}$    & 2.9 &  1.9 	 &$ 16.5 \pm  5.2 $ &  0.5  &  3.2 &           $ 24.1 \pm 6.3 $ & 2.9 & 3.8    \\   
   	  \hline
\end{tabular}  
  \label{tab.propertiesngc3310}
  \smallskip
{\az{  Column 1: the source identification number; columns 2 and 3: Right Ascension and Declination (J2000); columns 4 and 5: the major and minor radius of the elliptical source apertures; column 6: net source counts (and corresponding errors) for the longer (OBSID 2939) exposure; column 7: the estimated background counts within the extraction aperture of each source for the longer (OBSID 2939) exposure; column 8: the signal to noise ratio for the longer (OBSID 2939) exposure; columns 9, 10, 11: same as columns 6, 7, and 8 respectively but for the shorter exposure (OBSID 19891), column 12, 13, 14: same as columns 6, 7, and 8 respectively but for the co-added exposure.}}
 \end{minipage}
\end{table*}

\begin{table*}
	\centering
	\begin{minipage}{140mm}
		\caption{Broad-band ($\mathrm{0.5-7.0\ keV}$) photometry of the discrete sources in NGC\,2276.}
		\hskip-2.0cm
		\setlength{\tabcolsep}{4pt}
		\begin{tabular}[!htbp]{@{}lcccc|ccc|ccc|ccc@{}}
			\hline   
			Src     &  RA      & Dec  & $\mathrm{r_1}$  &   $\mathrm{r_2}$  &      Net counts$\pm$error& Bkg & S/N & Net counts$\pm$error &Bkg & S/N&Net counts$\pm$error&Bkg &S/N\\
			ID &h m s &$^{\circ}$ $'$ $''$   & $''$ & $''$&OBSID 4968 &&&OBSID 15648&&&co-added OBSIDs&&\\
			(1) &  (2)      & (3)   &       (4)& (5) & (6) &(7)  &   (8)  & (9)&(10)&(11)&(12)&(13)&(14)\\	
			\hline		
			1  & 7:26:36.6  &  +85:45:36.9       & 0.8 &    0.6& 	$19.6\pm{5.7}$&	1.4	&	4.0	& $8.4\pm{4.1}$ &	   0.6	&	2.5 	&  $27.9\pm{6.5}$ &  	2.1	&	4.8		\\
			2  &  7:26:37.4  &  +85:45:34.4    & 1.6 &    1.5&  	$415.5\pm{21.5}$& 4.5	&	20.1	& $90.9\pm{10.7}$ &	   2.1	&	9.2 	&  $504.5\pm{23.6}$ & 	6.5	&	22.1		\\
			3  &  7:26:43.3  &  +85:44:59.8     & 1.8 &    1.7&  	$38.2\pm{7.6}$&	4.8	&	5.4	& $5.8\pm{4.0}$ &	   2.2	&	1.6 	&  $43.7\pm{8.2}$ &  	7.3	&	5.6		\\
			4  & 7:26:46.1   &  +85:45:47.9      & 1.1 &     0.9& $3.2\pm{3.6}$&	2.8	&	1.0	& $21.7\pm{5.9}$ &	   1.3	&	4.2 	&  $24.6\pm{6.5}$ &  	4.4	&	4.1		\\
			5  &  7:26:46.3  &  +85:45:38.2     & 1.6 &    1.4&  	$34.9\pm{7.3}$&	4.1	&	5.2	& $9.3\pm{4.4}$ &	   1.7	&	2.4 	&  $45.3\pm{8.2}$ &  	5.7	&	5.9		\\
			6\footnote{IMBH candidate; \citet{mezcua15}}   &  7:26:47.9  &  +85:45:52.2    & 1.4 &    1.3& 	$183.2\pm{14.9}$&9.8	&	12.8	& $80.3\pm{10.2}$ &	   3.7	&	8.5 	&  $261.2\pm{17.6}$ & 	14.8	&	15.3	 \\
			7  &  7:26:48.2  &  +85:45:48.9    & 1.5 &    1.2&  	$107.2\pm{11.7}$&6.8	&	9.7	& $161.4\pm{13.8}$ &	   1.6	&	12.5	&  $270.5\pm{17.7}$ & 	7.5	&	15.9		\\
			8  &  7:26:48.2  &  +85:45:54.7    & 1.2 &    1.1&  	$323.4\pm{19.2}$&7.6	&	17.5	& $143.1\pm{13.1}$ &	   1.9	&	11.7	&  $469.4\pm{22.9}$ & 	10.6	&	21.1		\\
			9\footnote{Nucleus of the galaxy}  &  7:27:13.0  &  +85:45:16.8     & 1.5 &    1.6&   $19.5\pm{6.6}$&	11.5	&	2.9	& $10.7\pm{5.1}$ &	   5.3	&	2.2 	&  $32.6\pm{8.1}$ &  	17.4	&	3.9	 \\
			10  &  7:27:14.8  &  +85:46:11.4    & 1.0 &    0.9&  	$10.6\pm{4.4}$&	0.4	&	2.9	& $3.7\pm{3.2}$ &	   0.3	&	1.5 	&  $14.3\pm{5.0}$ &  	0.7	&	3.4		\\
			11  & 7:27:15.5   &  +85:45:54.7    &  1.7 &   1.2& 	$7.0\pm{4.0}$&	1.0	&	2.1	& $12.4\pm{4.7}$ &	   0.6	&	3.2 	&  $21.4\pm{5.9}$ &  	1.6	&	4.2		\\
			12 &  7:27:19.8  &  +85:46:32.6     & 1.5 &    1.1& 	$14.6\pm{5.0}$&	0.4	&	3.5	& $75.6\pm{9.8}$ &	   0.4	&	8.6 	&  $89.2\pm{10.5}$ &  	0.8	&	9.3		 \\
			13 &  7:27:25.6  &  +85:45:25.8     & 1.2 &    1.2&  	$58.9\pm{8.8}$&	1.1	&	7.4	& $37.8\pm{7.2}$ &	   0.2	&	6.0 	&  $96.6\pm{10.9}$ &  	1.4	&	9.6		 \\
			14 &  7:27:28.7  &  +85:45:11.0     & 1.2 &    1.1& 	$15.3\pm{5.2}$&	1.7	&	3.4	& $5.8\pm{3.8}$ &	   1.2	&	1.8 	&  $21.1\pm{6.0}$ &  	2.9	&	3.9		  \\
			15 &  7:27:34.4  &  +85:44:57.7     & 1.1 &    0.8&  	$9.7\pm{4.3}$&	0.3	&	2.8	& $1.7\pm{2.7}$ &	   0.3	&	0.8 	&  $10.5\pm{4.4}$ &  	0.5	&	2.9		  \\
			16 &  7:27:53.5  &  +85:46:09.0     & 1.3 &    1.5&  	$51.5\pm{8.3}$&	0.5	&	7.0	& $20.7\pm{5.7}$ &	   0.3	&	4.3 	&  $72.2\pm{9.6}$ &  	0.8	&	8.3		  \\
			17 &  7:27:58.5  &  +85:44:37.7     & 1.5 &    1.2& 	$22.5\pm{5.9}$&	0.5	&	4.5	& $0.8\pm{2.3}$ &	   0.2	&	0.5 	&  $23.3\pm{6.0}$ &  	0.7	&	4.5		   \\
			18 &  7:28:15.9  &  +85:44:36.1     & 1.1 &    1.2& 	$29.7\pm{6.5}$&	0.3	&	5.2	& $38.6\pm{7.3}$ &	   0.4	&	6.0 	&  $66.5\pm{9.2}$ &  	0.5	&	8.0		    \\
			19 &  7:28:19.7  &  +85:44:28.0    & 1.5 &    0.9&  	$13.2\pm{4.8}$&	0.8	&	3.2	& $5.5\pm{3.6}$ &          0.5	&	1.9 	&  $18.9\pm{5.6}$ &  	1.1	&	3.9		    \\
			\hline
		\end{tabular} 	
		\label{tab.propertiesngc2276}
        {\az{  Column 1: the source identification number; columns 2 and 3: Right Ascension and Declination (J2000); columns 4 and 5: the major and minor radius of the elliptical source apertures; column 6: net source counts (and corresponding errors) for the longer (OBSID 4968) exposure; column 7: the estimated background counts within the extraction aperture of each source for the longer (OBSID 4968) exposure; column 8: the signal to noise ratio for the longer (OBSID 4968) exposure; columns 9, 10, 11: same as columns 6, 7, and 8 respectively but for the shorter exposure (OBSID 15648), column 12, 13, 14: same as columns 6, 7, and 8 respectively but for the co-added exposure.}}
	\end{minipage}
\end{table*}

We used the \textit{dmextract} tool to perform the photometry on all the sources within the {\az{outline of each galaxy}} in the individual, co-added exposures and each of the broad, hard, medium, and soft bands. 
In more detail, following \citet{zezas06}, we measured the number of counts for each source in each image, the corresponding background counts from a ``swiss-cheese" image from which all sources were removed, and the relative effective area at the location of each source with respect to the galaxy centre from the normalised exposure map.
{\az{The SNR for each source is calculated by propagating the errors on the number of counts in the source and background areas \citep[see also][]{konna16}. Throughout this paper we adopt the \citep{gehrels} approximation for the errors on the number of counts. }}
{\az{In Tables \ref{tab.propertiesngc3310} and \ref{tab.propertiesngc2276} we present the results of the photometric analysis in the broad (0.5-7.0\,keV) band for the 31 sources in NGC\,3310 and the 19 sources in NGC\,2276 with $\textrm{SNR}\geq 3.0$. }}
 respectively.

\subsection{Spectral Analysis}\label{spectralanalysis}
The \textit{specextract} tool was used to extract source and background spectra of the discrete sources. {\az{The spectra for sources with more than 50 net counts were grouped to have at least 20 total counts per spectral bin in order to allow for $\chi^2$ fitting.}} 
{\az{The spectral fits were performed with the XSPEC v12.9.0 package \citep{xspec}. In our analysis we considered only events in the 0.4 --8.0\,keV range since events at lower or higher energies are dominated by the background.}}
For sources with less than 50 net counts we grouped the spectra to have at least 2-5 total counts per spectral bin and the spectral fitting was performed using \textit{sherpa} \citep{sherpa} with the \textit{wstat} statistic. This is equivalent to the XSPEC implementation of the Cash statistic where the observed background data is added to the model and do not have to be modelled

We fitted simultaneously the spectra from the two separate observations (OBSID 2939 and OBSID 19891 for NGC\,3310; OBSID 4968 and OBSID 15648 for NGC\,2276) with {\az{ all the model parameters apart from the normalization tied together. The normalization for the spectrum from each observation was left free to vary independently in order investigate for source variability.}}

For NGC\,3310 we used only the longer exposure (OBSID 2939) for sources 6, and 23. These two sources are diffuse emission clumps and 
{\az{ their significance in the shorter observation was very low (Table \ref{tab.spectralparam_ngc3310}).}}
Additionally fitting sources 1, 2, 21, and 26 simultaneously did not result in a good fit statistic {\az{and left significant residuals indicating spectral variability.}} Therefore we fitted the spectra for each OBSID separately.  
 
For NGC\,2276, in nine cases (Sources 1, 3, 5, 9, 10, 14, 15, 17, and 19) we used only the longer exposure since the very few counts of the shorter observation would not allow for spectral fitting (Table \ref{tab.spectralparam_ngc2276}).
All {\az{sources in}} NGC\,2276, and all but {\az{4 sources in}} NGC\,3310 were fitted well with a single power-law model with photoelectric absorption (model in XSPEC: \texttt{phabs$\times$po}). This model is generally used to fit the spectra of X-ray binaries (XRBs). 
For the majority of the sources, the best-fit photon indices are $\sim1.7-2.0$, consistent with those of XRBs, while the hydrogen column density is typically greater than the Galactic (NGC\,3310: $N_H^{Gal}\sim 5.52\times 10^{20} \mathrm{cm^{-2}}$; NGC\,2276: $N_H^{Gal}\sim 1.11\times 10^{20} \mathrm{cm^{-2}}$; using the \textit{Colden} tool\footnote{\url{http://cxc.harvard.edu/toolkit/colden.jsp}}).
For NGC\,2276 sources 4 and 11 would not allow for spectral fitting since they had very few counts. The fit parameters for the 17 out of 19 sources of NGC\,2276 {\az{for which we could perform spectral analysis}}  are reported in Table \ref{tab.spectralparam_ngc2276}.v

For NGC\,3310 source 16, which is the nucleus of the galaxy, an absorbed power-law model gave a good fit ($\mathrm{\chi^2/dof=123.2/74}$) and there was no sign for a FeK$\alpha$ emission line at 6.4 keV, in contrast to \citet{tzanavaris07} who report a line at $6.4\pm 0.1\ \mathrm{keV}$ with equivalent width (EW) of 0.3 keV. In order to test if our results are consistent with their work, we added a Gaussian line to the model with energy and width fixed at 6.4 keV and 0.1 keV respectively and normalisation free to vary. We found an EW of $0.23_{-0.23}^{+0.43}$ keV which is consistent with the EW of 0.3 keV reported in \citet{tzanavaris07}.
Sources 5, 6, 9, and 23 gave a good fit only with an absorbed thermal plasma model. Therefore we consider these sources as diffuse emission clumps within the galaxy. 
The fit results for the 31 sources in NGC\,3310 are reported in Table \ref{tab.spectralparam_ngc3310}.

\begin{table*}
	\centering
	\begin{minipage}{120mm}
		\caption{NGC\,3310 spectral parameters based on spectral fits.}
		\begin{tabular}[!htbp]{@{\extracolsep{5pt}}ccccccc@{}}
			\hline 
			Src ID   & $\mathrm{\Gamma}$ & kT & $\mathrm{N_H}$ & $\chi^2$ (dof) &  Binning & Binning \\
		   	& & keV&$\mathrm{10^{22} cm^{-2}}$ & & OBSID 2939& OBSID 19891\\
		   			(1)&(2)&(3)& (4)& (5)&(6)&(7)\\ 
		   			\hline
		  1   &  $1.76_{-0.37}^{+0.84}$    & -  &  $0.06_{-0.05}^{+0.25}$ &  0.4	(1)    & 20& -\\[5pt]		  		
		   	1 &  $3.26^{+2.88}_{-1.73}$ &- & $0.69\leq{1.2}$ &  6.2 (6)        & - & 2\\[5pt]
		   2   &  $2.78_{-1.09}^{+2.66}$    &   &  $0.26_{-0.26}^{+0.58}$ &  1.1	(1)   &   20 & -\\[5pt]	
		   	2 &  $3.51^{+2.85}_{-1.88}$ &- & $0.96 \leq 2.12$ & 13.6 (6) &  -&2 \\[5pt]			   
		   3 &  $2.21^{+1.45}_{-1.1}$ &- & $0.45\leq 0.68$ & 4.4 (7)	  & 5 & 5\\[5pt]
		   4 &  $2.35^{+0.2}_{-0.18}$ &- & $0.1^{+0.05}_{-0.05}$ & 60.0 (57)& 20  & 2  \\[5pt]
		   5 &  - &$2.06^{+2.36}_{-1.71}$ & $0.12\leq{1.52}$ & 18.2 (16)	 &  5 & 2\\[5pt]
		   6   &  	-	          	 &  $0.99_{-0.90}^{+0.37}$  &  $0.11\leq1.25$                      &     1.5 	(4)   & 2& -\\[5pt]
		   7 &  $2.1^{+0.35}_{-0.31}$ &- & $0.35^{+0.15}_{-0.13}$ & 12.5 (16)	 &  20 & 20 \\[5pt]
		   8 &  $2.49^{+0.44}_{-0.38}$ &- & $0.9^{+0.32}_{-0.24}$ & 13.5 (15)	 & 20 & 20 \\[5pt]
		   9 &  - &$1.45^{+5.8}_{-0.46}$ & $1.19^{+0.71}_{-1.08}$ & 31.8 (13)	 & 5 & 2 \\[5pt]
		   10 &  $1.44^{+0.29}_{-0.26}$ &- & $0.74^{+0.27}_{-0.21}$ & 17.6 (26)	  &  20 & 20 \\[5pt]
		   11 &  $1.6^{+0.4}_{-0.36}$ & - &$0.2^{+0.17}_{-0.14}$ & 6.7 (10)		  &  20 & 20 \\[5pt]
		   12 &  $2.45^{+0.53}_{-0.48}$ &- & $0.28^{+0.19}_{-0.15}$ & 60.5 (31)	  &  20 & 5\\[5pt]
		   13 &  $4.26^{+3.17}_{-1.78}$ &- & $0.45\leq{0.72}$ & 15.0 (14)	  &  20 & 5 \\[5pt]
		   14 &  $2.94^{+3.62}_{-1.22}$ & - &$2.5^{+5.96}_{-1.9}$ & 25.3 (18) & 2 & 2 \\[5pt]
		   15 &  $3.11^{+1.57}_{-1.01}$ &- & $1.84^{+1.76}_{-1.2}$ & 51.3 (39)	 & 20  & 2 \\[5pt]
		   16 &  $1.4^{+0.16}_{-0.15}$ &- & $0.54^{+0.11}_{-0.09}$ & 123.2 (74)	  &  20 & 20 \\[5pt]
		   17 &  $1.94^{+0.17}_{-0.16}$ &- & $0.61^{+0.11}_{-0.09}$ & 69.2 (65)	  & 20 & 20 \\[5pt]
		   18 &  $1.86^{+0.23}_{-0.21}$ &- & $0.29^{+0.12}_{-0.1}$ & 51.1 (31)	 & 20 & 20 \\[5pt]
		   19 &  $1.85^{+0.5}_{-0.43}$ & - &$0.6^{+0.32}_{-0.23}$ & 9.5 (11)	 & 20  & 20 \\[5pt]
		   20 &  $1.8^{+0.44}_{-0.39}$ & - &$1.8^{+0.55}_{-0.44}$ & 15.9 (20)	 & 20  & 20 \\[5pt]
		   21  &  $2.42_{-0.52}^{+1.02}$    &   &  $0.11_{-0.08}^{+0.19}$ &  2.4	(6)   & 20  & - \\[5pt]
		   21  &  $1.77_{-0.23}^{+0.25}$    &   &  $0.36_{-0.14}^{+0.16}$ &  20.5	(32)   & -  & 20 \\[5pt]		   
		   22 &  $1.96^{+0.61}_{-0.52}$ &- & $0.22^{+0.28}_{-0.21}$ & 8.3 (5)	 & 20 & 15 \\[5pt]
		   23  &  		 -         	 & $1.85_{-1.81}^{+1.45}$  &   $0.11\leq1.89$                      &      12.2 	(5)   & 5 & -\\[5pt]		   
		   24 &  $2.0^{+0.18}_{-0.17}$ & - &$0.29^{+0.08}_{-0.08}$ & 61.9 (51)	 & 20 & 20 \\[5pt]
		   25 &  $3.89^{+1.57}_{-1.2}$ &- & $3.12^{+1.7}_{-1.25}$ & 46.4 (27)	 & 5 & 2 \\[5pt]
		   26  &  	$2.83_{-1.62}^{+3.00}$		          	 &  - &    $0.50\leq1.35$                    &       0.7	(3) & 5&- \\[5pt]
		   26  &  	$2.24_{-1.13}^{+1.50}$		          	 &  - &    $0.11\leq0.45$                    &       7.1	(3) & -& 2 \\[5pt]
		   27 &  $2.29^{+0.67}_{-0.54}$ &- & $0.16^{+0.26}_{-0.16}$ & 8.5 (5)	 & 20 & 15\\[5pt]
		   28 &  $2.49^{+1.6}_{-0.81}$ &- & $0.18\leq{0.48}$ & 7.0 (8)	 & 5 & 2 \\[5pt]
		   29 &  $1.79^{+0.18}_{-0.17}$ &- & $0.13^{+0.07}_{-0.06}$ & 40.4 (47)	  &  20  & 20 \\[5pt]
		   30 &  $3.39^{+1.37}_{-1.1}$ &- & $0.47^{+0.39}_{-0.32}$ & 23.5 (12)	 &  5& 2\\[5pt]
		   31 &  $2.5^{+0.93}_{-0.57}$ &- & $0.11\leq{0.17}$ & 16.1 (13)       & 2 & 2 \\[5pt]	   			
			\hline				
			\end{tabular} 	
				\label{tab.spectralparam_ngc3310}
                \smallskip
 \az{               Column 1: source ID; column 2: photon index $\Gamma$ from spectral fitting; column 3: thermal plasma model temperatures (kT); column 4: line-of-sight hydrogen column density (\nh); column 5: \chisq of the spectral fit and corresponding degrees of freedom (d.o.f); and columns 6 and 7: spectral binning for OBSIDs 2939 and 19891 respectively.}
				\end{minipage}
				\end{table*}

\begin{table*}
	\centering
	\begin{minipage}{110mm}
		\caption{NGC\,2276 spectral parameters based on spectral fits.}
		\begin{tabular}{@{\extracolsep{5pt}}cccccc@{}}
			\hline 
            & $\mathrm{\Gamma}$ & $\mathrm{N_H}$ & $\chi^2$ (dof) &  Binning & Binning\\
			&& $\mathrm{10^{22} cm^{-2}}$ & & OBSID 4968& OBSID 15648\\
					   			(1)&(2)&(3)& (4)& (5)&(6)\\ 
			
		\hline
		1  &  $6.43_{-3.96}^{+9.09}$  &  $0.78\leq{0.96}$  & 0.009 (2) & 5& - \\ [5pt]
		2       &  $2.18_{-0.31}^{+0.28}$   &  $0.60_{-0.12}^{+0.14}$  &  21.7 (21)   & 20 & 20\\ [5pt]
		3       &  $1.16_{-0.50}^{+0.60}$   &  $0.05\leq{0.17}$        &  4.5  (6)    & 5 & 2 \\ [5pt]	
		4       &  -                        &  -                       &  -    	      & - & - \\ [5pt]
		5  &  $3.00_{-1.33}^{+2.27}$  &  $0.45_{-0.39}^{+0.62}$  & 2.3  (5) & 5 & 2 \\ [5pt]		
		6       &  $1.81_{-0.35}^{+0.40}$   &  $0.21_{-0.13}^{+0.16}$  &  9.7  (11)   & 20 & 20\\ [5pt]
		7       &  $1.73_{-0.45}^{+0.52}$   &  $1.17_{-0.43}^{+0.55}$  &  15.4 (11)   & 20 & 20\\ [5pt]
		8       &  $2.22_{-0.28}^{+0.25}$   &  $0.56_{-0.13}^{+0.16}$  &  39.6 (21)   & 20 & 20\\ [5pt]
	    9  &  $4.39_{-2.68}^{+7.20}$  &  $0.62\leq{1.38}$     & 7.6  (4) & 5 & -\\ [5pt]			
		10      &  $8.50_{-5.70}^{+27.66}$  &  $1.09\leq{0.64}$        &  3.3  (3)    & 2  & -  \\ [5pt]
		11      &  -                        &  -                       &  -    	      & - & -\\ [5pt]
		12      &  $2.52_{-0.57}^{+0.60}$   &  $0.11\leq{0.16}$        &  91   (14)    & 5  & 2\\ [5pt]
		13      &  $4.68_{-1.16}^{+1.40}$   &  $0.50_{-0.24}^{+0.29}$  &  9.3  (16)   & 5 & 2\\ [5pt]
		14 &  $2.27_{-1.30}^{+2.28}$  &  $0.29\leq{0.94}$  & 0.9  (1)	 & 5 &-\\ [5pt]
		15      &  $5.04_{-3.16}^{+6.36}$   &  $0.74\leq{1.42}$        &  3.3  (3)    & 2 & -\\ [5pt]
		16      &  $2.59_{-0.75}^{+0.86}$   &  $0.43_{-0.26}^{+0.31}$  &  22.2 (12)   & 5 & 2\\ [5pt]
		17      &  $0.35_{-1.41}^{+1.32}$   &  $1.01\leq{4.09}$        &  0.7  (2)    & 5 & -\\ [5pt]
		18      &  $1.69_{-0.36}^{+0.71}$   &  $0.06\leq{0.26}$        &  16.8 (10)   & 5 & 2\\ [5pt]		
		19 &  $1.62_{-1.41}^{+1.82}$  &	 $0.26\leq{0.81}$&	   1.4(4) &	2 & -\\ [5pt]
		\hline				
				\end{tabular} 	
				\label{tab.spectralparam_ngc2276}
\smallskip
 {\az{     Column 1: source ID; column 2: photon index $\Gamma$ from spectral fitting; column 3: thermal plasma model temperatures (kT); column 4: line-of-sight hydrogen column density (\nh); column 5: \chisq of the spectral fit and corresponding degrees of freedom (d.o.f); and columns 6 and 7: spectral binning for OBSIDs 4968 and 15648  respectively.}}
\end{minipage}
				\end{table*}

\subsection{X-ray colours}\label{hardnessratios}

{\az{In order to characterize the spectra of the sources with too few counts for spectral analysis we calculated their X-ray colours. For a consistency check with the results from the spectral analysis we also calculated the X-ray colours for the sources for which we performed spectral fits.}}
The X-ray colours are defined as $\mathrm{C_1\equiv\log_{10} ({S}/{M})}$, $\mathrm{C_2\equiv\log_{10} ({M}/{H})}$, $\mathrm{C_3\equiv\log_{10} ({S}/{H})}$ where S, M, and H  are the net counts in the soft (0.5-1.2 \textrm{keV}), medium (1.2-2.0 \textrm{keV}), and hard (2.0-7.0 \textrm{keV}) bands.
 
We calculated the X-ray colours and their uncertainties for the 31 sources in NGC\,3310 and the 19 sources in NGC2276, using the Bayesian Estimation of Hardness Ratios\footnote{http://hea-www.harvard.edu/astrostat/behr/} tool \citet[][; BEHR]{park}. This tool evaluates the posterior probability distribution of the X-ray colours given the measured number of counts in the source and background apertures. {\az{As a result it}} provides reliable estimates and confidence limits even when either or both soft and hard counts are very low.  {\az{In addition }} it can account for effective area differences between the sources {\az{or between observations. For this correction we used the average exposure within the aperture of each source  (based on the broad-band exposure maps) normalized to the value at the center of each galaxy.}} The resulting X-ray colours and their corresponding  90\% confidence intervals, are presented in Tables  \ref{tab.colours_ngc3310} and \ref{tab.colours_ngc2276} for NGC\,3310 and NGC\,2276 respectively.

In order to estimate the spectral parameters of the X-ray sources {\az{from their X-ray colours}} we created grids on {\az{colour-colour}} plots and placed our sources on them. The grids were calculated by simulating absorbed power-law spectra for different values of the photon index ($\Gamma$) and the hydrogen column density ($\mathrm{N_H}$) for a fiducial source at the centre of each galaxy, {\az{the reference position we normalized the source counts to.}} Fig. \ref{fig.grid_ngc3310} and \ref{fig.grid_ngc2276} show the location of the sources on the $\mathrm{C_2-C_1}$ grid for NGC\,3310 and NGC\,2276 respectively.

We see that the majority of the sources for OBSID 2939 of NGC\,3310 fall on the region corresponding to $\mathrm{\Gamma\sim 1.5-2.5}$ and $\mathrm{N_H<0.85\times 10^{22} cm^{-2}}$, and for NGC\,2276 on the region of $\mathrm{\Gamma\sim 1.5-4.0}$ and $\mathrm{N_H<0.85\times 10^{22} cm^{-2}}$. However there are 4 sources for OBSID 2939 of NGC\,3310 (6,9,23,25), 11 sources for OBSID 19891 of NCG\,3310 (1, 5, 6, 11, 13, 15, 25, 26, 28, and 31), 7 sources for OBSID 4968 of NGC\,2276 (3, 9, 11, 13, 15, 17, and 19), and 9 sources for OBSID 15648 of NGC\,2276 (1, 3, 5, 11, 12, 13, 15, 17, and 18) that fall outside the grid. 

In the case of NGC\,3310 sources 5, 6, 9, and 23 are diffuse emission sources and therefore are not expected to be consistent with a grid based on a power-law model with $\mathrm{\Gamma\sim 0-4}$. For OBSID 2939, source 25 appears also to be very soft and with a high hydrogen column density and its spectrum, although of low quality, agrees with that values. For OBSID 2939 the remaining sources that are out of the grid agree within the errors with the expected values from the spectral parameters (see Table \ref{tab.specparam_galaxy3310}).

For NGC\,2276 sources 11, 13, and 15 for both OBSIDS, and sources 1, 5 for OBSID  15648 are positioned above the grid indicating relatively soft spectra ($\mathrm{\Gamma\geq3}$), after accounting for the uncertainties on the colours.
Source 17 for both OBSIDs and source 19 for OBSID 4968 appear to have hard spectra and moderate absorption, therefore, their position at the lower right (``softer colour") corner of the diagram could be attributed to an additional soft component (e.g. local diffuse emission) which cannot be recovered spectrally due to the poor quality of their spectra. 
Sources 3 and 9 for OBSID 4968 seem to have photon indices of about 0.7 and 2.5 respectively but very low absorption. Very low absorption also show sources 3, 12, and 18 with photon indices of 1.5, 1.7, and 0.8 respectively. For the rest of the sources we estimate their spectral parameters based on their location on the grid and we present these estimates in Tables \ref{tab.spectralparam_ngc3310} and \ref{tab.spectralparam_ngc2276}. 

We compared the {\az{X-ray colour based}} spectral parameters, 
to those calculated from the spectral fits (Fig. \ref{fig.gridspectrum3310} and Fig.\ref{fig.gridspectrum2276}), and we found that they agree well within the errors. We did not include sources not fitted with a single absorbed power-law model as well as for the sources not falling on the grids.

\begin{table*}
	\centering
	\begin{minipage}{140mm}
		\caption{NGC\,3310 X-ray colours of the discrete sources.}
		\hskip-2.5cm
		\setlength{\tabcolsep}{3pt}
		\begin{tabular}[!htbp]{@{\extracolsep{5pt}}crrrccrrrcc@{}}
			\hline 
			&\multicolumn{5}{c}{OBSID 2939}&\multicolumn{5}{c}{OBSID 19891}\\
			\cline{2-6}
			\cline{7-11}
			&&&&&&&&&&\\
			Src ID 
            & $\mathrm{C_1}$ &  $\mathrm{C_2}$ & $\mathrm{C_3}$ & $\mathrm{\Gamma}$  &  $\mathrm{N_{H}}$ & $\mathrm{C_1}$ &  $\mathrm{C_2}$ & $\mathrm{C_3}$ & $\mathrm{\Gamma}$  &  $\mathrm{N_{H}}$ \\
			&          &&& &$\mathrm{10^{22} cm^{-2}}$ &         &    & & &$\mathrm{10^{22} cm^{-2}}$ \\
			(1)&(2)&(3)& (4)& (5)& (6)& (7) &(8)& (9)&(10)&(11)\\ 
	\hline	

1 & $ 0.04 \pm 0.19$ & $ 0.18 \pm 0.22$ & $ 0.23 _{-0.21}^{+0.22}$     &    2.2  & 0.20     &  $ -0.55 _{-0.47}^{+0.44}$ & $ 0.56 _{-0.46}^{+0.5}$ & $ 0.01 _{-0.59}^{+0.6}$           &       4.5  		& 1.0   \\ [5pt]	    
2 & $ 0.31 \pm 0.21$ & $ 0.19 \pm 0.28$ & $ 0.51 _{-0.24}^{+0.25}$	   &	 2.0 &   0.011  &  $ -0.76 _{-0.84}^{+0.76}$ & $ -0.12 \pm 0.39$ & $ -0.88 _{-0.84}^{+0.73}$       &       2.0		& 0.9   \\ [5pt]		    
3 & $ -0.3 _{-0.4}^{+0.38}$ & $ 0.16 _{-0.35}^{+0.36}$ & $ -0.13 _{-0.43}^{+0.42}$	   &	 2.5 &   0.60   &  $ -0.65 _{-0.8}^{+0.73}$ & $ 0.11 _{-0.48}^{+0.5}$ & $ -0.54 _{-0.84}^{+0.76}$	        &        2.7  	& 0.9   \\ [5pt]		    
4 & $ 0.29 \pm 0.05$ & $ 0.24 \pm 0.07$ & $ 0.53 \pm 0.07$	   &	  2.2&    0.1   &  $ -0.12 \pm 0.07$ & $ 0.3 _{-0.08}^{+0.09}$ & $ 0.18 \pm 0.09$	        &       2.7  		& 0.30  \\ [5pt]	    
5 & $ 0.03 \pm 0.33$ & $ 0.58 _{-0.5}^{+0.56}$ & $ 0.62 _{-0.54}^{+0.56}$	   &	4.0  &  0.60    &  $ 0.28 _{-0.52}^{+0.56}$ & $ -0.21 _{-0.54}^{+0.54}$ & $ 0.08 \pm 0.43$	        &     $\sim$ 0.7 	& -     \\ [5pt]	    
6 & $ 1.03 _{-0.8}^{+1.07}$ & $ -0.27 _{-1.38}^{+1.15}$ & $ 0.74 _{-0.64}^{+0.75}$	   &	-    &	-	&  $ 0.0 \pm 2.22$ & $ 0.0 \pm 2.22$ & $ 0.0 \pm 2.22$	        &      $\sim$ 1.5 	& -     \\ [5pt]		    
7 & $ -0.09 \pm 0.11$ & $ 0.11 \pm 0.1$ & $ 0.02 \pm 0.11$	   &	2.0  &  0.30    &  $ -0.23 _{-0.22}^{+0.21}$ & $ 0.18 \pm 0.21$ & $ -0.04 \pm 0.24$        &      2.5   		& 0.3   \\ [5pt]	    
8 & $ -0.52 \pm 0.16$ & $ 0.11 _{-0.09}^{+0.1}$ & $ -0.4 \pm 0.16$	   &	2.5  &  0.90	&  $ -1.18 _{-1.28}^{+0.72}$ & $ -0.01 \pm 0.16$ & $ -1.18 _{-1.29}^{+0.71}$       &     3.5		& 2.0   \\ [5pt]		    
9 & $ -0.79 _{-1.28}^{+0.75}$ & $ 0.54 _{-0.39}^{+0.42}$ & $ -0.27 _{-1.36}^{+0.89}$   &	-    & - 	&  $ 0.0 \pm 1.21$ & $ -0.63 _{-0.83}^{+0.62}$ & $ -0.63 _{-0.83}^{+0.62}$	        &     -		& -     \\ [5pt]		    
10 & $ -0.51 _{-0.14}^{+0.13}$ & $ -0.22 \pm 0.07$ & $ -0.73 \pm 0.13$ &	1.2  &  0.60 	&   $ -0.81 _{-0.47}^{+0.39}$ & $ -0.21 \pm 0.14$ & $ -1.01 _{-0.47}^{+0.39}$      &     1.7		& 0.9   \\ [5pt]		    
11 & $ 0.05 \pm 0.14$ & $ -0.04 _{-0.13}^{+0.14}$ & $ 0.0 \pm 0.14$	   &	1.2  &  0.011	&   $ -0.07 \pm 0.22$ & $ -0.06 _{-0.19}^{+0.2}$ & $ -0.13 \pm 0.21$       &     $\sim$ 1.5	& -     \\ [5pt]		    
12 & $ 0.06 \pm 0.15$ & $ 0.21 \pm 0.16$ & $ 0.27 \pm 0.17$	   &	2.2  &  0.25 	&   $ -0.25 _{-0.44}^{+0.39}$ & $ 0.26 _{-0.29}^{+0.29}$ & $ 0.01 _{-0.46}^{+0.41}$        &     2.7  		& 0.4   \\ [5pt]		    
13 & $ 0.19 _{-0.43}^{+0.42}$ & $ 0.43 _{-0.52}^{+0.56}$ & $ 0.62 _{-0.56}^{+0.6}$	   &	3.0  &  0.30 	&   $ 0.85 _{-0.79}^{+1.29}$ & $ 0.04 _{-1.5}^{+1.13}$ & $ 0.9 _{-0.63}^{+0.7}$	        &     $\sim$ 1.5 	& -     \\ [5pt]	    
14 & $ -0.27 _{-1.49}^{+0.99}$ & $ -0.43 _{-0.52}^{+0.47}$ & $ -0.71 _{-1.3}^{+0.81}$  &	0.25 &  0.1 	&   $ -1.11 _{-1.64}^{+1.15}$ & $ 0.11 _{-0.54}^{+0.55}$ & $ -0.99 _{-1.64}^{+1.15}$       &     4.0  		& 2.0   \\ [5pt]		    
15 & $ -0.19 _{-0.41}^{+0.34}$ & $ 0.03 \pm 0.19$ & $ -0.16 _{-0.41}^{+0.35}$  &	1.7  &  0.35	&   $ -0.04 _{-0.63}^{+0.56}$ & $ -0.28 _{-0.35}^{+0.32}$ & $ -0.32 _{-0.52}^{+0.41}$      &     $\sim$ 0.7 	& -     \\ [5pt]		    
16 & $ -0.39 \pm 0.06$ & $ -0.09 \pm 0.04$ & $ -0.48 \pm 0.06$ &	1.5  &  0.5  	&   $ -0.6 _{-0.33}^{+0.3}$ & $ -0.63 _{-0.12}^{+0.11}$ & $ -1.23 _{-0.32}^{+0.27}$        &     0.0  		& 0.13  \\ [5pt]		    
17 & $ -0.32 \pm 0.06$ & $ -0.04 \pm 0.05$ & $ -0.37 \pm 0.06$ &	1.7  &  0.45 	&   $ -0.42 _{-0.12}^{+0.11}$ & $ 0.02 _{-0.08}^{+0.09}$ & $ -0.4 \pm 0.12$        &     2.0		& 0.45  \\ [5pt]		    
18 & $ -0.07 \pm 0.1$ & $ 0.11 \pm 0.1$ & $ 0.04 _{-0.11}^{+0.1}$	   &	2.0  &  0.30 	&   $ -0.48 \pm 0.11$ & $ -0.05 _{-0.07}^{+0.08}$ & $ -0.53 \pm 0.11$      &     1.7		& 0.45  \\ [5pt]		    
19 & $ -0.36 \pm 0.18$ & $ 0.02 \pm 0.12$ & $ -0.34 \pm 0.18$  &	2.0  &  0.60 	&   $ -1.25 _{-1.29}^{+0.72}$ & $ -0.04 \pm 0.16$ & $ -1.29 _{-1.29}^{+0.72}$      &     3.5		& 2.0   \\ [5pt]		    
20 & $ -1.13 _{-0.58}^{+0.42}$ & $ -0.42 _{-0.09}^{+0.08}$ & $ -1.54 _{-0.59}^{+0.41}$ &	1.0  &  1.6  	&   $ -1.07 _{-0.91}^{+0.56}$ & $ -0.2 \pm 0.14$ & $ -1.28 _{-0.91}^{+0.56}$       &     2.2		& 1.5   \\ [5pt]		    
21 & $ 0.35 \pm 0.16$ & $ 0.3 _{-0.24}^{+0.25}$ & $ 0.65 _{-0.22}^{+0.24}$	   &	2.4  &  0.011	&   $ -0.41 _{-0.09}^{+0.08}$ & $ -0.03 \pm 0.06$ & $ -0.43 \pm 0.08$      &     1.7		& 0.35  \\ [5pt]		    
22 & $ 0.12 _{-0.2}^{+0.21}$ & $ -0.02 \pm 0.21$ & $ 0.1 \pm 0.2$	   &	1.2  &  0.011	&   $ -0.51 _{-0.46}^{+0.42}$ & $ -0.04 _{-0.26}^{+0.24}$ & $ -0.55 _{-0.46}^{+0.4}$       &     1.7		& 0.45  \\ [5pt]		    
23 & $ 0.51 _{-0.44}^{+0.5}$ & $ 0.8 _{-1.07}^{+1.64}$ & $ 1.34 _{-0.92}^{+1.45}$	   &	4.5  &  0.23	&   $ 0.3 _{-1.6}^{+1.84}$ & $ 0.5 _{-1.95}^{+2.07}$ & $ 0.8 _{-1.68}^{+2.03}$	        &     3.2		& 0.01  \\ [5pt]		    
24 & $ -0.04 \pm 0.06$ & $ 0.1 \pm 0.06$ & $ 0.06 \pm 0.06$	   &	1.8  &  0.23 	&   $ -0.36 _{-0.11}^{+0.1}$ & $ 0.07 \pm 0.08$ & $ -0.28 \pm 0.11$        &     2.2		& 0.35  \\ [5pt]		    
25 & $ -1.76 _{-1.57}^{+1.07}$ & $ 0.05 _{-0.21}^{+0.2}$ & $ -1.72 _{-1.57}^{+1.11}$   &	-    &	-	&   $ -0.08 _{-1.77}^{+1.81}$ & $ -0.85 _{-1.32}^{+0.79}$ & $ -0.93 _{-1.26}^{+0.78}$      &     -		& -     \\ [5pt]		    
26 & $ -0.3 _{-0.89}^{+0.66}$ & $ 0.35 _{-0.52}^{+0.54}$ & $ 0.05 _{-1.02}^{+0.83}$	   &	3.5  &  0.84 	&   $ 0.11 _{-1.11}^{+1.01}$ & $ -0.34 _{-0.78}^{+0.74}$ & $ -0.23 _{-1.02}^{+0.83}$       &     $\sim$ 0.5	& -     \\ [5pt]		    
27 & $ -0.03 _{-0.17}^{+0.16}$ & $ 0.42 \pm 0.22$ & $ 0.38 \pm 0.22$   &	3.2  &  0.55 	&   $ -0.08 \pm 0.26$ & $ 0.06 _{-0.26}^{+0.25}$ & $ -0.03 _{-0.27}^{+0.26}$       &     1.7  		& 0.01  \\ [5pt]		    
28 & $ 0.08 _{-0.44}^{+0.43}$ & $ 0.11 _{-0.5}^{+0.48}$ & $ 0.17 _{-0.47}^{+0.48}$	   &	1.7  &  0.13 	&   $ 0.24 _{-0.44}^{+0.46}$ & $ 0.01 _{-0.52}^{+0.51}$ & $ 0.26 \pm 0.46$	        &     $\sim$ 1.5	& -     \\ [5pt]	    
29 & $ 0.08 \pm 0.06$ & $ 0.12 \pm 0.07$ & $ 0.2 \pm 0.07$	   &	1.7  &  0.13 	&   $ -0.21 \pm 0.09$ & $ -0.06 \pm 0.08$ & $ -0.26 \pm 0.09$      &     1.5  		& 0.13  \\ [5pt]		    
30 & $ 0.03 _{-0.28}^{+0.27}$ & $ 0.42 _{-0.38}^{+0.4}$ & $ 0.46 _{-0.38}^{+0.39}$	   &	3.2  &  0.45 	&   $ 0.07 _{-0.39}^{+0.4}$ & $ 0.32 _{-0.54}^{+0.55}$ & $ 0.39 _{-0.52}^{+0.54}$	        &     2.7		& 0.13  \\ [5pt]	    
31 & $ 0.05 _{-0.59}^{+0.58}$ & $ 0.56 _{-0.91}^{+1.26}$ & $ 0.62 _{-0.97}^{+1.33}$    &	 3.7 &	0.50	&   $ 0.12 _{-0.43}^{+0.44}$ & $ -0.01 \pm 0.46$ & $ 0.12 \pm 0.43$        &     $\sim$ 1.5	& -     \\ [5pt]                      
   \hline			
		\end{tabular} 	
		\label{tab.colours_ngc3310}
        \smallskip
{\az{       Column 1: Source ID; columns 2, 3, 4, 7, 8, and 9: X-ray colours, and their corresponding uncertainties for OBSIDs 2939 and 19891, defined as $C_1=\log_{10} ({S}/{M})$, $C_2=\log_{10} ({M}/{H})$, $C_3=\log_{10} ({S}/{H})$ where S, M, and H are the net counts in the soft (0.5-1.2 \textrm{keV}), medium (1.2-2.0 \textrm{keV}), and hard (2.0-7.0 \textrm{keV}) bands respectively (see text for details). Columns 5, 6, 10, and 11: photon index $\mathrm{\Gamma}$ and hydrogen column density \nh\ based on the X-ray colours for OBSIDs 2939 and 19891.}}
	\end{minipage}
\end{table*}

\begin{table*}
	\centering
	\begin{minipage}{140mm}
		\caption{NGC\,2276 X-ray colours of the discrete sources.}
		\hskip-2.5cm
		\setlength{\tabcolsep}{3pt}
		\begin{tabular}[!htbp]{@{\extracolsep{5pt}}crrrccrrrcc@{}}
			\hline 
			&\multicolumn{5}{c}{OBSID 4968}&\multicolumn{5}{c}{OBSID 15648}\\
			\cline{2-6}
			\cline{7-11}
			&&&&&&&&&&\\
			Src ID
            & $\mathrm{C_1}$ &  $\mathrm{C_2}$ & $\mathrm{C_3}$ & $\mathrm{\Gamma}$  &  $\mathrm{N_{H}}$ & $\mathrm{C_1}$ &  $\mathrm{C_2}$ & $\mathrm{C_3}$ & $\mathrm{\Gamma}$  &  $\mathrm{N_{H}}$ \\
			&          &&& &$\mathrm{10^{22} cm^{-2}}$ &         &    & & &$\mathrm{10^{22} cm^{-2}}$ \\
			(1)&(2)&(3)& (4)& (5)& (6)& (7) &(8)& (9)&(10)&(11)\\ 
	\hline	
	1  &  	 $0.31_{-0.39}^{+0.40}$    &       $0.31 _{-0.60}^{+0.63}$      &     $0.62_{-0.52}^{+0.58}$   &  2.3		& 0.1   & $-0.55_{-0.86}^{+0.72}$ &  $0.84_{-0.88}^{+1.11}$ &  $0.27_{-1.3}^{+1.45}$     &     -    &   1.0   \\ [5pt]
	2  &     $-0.33_{-0.1}^{+0.09}$    &       $0.11 \pm 0.08$      &     $-0.22\pm 0.1$  &  2.2 		& 0.5 	& $-0.31_{-0.21}^{+0.22}$ &  $-0.02\pm 0.17$ &  $-0.32_{-0.22}^{+0.21}$  &    1.7   &   0.3   \\ [5pt]
	3  &     $0.2  _{-0.35}^{+0.34}$    &       $-0.24_{-0.34}^{+0.31}$      &     $-0.04_{-0.29}^{+0.28}$  &  0.7 		& -	& $0.11_{-1.41}^{+1.26}$ &  $0.05_{-1.26}^{+1.34}$ &  $0.15_{-1.46}^{+1.42}$	   &  1.5   &   -     \\ [5pt]  
	4  &     $-0.34_{-1.68}^{+1.22}$    &       $0.42 _{-1.18}^{+1.56}$      &     $0.07_{-1.98}^{+1.87}$   &  4.0 		& 0.85 	& $-0.62_{-0.77}^{+0.6}$ &  $0.15\pm 0.34$ &  $-0.46_{-0.79}^{+0.63}$	   & 3.0	   &   0.9   \\ [5pt]  
	5  &  	 $-0.21_{-0.31}^{+0.3}$    &       $0.47 _{-0.36}^{+0.39}$      &     $0.26 \pm 0.42$  &  4.0 		& 0.8 	& $-0.36_{-0.91}^{+0.72}$ &  $0.8_{-0.88}^{+1.15}$ &  $0.42_{-1.34}^{+1.53}$	   & -	   &   0.9   \\ [5pt]  
	6  &     $0.0  \pm 0.13$    &       $0.03 _{-0.13}^{+0.14}$      &     $0.04 \pm 0.14$  &  1.5 		& 0.13	& $-0.23_{-0.22}^{+0.21}$ &  $0.04\pm 0.19$ &  $-0.19\pm 0.22$   &   1.8	   &   0.3   \\ [5pt]
	7  &  	 $-0.98_{-0.95}^{+0.58}$    &       $-0.25\pm 0.15$      &     $-1.23_{-0.94}^{+0.58}$  &  2.0 		&  1.2 	& $-0.97_{-0.32}^{+0.31}$ &  $-0.27\pm 0.12$ &  $-1.23_{-0.32}^{+0.3}$   &   1.7	   &   1.0   \\ [5pt]
	8  & 	 $-0.27\pm 0.11$    &       $0.02 \pm 0.09$      &     $-0.26\pm 0.11$  &  1.9 		& 0.45 	& $-0.37\pm 0.17$ &  $0.03\pm 0.13$ &  $-0.34\pm 0.17$   &   2.0	   &   0.45  \\ [5pt]
	9  &  	 $0.58 _{-0.56}^{+0.6}$    &       $0.31 _{-0.96}^{+0.99}$      &     $0.88 _{-0.76}^{+0.88}$  &  2.2 		& -	& $0.16_{-0.79}^{+0.81}$ &  $0.54_{-1.1}^{+1.56}$ &  $0.7_{-1.07}^{+1.56}$	   & 3.5	   &   0.1   \\ [5pt]  
	10  &  	 $0.0_{-0.6}^{+0.62}$      &       $0.5  _{-0.92}^{+1.15}$      &     $0.5_{-0.92}^{+1.15}$    &  3.5 		& 0.5 	&  $-0.31_{-1.15}^{+0.99}$ &  $0.34_{-1.03}^{+1.26}$ &  $0.04_{-1.42}^{+1.49}$   &   3.5	   &   0.6   \\ [5pt]
	11  &    $-0.27_{-0.84}^{+0.73}$    &       $0.68_{-0.91}^{+1.30}$        &     $-0.42_{-1.26}^{+1.57}$ &  $\sim$4.5 	& 0.85	&  $-0.96_{-1.15}^{+0.84}$ &  $0.34_{-0.47}^{+0.5}$ &  $-0.61_{-1.22}^{+0.96}$   &   -	   &   2.0   \\ [5pt]
	12 &  	 $-0.16\pm 0.23$    &       $0.26 \pm 0.25$      &     $0.09 _{-0.27}^{+0.28}$  &  2.6 		&0.45	&  $0.27\pm 0.19$ &  $0.12\pm 0.24$ &  $0.39\pm 0.21$	   & 1.7	   &   -     \\ [5pt]  
	13 & 	 $0.19 \pm 0.19$    &       $0.9  _{-0.47}^{+0.55}$      &     $1.1  _{-0.47}^{+0.54}$  &  $\sim$5.0	& 0.5 	&  $0.09\pm 0.24$ &  $0.9_{-0.52}^{+0.59}$ &  $0.99_{-0.52}^{+0.58}$	   & -	   &   0.6   \\ [5pt]  
	14 &     $-0.11_{-0.54}^{+0.51}$    &       $0.09 \pm 0.46$      &     $-0.03\pm 0.55$  &  2.0 		& 0.3 	&  $-0.36_{-1.25}^{+0.95}$ &  $0.28_{-0.89}^{+1.01}$ &  $-0.09_{-1.53}^{+1.37}$  &   3.0	   &   0.6   \\ [5pt]
	15 &     $-0.09_{-0.5}^{+0.48}$    &       $0.73 _{-0.84}^{+1.03}$      &     $0.61 _{-0.84}^{+1.07}$  &  $\sim$4.5 	& 0.8 	&  $-0.04_{-1.3}^{+1.26}$ &  $0.65_{-1.53}^{+1.91}$ &  $0.61_{-1.61}^{+1.99}$	   & -	   &   0.6   \\ [5pt]  
	16 &     $-0.15_{-0.58}^{+0.56}$    &       $0.57 _{-0.76}^{+0.92}$      &     $0.42 _{-0.84}^{+0.96}$  &  $\sim$4.0 	& 0.8 	&  $-0.17_{-0.38}^{+0.38}$ &  $0.2\pm 0.39$ &  $0.03_{-0.43}^{+0.42}$	   & 2.0	   &   0.3   \\ [5pt]  
	17 &     $0.57 _{-0.88}^{+0.96}$    &       $-1.22_{-0.92}^{+0.69}$      &     $-0.67_{-0.44}^{+0.4}$  &  2.0  	& - 	&  $0.0\pm 2.22$ &  $-0.61_{-1.99}^{+1.61}$ &  $-0.61_{-1.99}^{+1.61}$   &   -	   &   -     \\ [5pt]
	18 &     $-0.11_{-0.31}^{+0.3}$    &       $0.27 \pm 0.34$      &     $0.15 _{-0.34}^{+0.36}$  &  2.6 		& 0.45	&  $-0.04\pm 0.32$ &  $-0.22_{-0.28}^{+0.27}$ &  $-0.26\pm 0.28$ &   0.8	   &	-    \\ [5pt]
	19 &     $1.03 _{-1.11}^{+1.57}$    &       $-0.8 _{-1.64}^{+1.11}$      &     $0.21 \pm 0.62$  &  -  		& -   	&  $0.0_{-0.81}^{+0.81}$ &  $0.19_{-1.05}^{+1.28}$ &  $0.19_{-1.05}^{+1.28}$     &   2.2   &	0.1  \\ [5pt]	
	\hline			
		\end{tabular} 	
		\label{tab.colours_ngc2276}
        \smallskip
        {\az{       Column 1: Source ID; columns 2, 3, 4, 7, 8, and 9: X-ray colours, and their corresponding uncertainties for OBSIDs 4968 and 15648, defined as $C_1=\log_{10} ({S}/{M})$, $C_2=\log_{10} ({M}/{H})$, $C_3=\log_{10} ({S}/{H})$ where S, M, and H are the net counts in the soft (0.5-1.2 \textrm{keV}), medium (1.2-2.0 \textrm{keV}), and hard (2.0-7.0 \textrm{keV}) bands respectively (see text for details). Columns 5, 6, 10, and 11: photon index $\mathrm{\Gamma}$ and hydrogen column density \nh\ based on the X-ray colours for OBSIDs 4968 and 15648.}}
	\end{minipage}
\end{table*}

\begin{figure*}	
\begin{minipage}{130mm}
	\begin{subfigure}[!hb]{0.475\textwidth}
		\resizebox{\hsize}{!}{\includegraphics[scale=1.0]{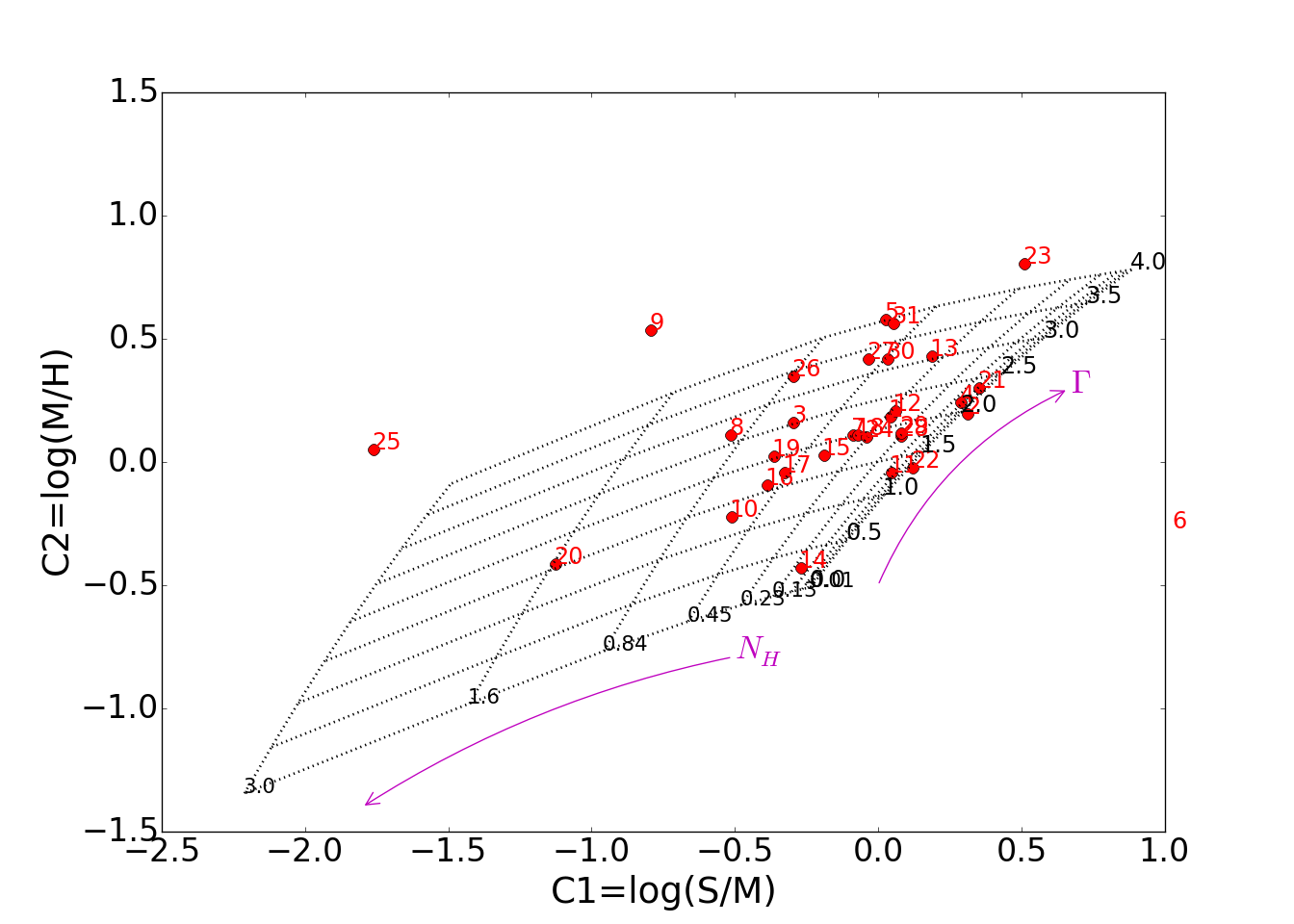}}
		\end{subfigure}
		        \hfill
	\begin{subfigure}[!hb]{0.475\textwidth}       		\resizebox{\hsize}{!}{\includegraphics[scale=1.0]{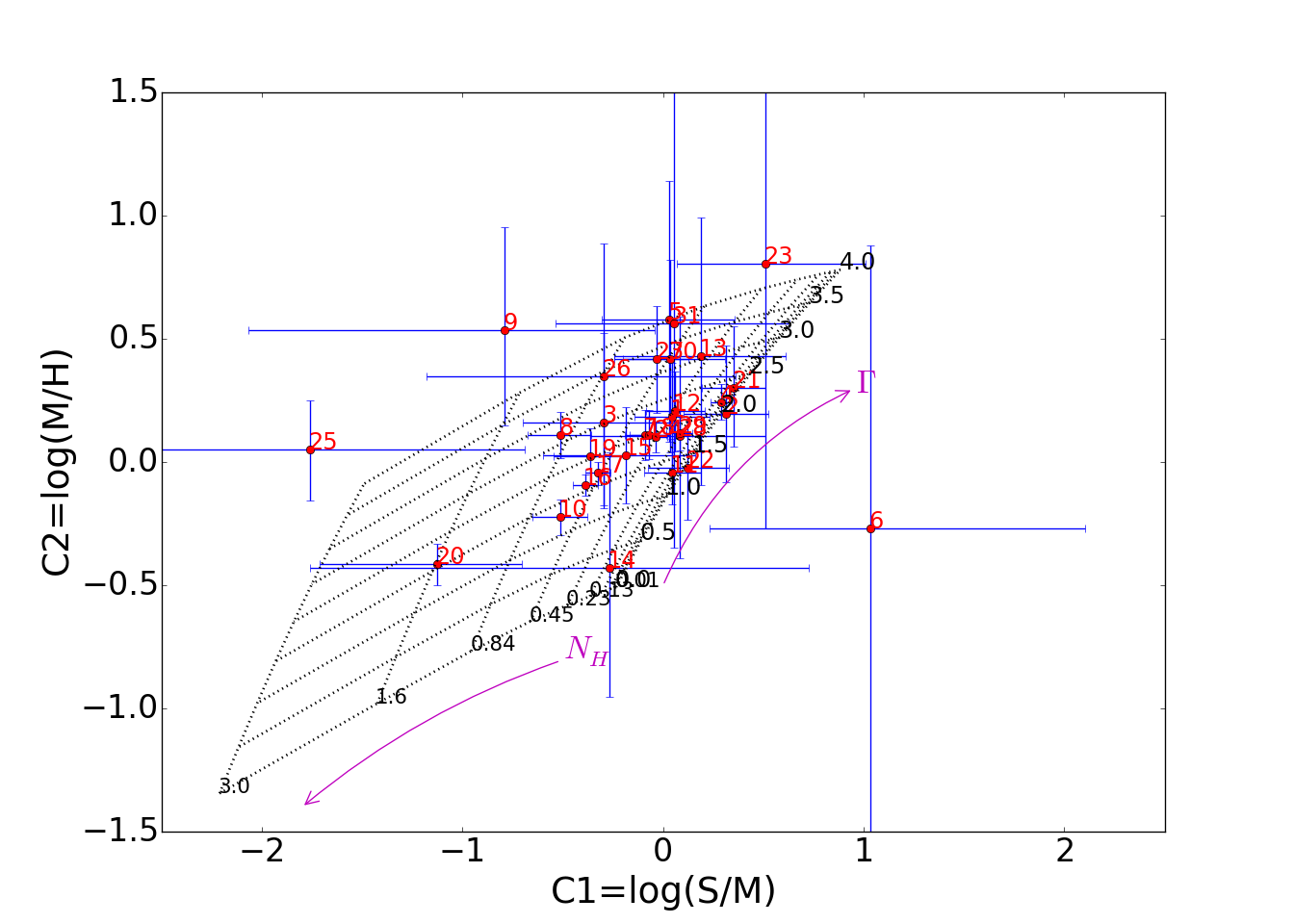}}
	\end{subfigure}
		        \vskip\baselineskip
	\begin{subfigure}[!hb]{0.475\textwidth}      
		\resizebox{\hsize}{!}{\includegraphics[scale=1.0]{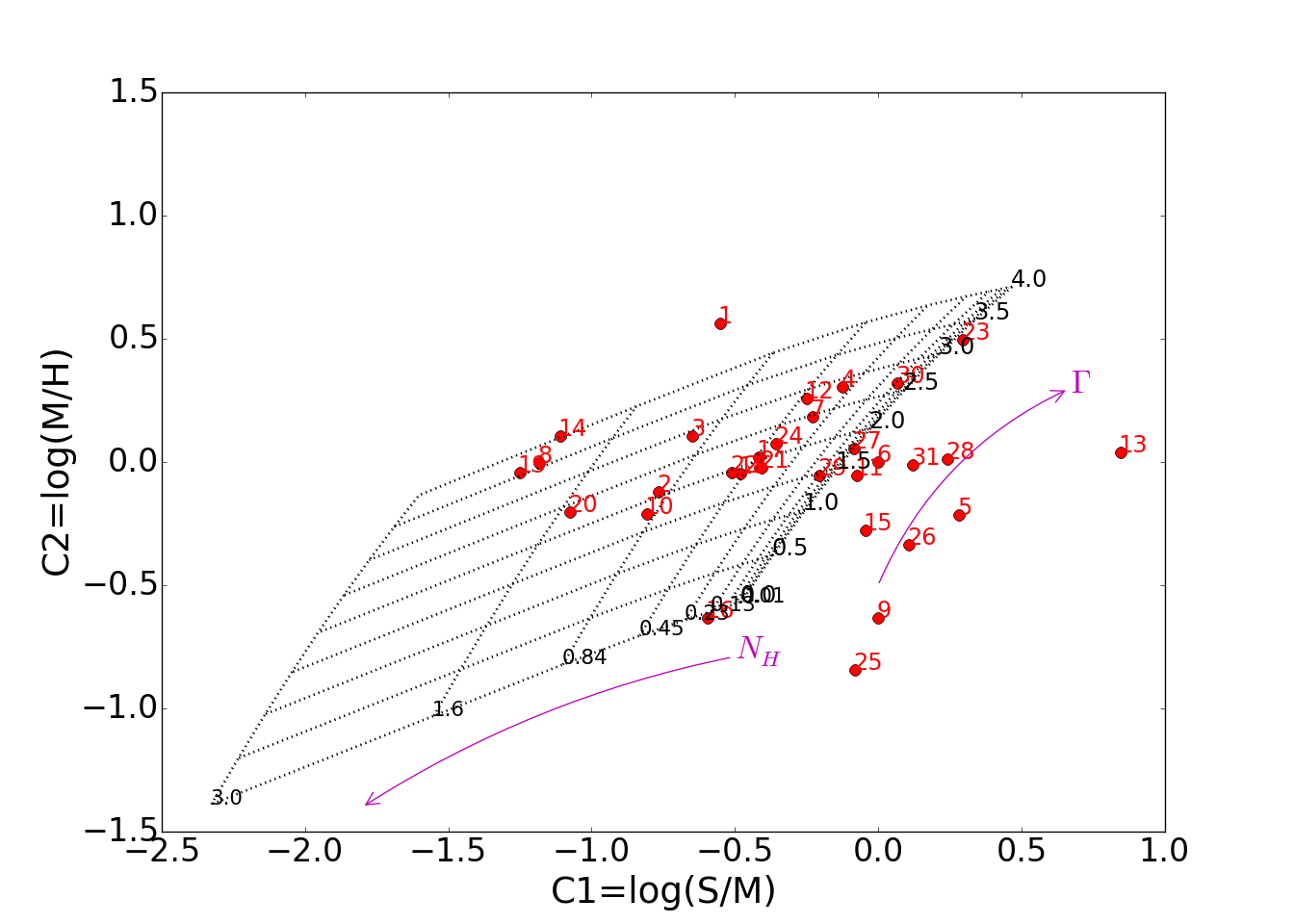}}
	 \end{subfigure}
		        \quad
	 \begin{subfigure}[!hb]{0.475\textwidth}   
		\resizebox{\hsize}{!}{\includegraphics[scale=1.0]{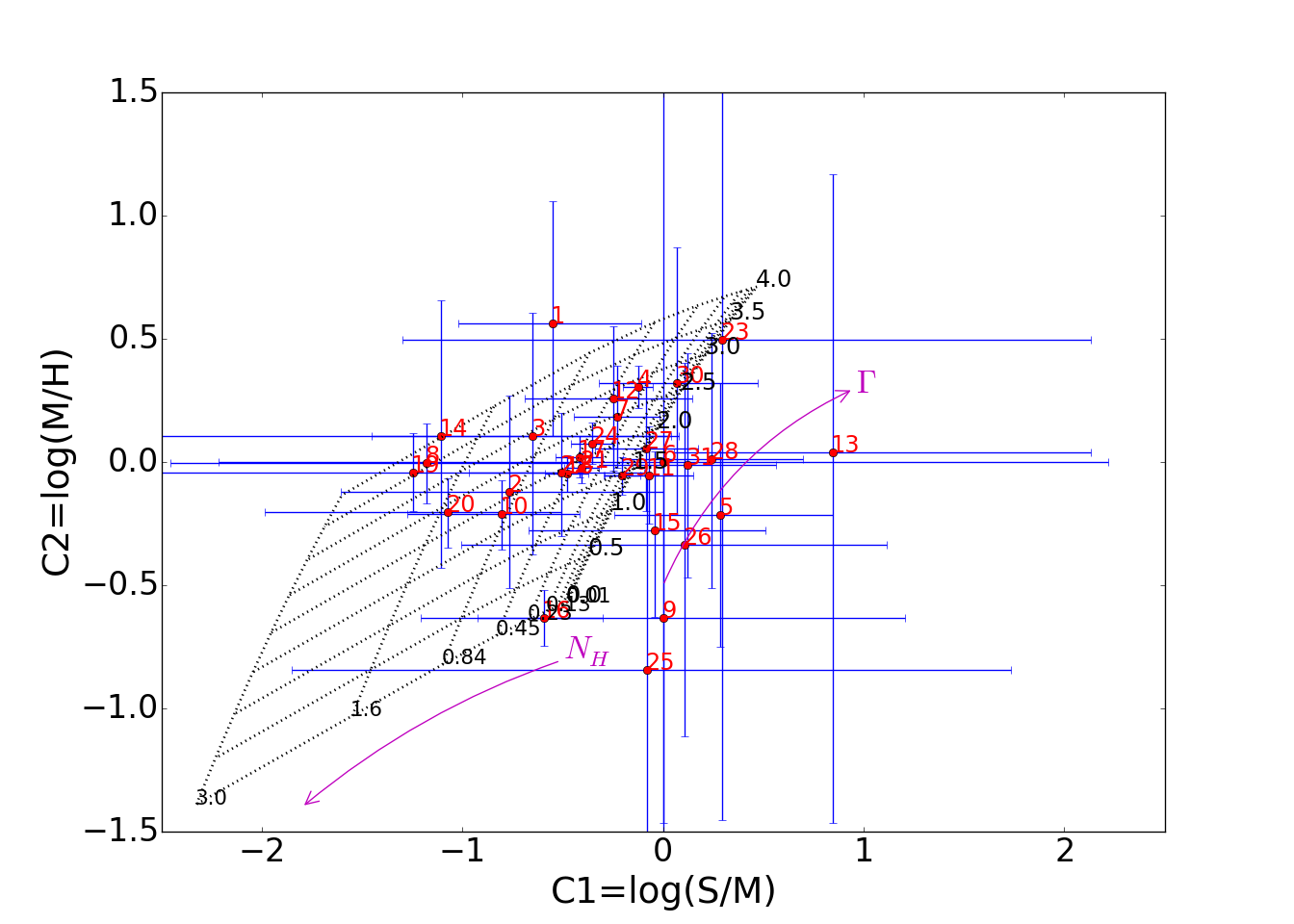}}	
		\end{subfigure}
		\caption{Colour-colour diagram for the X-ray sources detected in NGC\,3310 showing the colours (left) and corresponding errors (right) for OBSID 2939 (top) and OBSID 19891 (bottom). The source numbers refer to the source ID in Table \ref{tab.propertiesngc3310}. A grid showing the expected colours for absorbed power-law spectra is also overlaid. The value of the photon index and the hydrogen column density (in units of $\mathrm{10^{22}\,cm^{-2}}$) are shown at the edge of the grid.}
	\label{fig.grid_ngc3310}	
\end{minipage}
\end{figure*}

\begin{figure*}
\begin{minipage}{130mm}
	\begin{subfigure}[!hb]{0.475\textwidth}
		\resizebox{\hsize}{!}{\includegraphics[scale=1.0]{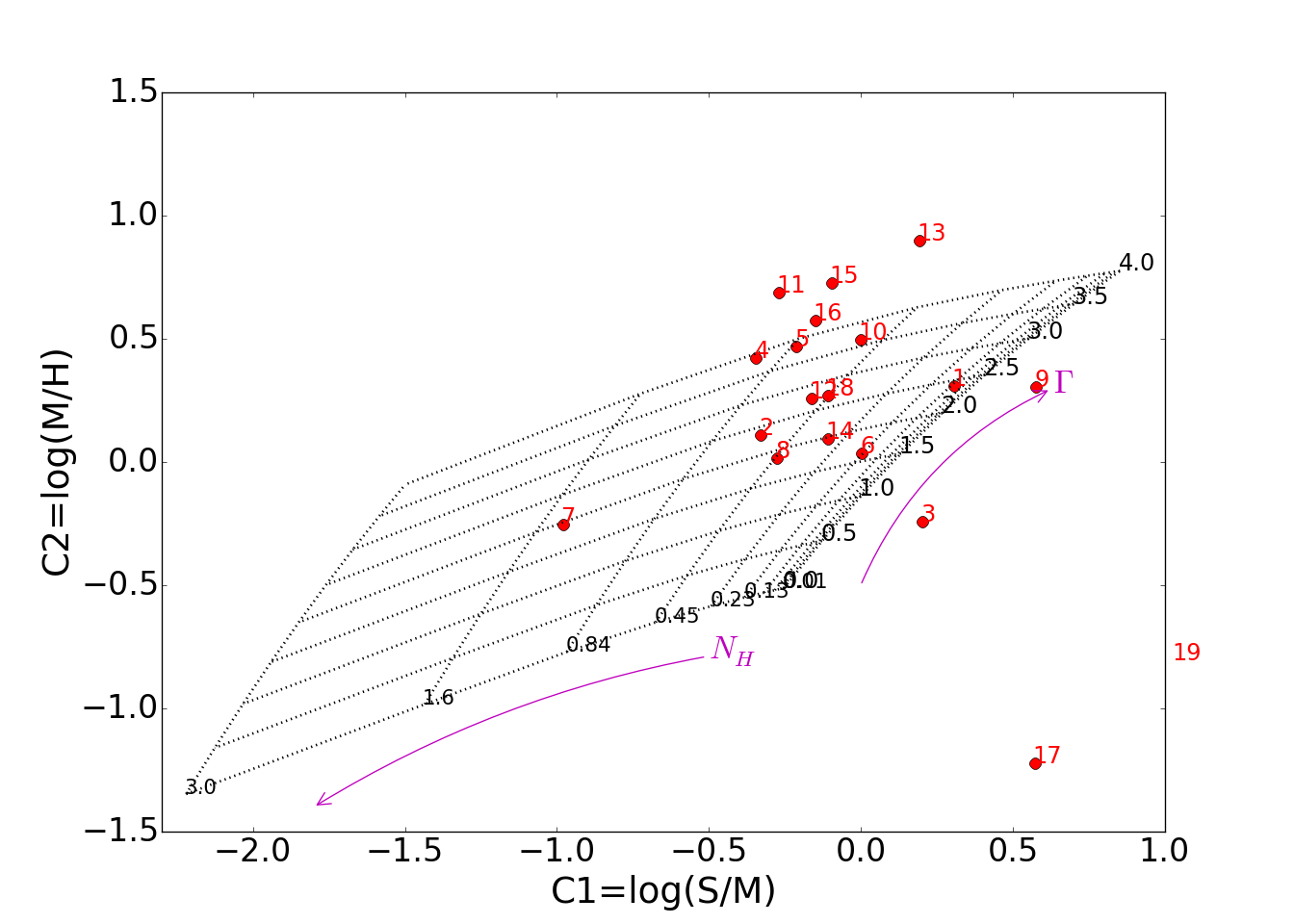}}
		\end{subfigure}
		        \hfill
	\begin{subfigure}[!hb]{0.475\textwidth}       		\resizebox{\hsize}{!}{\includegraphics[scale=1.0]{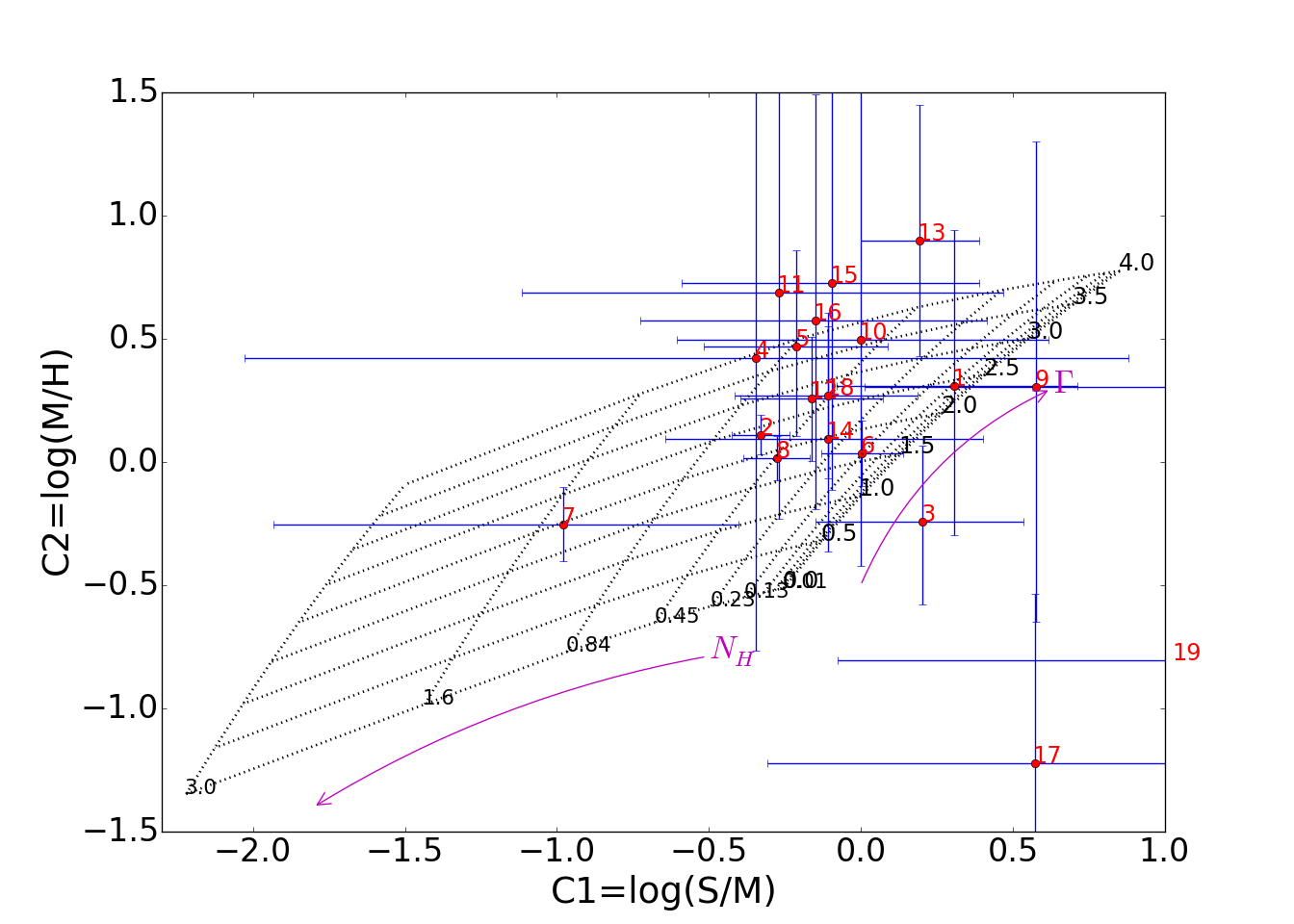}}
	\end{subfigure}
		        \vskip\baselineskip
	\begin{subfigure}[!hb]{0.475\textwidth}      
		\resizebox{\hsize}{!}{\includegraphics[scale=1.0]{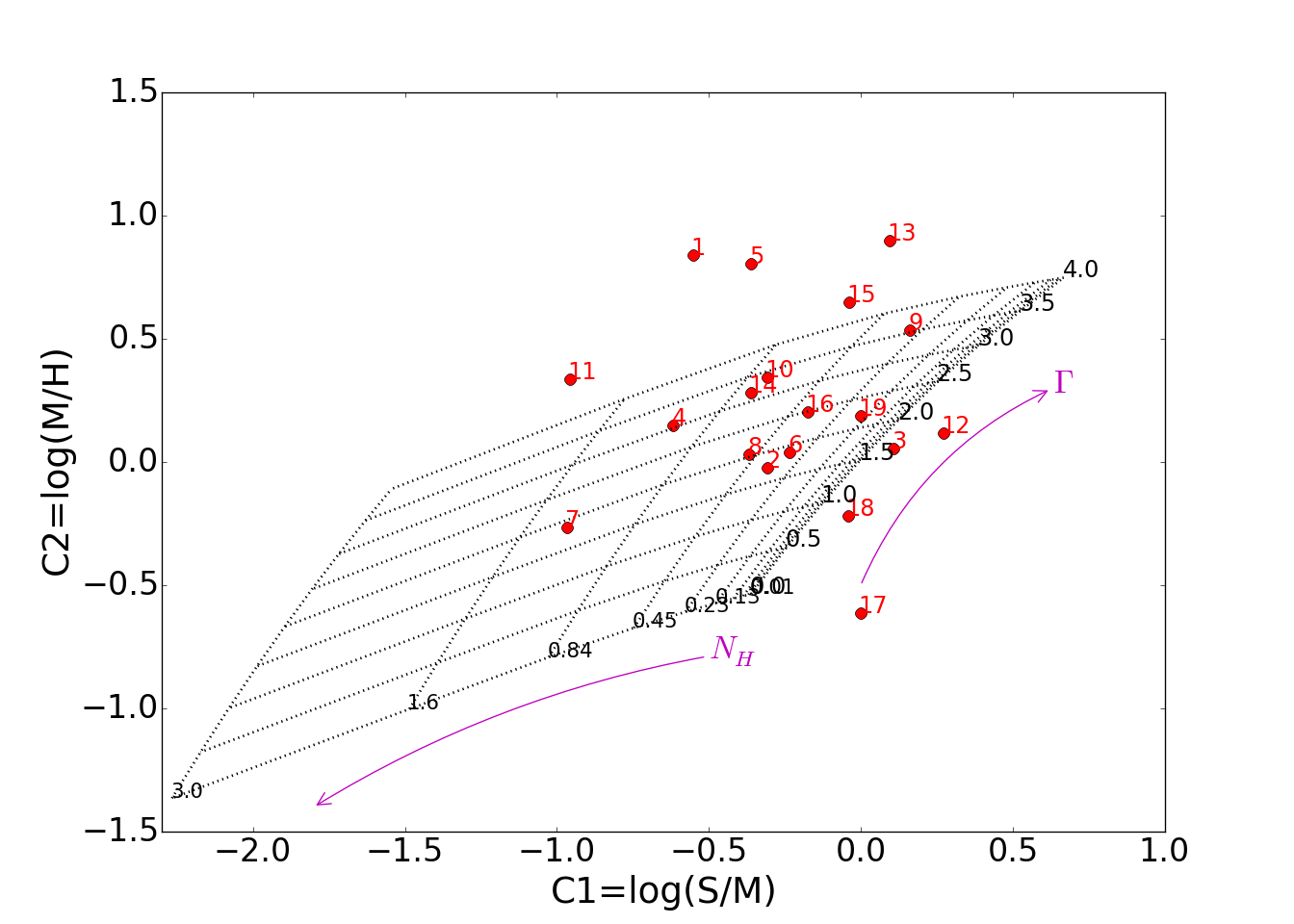}}
	 \end{subfigure}
		        \quad
	 \begin{subfigure}[!hb]{0.475\textwidth}   
		\resizebox{\hsize}{!}{\includegraphics[scale=1.0]{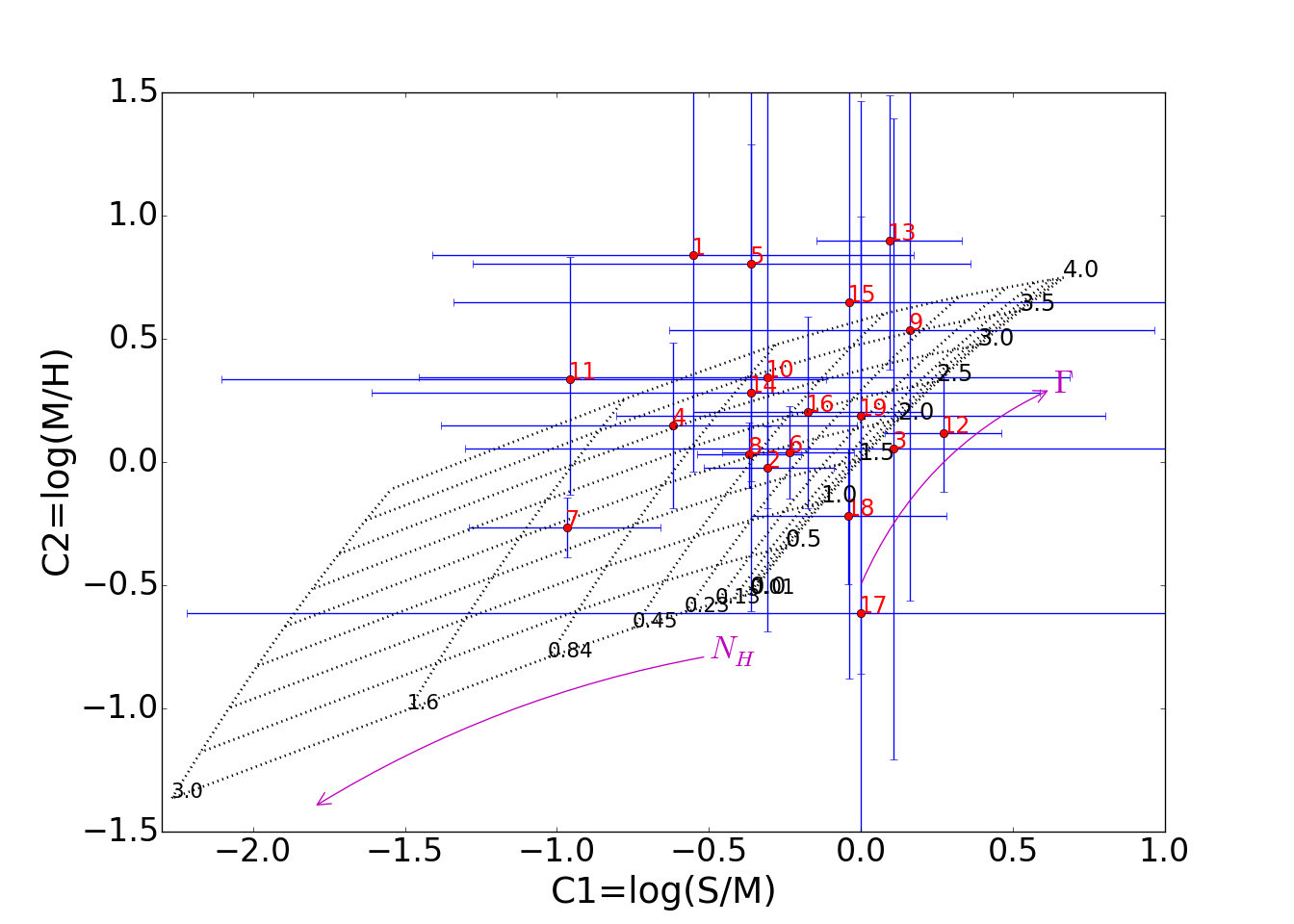}}	
		\end{subfigure}
		\caption{Colour-colour diagram for the X-ray sources detected in NGC\,2276 showing the colours (left) and corresponding errors (right) for OBSID 4968 (top) and OBSID 15648 (bottom). The source numbers refer to the source ID in Table \ref{tab.propertiesngc2276}. A grid showing the expected colours for absorbed power-law spectra is also overlaid. The value of the photon index and the hydrogen column density (in units of $\mathrm{10^{22}\,cm^{-2}}$) are shown at the edge of the grid.}
	\label{fig.grid_ngc2276}	
\end{minipage}
\end{figure*}

\begin{figure*}	
\begin{minipage}{130mm}
	\begin{subfigure}[b]{0.475\textwidth}
	\resizebox{\hsize}{!}{\includegraphics[scale=1]{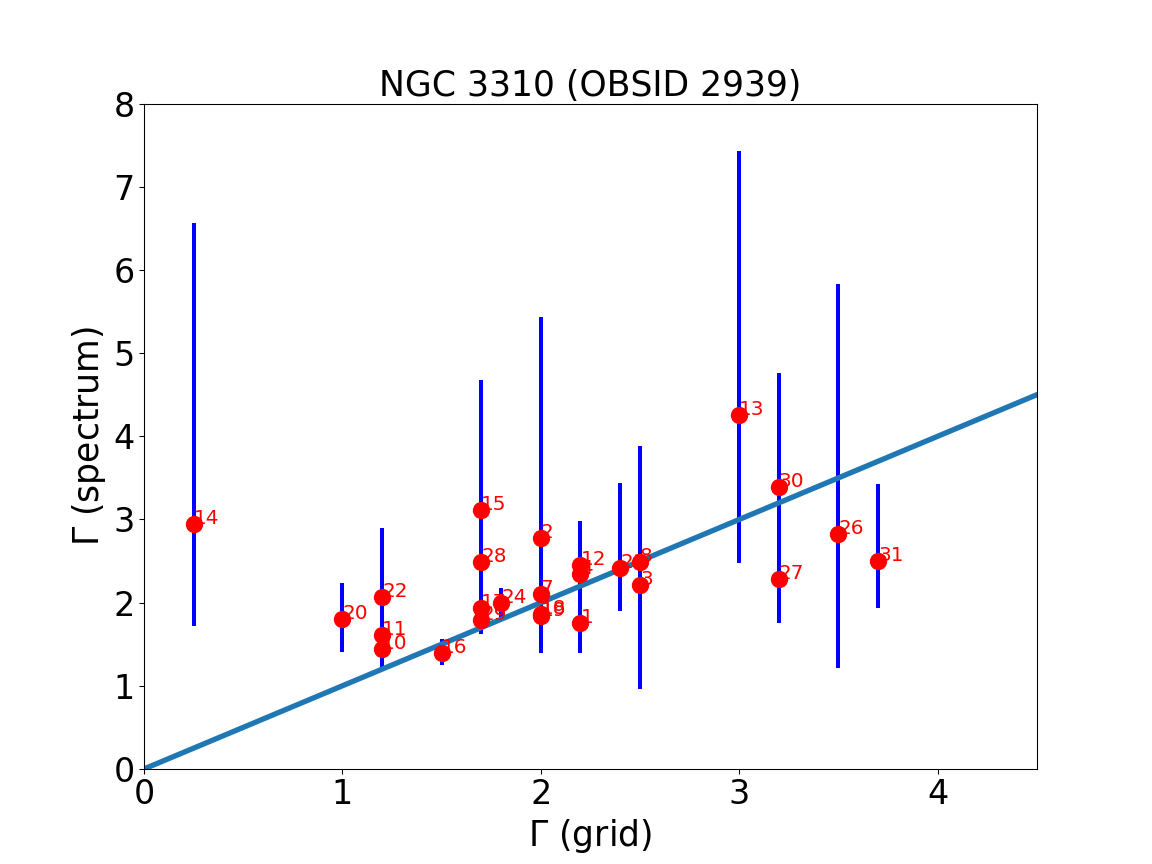}}
		\end{subfigure}
		 \hfill
			\begin{subfigure}[b]{0.475\textwidth}
	\resizebox{\hsize}{!}{\includegraphics[scale=1]{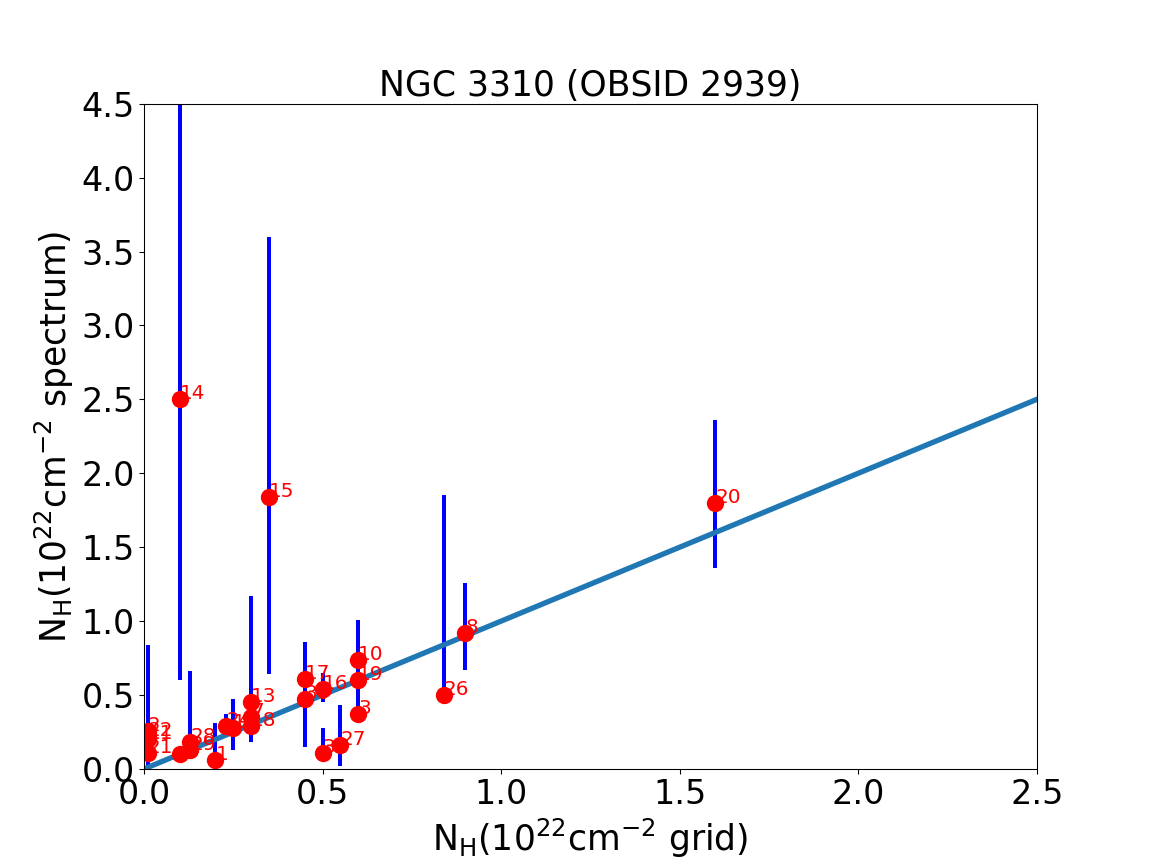}}	
	\end{subfigure}
			        \vskip\baselineskip
		\begin{subfigure}[b]{0.475\textwidth}  	        
	\resizebox{\hsize}{!}{\includegraphics[scale=1]{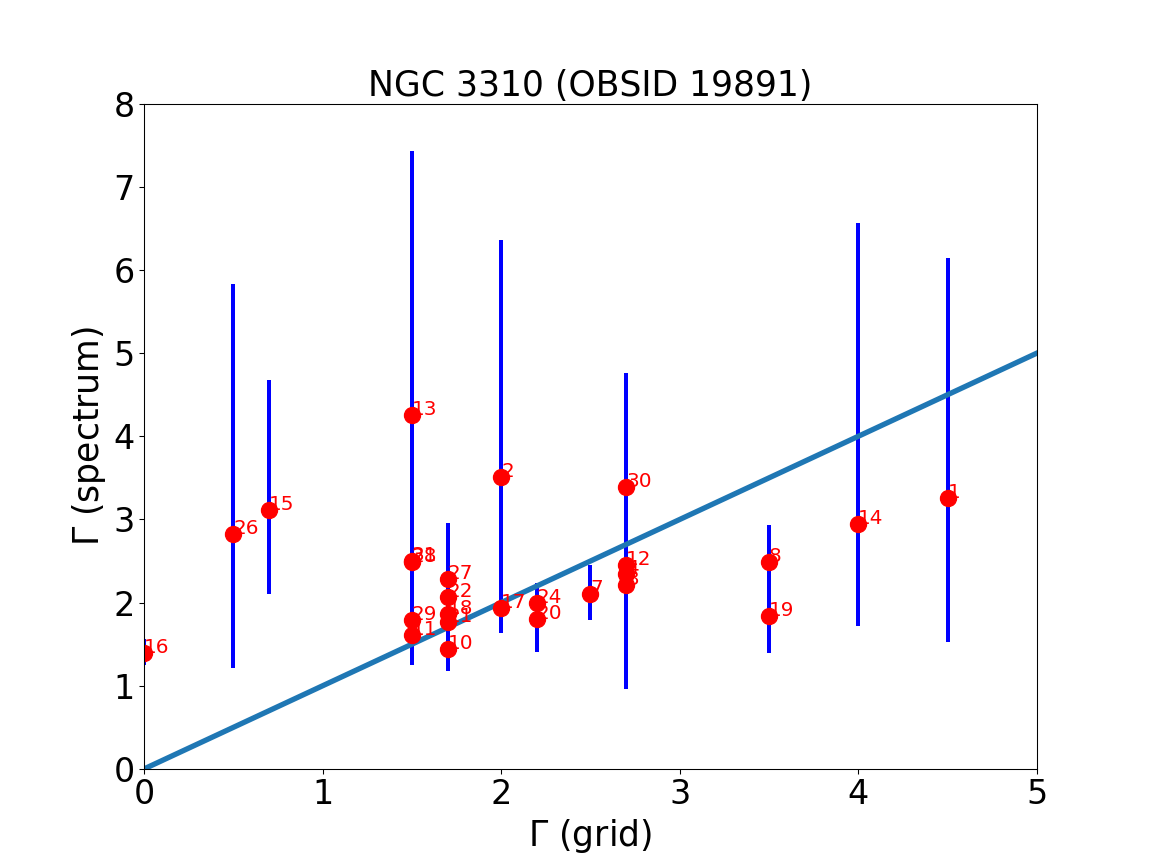}}
	\end{subfigure}
			        \quad
		 \begin{subfigure}[b]{0.475\textwidth}  	
	\resizebox{\hsize}{!}{\includegraphics[scale=1]{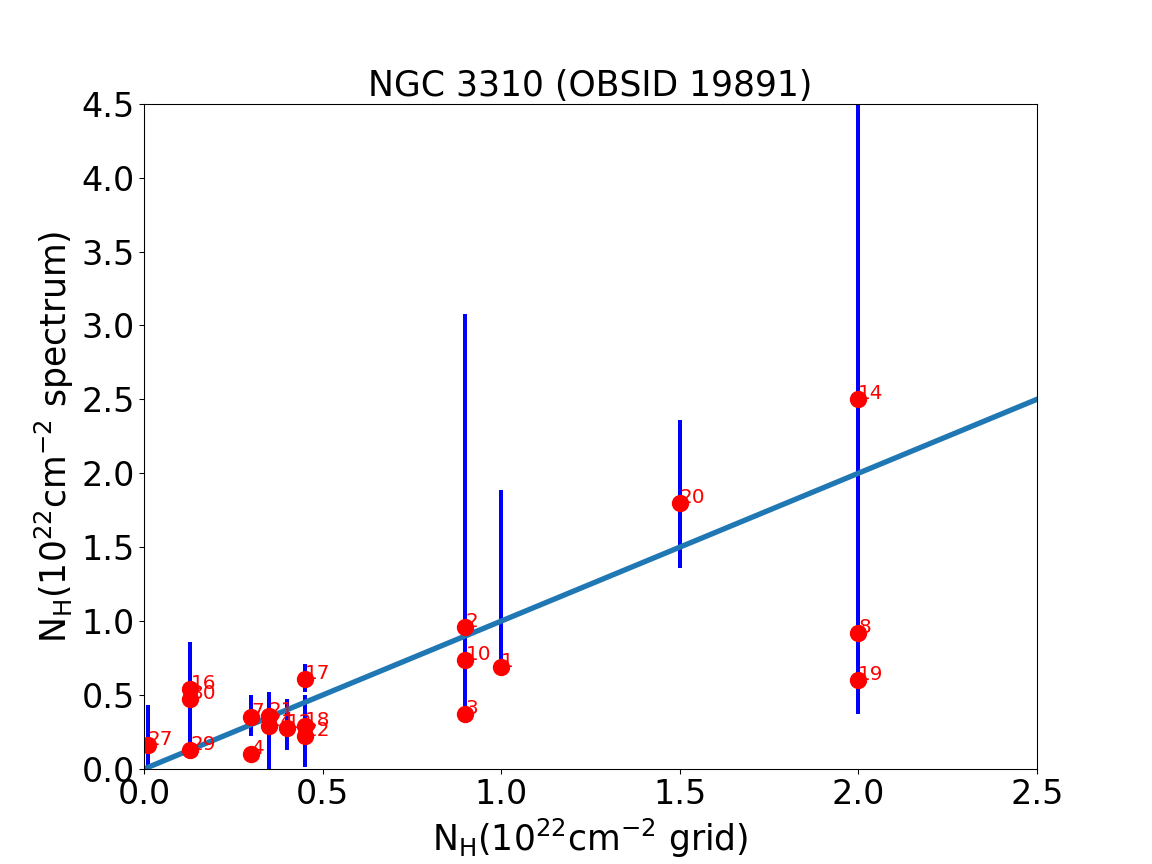}}	
	\end{subfigure}
	\caption{
{\az{    Comparison of the photon index (left) and \ion{H}{i} column density (\nh; right) determined from the X-ray colours and spectral fits with an absorbed power-law model. Top and bottom rows correspond to the long (OBSID 2939)  and short (OBSID 19891) observations of NGC\,3310. The blue solid line shows the 1:1 line. The numbers indicate the source IDs. }}
   }
	\label{fig.gridspectrum3310}
	\end{minipage}
\end{figure*}

\begin{figure*}	
\begin{minipage}{130mm}
	\begin{subfigure}[!hb]{0.475\textwidth}
		\resizebox{\hsize}{!}{\includegraphics[scale=1.0]{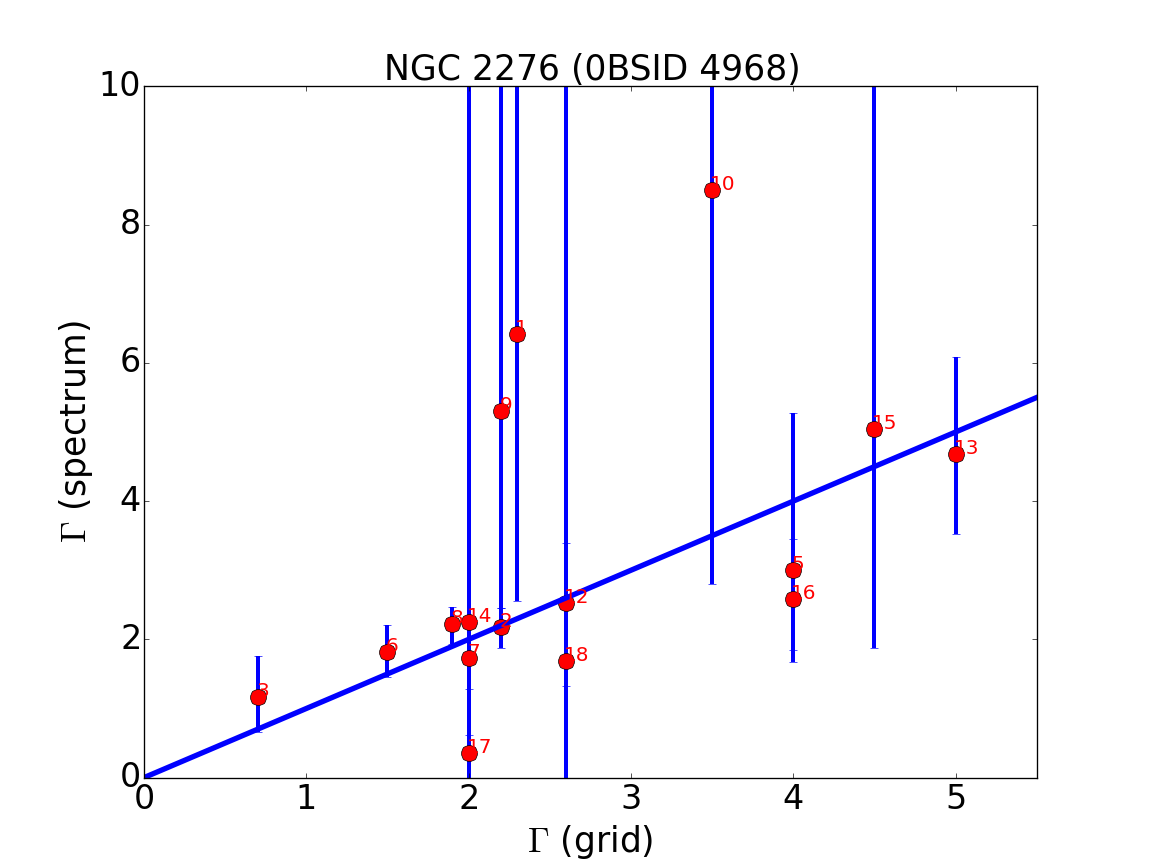}}
		\end{subfigure}
		        \hfill
	\begin{subfigure}[!hb]{0.475\textwidth}       
		\resizebox{\hsize}{!}{\includegraphics[scale=1.0]{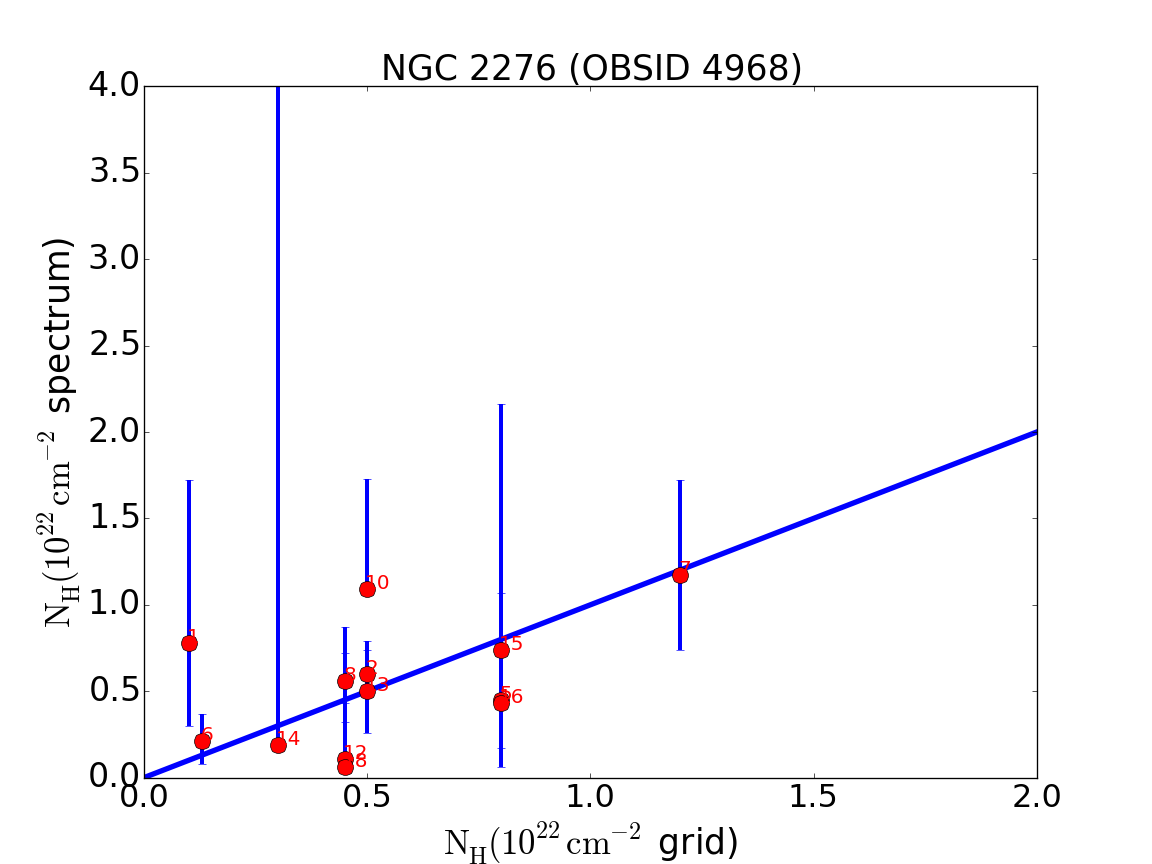}}
	\end{subfigure}
		        \vskip\baselineskip
	\begin{subfigure}[!hb]{0.475\textwidth}      
		\resizebox{\hsize}{!}{\includegraphics[scale=1.0]{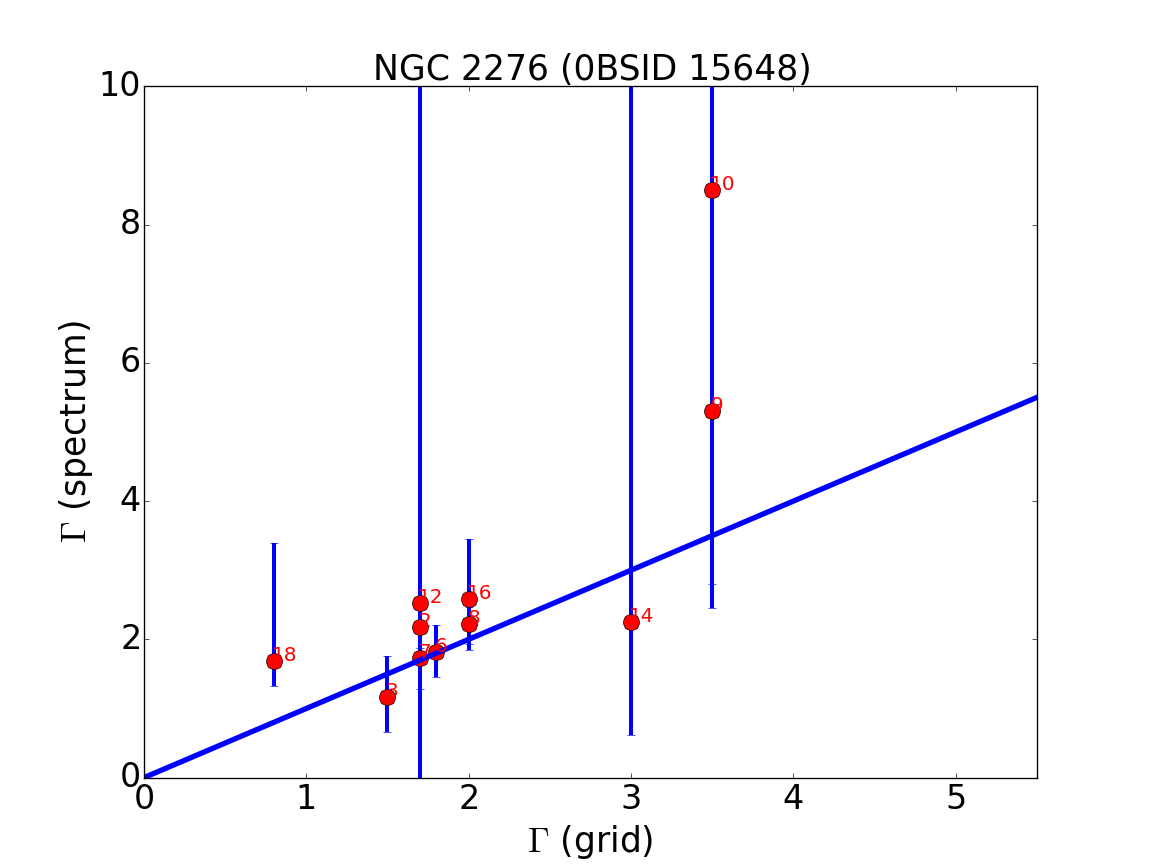}}
	 \end{subfigure}
		        \quad
	 \begin{subfigure}[!hb]{0.475\textwidth}   
		\resizebox{\hsize}{!}{\includegraphics[scale=1.0]{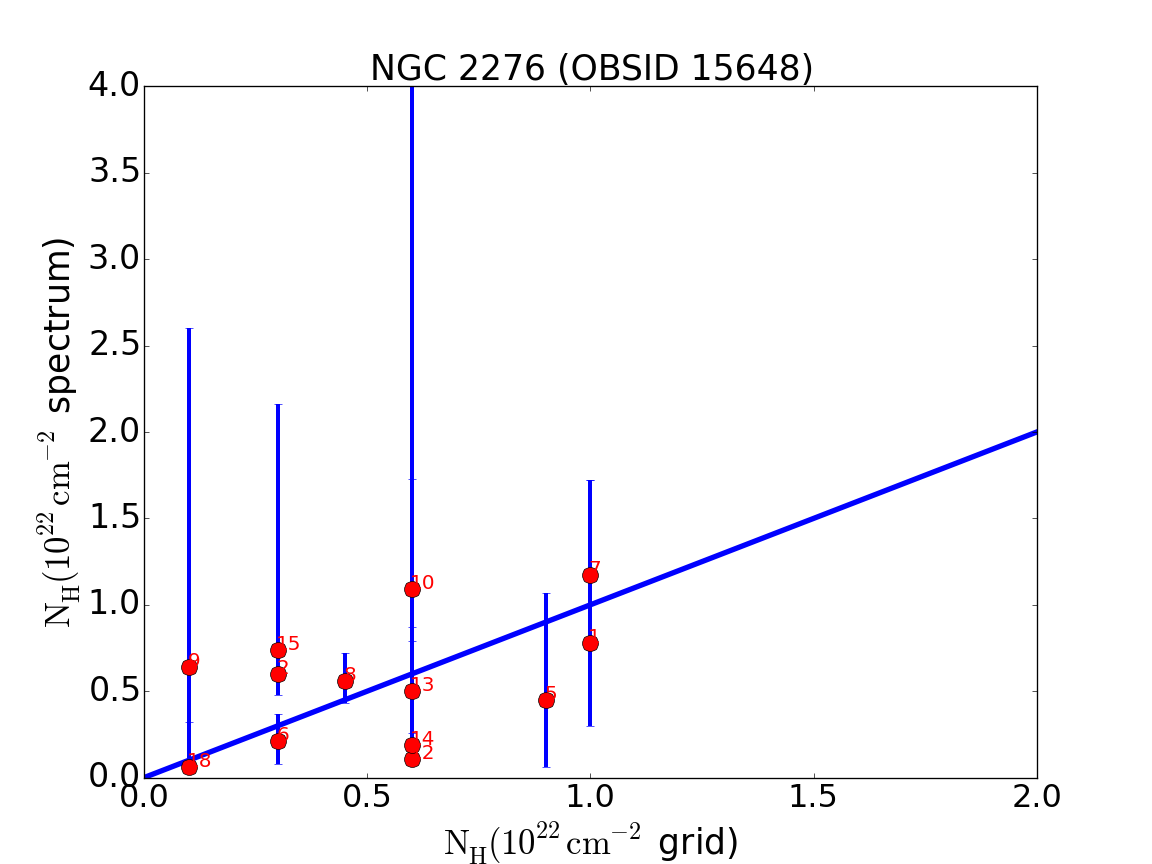}}	
		\end{subfigure}
		\caption{\az{    Comparison of the photon index (left) and \ion{H}{i} column density (\nh; right) determined from the X-ray colours and spectral fits with an absorbed power-law model. Top and bottom rows correspond to the long (OBSID 4968)  and short (OBSID 15648) observations of NGC\,2276. The blue solid line shows the 1:1 line. The numbers indicate the source IDs. 
        }}
	\label{fig.gridspectrum2276}	
\end{minipage}
\end{figure*}

\subsection{Fluxes and luminosities of the discrete sources.}\label{fluxesandluminositiesfthediscretesources}

Following our spectral analysis (Section \ref{spectralanalysis}) the flux and corresponding luminosity for each source was calculated by integrating the best-fit models (Tables \ref{tab.spectralparam_ngc3310} and \ref{tab.spectralparam_ngc2276}). The errors on the fluxes are estimated by drawing model parameter values from their best-fit joint distribution and calculating the flux corresponding to the model for each set of draws.  Then, the resulting fluxes are ordered and the central 90\% is selected to give the error range.
The errors on the luminosities were calculated by propagating the errors on the fluxes.

For sources 1, 3, 5, 9, 10, 14, 15, 17, and 19 of the shorter observation of NGC\,2276 (OBSID 15648), since they had very few counts, we calculated the fluxes  and luminosities (bottom section of \ref{tab.luminosities2276}) as follows. For each source we calculated the count-rate-to-flux conversion factor for each of the three bands based on the best-fit spectral model from the longer observation and the ancillary response file (ARF) from the second, shorter, observation. The errors on the corresponding flux and luminosity were calculated by simply propagating the errors on the number of observed counts in the second observation.

Also for the calculation of the luminosities of NGC\,2276 sources 4 and 11 for both OBSIDs, for which we could not perform spectral analysis, we used the best-fit model of source 6 which is a typical model for XRBs. 
We used the ARFs of each observation to calculate the count-rate-to-flux conversion. The flux and luminosity errors were calculated by propagating the count-rate errors. 

The observed and absorption corrected luminosities in the broad ($\mathrm{0.3-10.0\ keV}$), soft ($\mathrm{0.3-2.0\ keV}$), and hard ($\mathrm{2.0-10.0\ keV}$) bands  are presented in Tables \ref{tab.luminosities3310} and \ref{tab.luminosities3310_19891} for NGC\,3310 and Table \ref{tab.luminosities2276} for NGC\,2276. The limiting luminosity is $\mathrm{1.0\times 10^{38}\ erg\ s^{-1}}$  and for both galaxies. {\az{Here we only present data for the individual exposures because of the long time span between the observations of each galaxy which results in difference instrumental sensitivity. We note that there are no sources detected at above the $3\sigma$ level in the co-added exposure, therefore, Tables \ref{tab.luminosities3310} to  \ref{tab.luminosities2276} include all significant sources. }}
We find 14 ULXs for NGC\,3310 and 11 ULXs for NGC\,2276 reaching luminosities in the broad band ($\mathrm{0.3-10\ keV}$) of $\mathrm{1.5\times 10^{40}\ erg\ s^{-1}}$ and $\mathrm{2.1\times 10^{40}\ erg\ s^{-1}}$ respectively.
Adopting the distance of 32.9\,Mpc reported in \citet{wolter15}, for NGC\,2276, results in a luminosity change of $\sim 40\%$ and a total number of 8 ULXs. We estimated the contamination from background sources using the logN-logS distribution of \citet{kim2007}. We find that within the outline of each galaxy, at the detection limit of $\mathrm{f(0.5-8.0\ keV)=2.5\times 10^{-15}erg\ cm^{-2}\ s^{-1}}$, $\sim1.6$ sources are expected to be background sources.

\begin{table*}
	\centering                             
	\begin{minipage}{140mm}
		\caption{NGC\,3310 observed and absorption-corrected fluxes and luminosities (OBSID 2939).}
		\hskip-2.0cm\begin{tabular}[!htbp]{@{}lllllll@{}}
			\hline   
			Src & $\mathrm{fx^{obs}(f_x^{corr})}$ & $\mathrm{fx^{obs}(f_x^{corr})}$  & $\mathrm{fx^{obs}(f_x^{corr})}$  &  $\mathrm{L_x^{obs}(L_x^{corr})}$ & $\mathrm{L_x^{obs}(L_x^{corr})}$   & $\mathrm{L_x^{obs}(L_x^{corr})}$  \\
			ID & ($\mathrm{0.3-10.0\ keV}$) &($\mathrm{0.3-2.0\ keV}$) & ($\mathrm{2.0-10.0\ keV}$)& ($\mathrm{0.3-10.0\ keV}$) &($\mathrm{0.3-2.0\ keV}$) & ($\mathrm{2.0-10.0\ keV}$)\\
			& $\mathrm{10^{-14}\ erg\ s^{-1}\ cm^{-2}}$  &  $\mathrm{10^{-14}\ erg\ s^{-1}\ cm^{-2}}$&  $\mathrm{10^{-14}\ erg\ s^{-1}\ cm^{-2}}$& $\mathrm{10^{39}\ erg\ s^{-1}}$     & $\mathrm{10^{39}\ erg\ s^{-1}}$  & $\mathrm{10^{39}\ erg\ s^{-1}}$    \\
			(1) & (2) & (3) & (4) & (5) & (6) & (7)\\
			\hline	
			1 & $1.3_{-0.6}^{+0.4}$ (1.5) & $0.5_{-0.3}^{+0.1}$ (0.6) & $0.8_{-0.5}^{+0.4}$ (0.8) & $0.8_{-0.4}^{+0.2}$ (0.9) & $0.3_{-0.2}^{+0.1}$ (0.4) & $0.5\pm 0.3$ (0.5) \\[5pt] 
			2 & $0.7_{-0.8}^{+0.3}$ (2.0) & $0.4_{-0.4}^{+0.1}$ (1.7) & $0.3\pm 0.3$ (0.4) & $0.5_{-0.5}^{+0.2}$ (1.2) & $0.3_{-0.3}^{+0.1}$ (1.0) & $0.2\pm 0.2$ (0.2) \\[5pt] 
			3 &  $ 0.5 ^{+3.8}_{-0.4}$ (0.7)& $ 0.1 ^{+0.3}_{-0.1}$ (0.4) & $ 0.3 ^{+4.4}_{-0.3}$ (0.3)& $ 0.3 ^{+2.1}_{-0.2}$ (0.4)& $ 0.1 ^{+0.2}_{-0.1}$ (0.2) & $ 0.2 ^{+2.4}_{-0.2}$ (0.2)  \\[5pt]
			4\textasteriskcentered &  $12.0 ^{+1.2}_{-1.1}$ (17.3)& $6.7 ^{+0.6}_{-0.5}$ (11.8) & $5.4 \pm 0.8$ (5.4)& $6.7 ^{+0.7}_{-0.6}$ (9.7)& $3.7 \pm 0.3$ (6.6) & $3.0 \pm 0.5$ (3.1)  \\[5pt]
			5$\dagger$ &  $ 0.1 ^{+0.2}_{-0.1}$ (0.4)& $ 0.1 ^{+0.2}_{-0.1}$ (0.3) & 0.0 & $ 0.1 \pm 0.1$ (0.2)& $ 0.1 \pm 0.1$ (0.2) & 0.0  \\[5pt]
			6$\dagger$ & $0.04_{-0.03}^{+0.06}$ (0.09) & $0.03_{-0.03}^{+0.06}$ (0.09) & 0.0 & $0.02_{-0.02}^{+0.04}$ (0.05) & $0.02_{-0.01}^{+0.03}$ (0.05) & 0.0\\[5pt]
			7\textasteriskcentered &  $5.2 \pm 0.9$ (9.0)& $1.6 \pm 0.2$ (5.3) & $3.6 ^{+0.9}_{-0.8}$ (3.8)& $2.9 \pm 0.5$ (5.0)& $0.9 \pm 0.1$ (2.9) & $2.0 \pm 0.5$ (2.1)  \\[5pt]
			8\textasteriskcentered &  $5.5 ^{+0.8}_{-1.1}$ (17.4)& $1.4 ^{+0.1}_{-0.3}$ (12.8) & $4.1 ^{+0.7}_{-1.0}$ (4.6)& $3.1 ^{+0.5}_{-0.6}$ (9.7)& $0.8 \pm 0.1$ (7.2) & $2.3 ^{+0.4}_{-0.5}$ (2.6)  \\[5pt]
			9$\dagger$ &  $ 0.4 ^{+0.3}_{-0.2}$ (1.5)& $ 0.2 ^{+0.2}_{-0.1}$ (1.3) & $ 0.2 \pm 0.2$ (0.3)& $ 0.2 ^{+0.2}_{-0.1}$ (0.9)& $ 0.1 \pm 0.1$ (0.7) & $ 0.1 \pm 0.1$ (0.2)  \\[5pt]			
			10\textasteriskcentered &  $15.9 ^{+1.9}_{-2.5}$ (21.7)& $1.7 \pm 0.2$ (6.7) & $14.2 ^{+1.8}_{-2.3}$ (15.0)& $8.9 ^{+1.0}_{-1.4}$ (12.1)& $0.9 \pm 0.1$ (3.7) & $7.9 ^{+1.0}_{-1.3}$ (8.4)  \\[5pt]
			11\textasteriskcentered &  $4.2 \pm 0.1$ (5.2)& $1.0 \pm 0.2$ (1.9) & $3.2 \pm 0.9$ (3.3)& $2.4 \pm 0.5$ (2.9)& $0.6 \pm 0.1$ (1.1) & $1.8 \pm 0.5$ (1.8)  \\[5pt]	
			12\textasteriskcentered &  $ 2.1 ^{+2.1}_{-1.2}$ (4.3)& $ 0.9 ^{+0.7}_{-0.5}$ (3.1) & $ 1.1 ^{+1.7}_{-0.7}$ (1.2)& $ 1.2 ^{+1.2}_{-0.7}$ (2.4)& $ 0.5 ^{+0.4}_{-0.3}$ (1.8) & $ 0.6 ^{+0.9}_{-0.4}$ (0.7)  \\[5pt]
			13 &  $ 0.8 ^{+23.1}_{-0.8}$ (1.0)& $ 0.2 ^{+0.5}_{-0.2}$ (0.9) & $ 0.2 ^{+3.2}_{-0.2}$ (0.2)& $ 0.5 ^{+13.0}_{-0.4}$ (0.6)& $ 0.1 ^{+0.3}_{-0.1}$ (0.5) & $ 0.1 ^{+1.8}_{-0.1}$ (0.1)  \\[5pt]
			14 &  $ 0.4 ^{+85.4}_{-0.4}$ (4.0)& $ 0.2^{+0.3}_{-0.1}$ (3.5) & $ 0.2 ^{+32.8}_{-0.3}$ (0.5)& $ 0.2 ^{+47.8}_{-0.2}$ (2.2)& $ 0.1 ^{+0.2}_{-0.1}$ (1.9) & $ 0.1 ^{+18.3}_{-0.1}$ (0.3)  \\[5pt]
			15\textasteriskcentered &  $ 2.2 ^{+18.0}_{-2.0}$ (16.0)& $ 0.4 ^{+2.1}_{-0.4}$ (14.3) & $ 1.5 ^{+15.0}_{-1.4}$ (1.6)& $ 1.2 ^{+10.1}_{-1.1}$ (8.9)& $ 0.2 ^{+1.2}_{-0.2}$ (8.0) & $ 0.8 ^{+8.4}_{-0.8}$ (0.9)  \\[5pt]
			16\footnote{Nucleus of the galaxy.} &  $42.9 ^{+3.3}_{-3.7}$ (55.6)& $5.3 ^{+0.3}_{-0.4}$ (16.4) & $37.6 ^{+3.5}_{-3.6}$ (39.2)& $24.0 ^{+1.9}_{-2.1}$ (31.1)& $2.9 \pm 0.2$ (9.2) & $21.1 ^{+2.0}_{-2.0}$ (21.9)  \\[5pt]
			17\textasteriskcentered &  $25.9 ^{+2.0}_{-2.2}$ (45.3)& $5.2 \pm 0.3$ (23.3) & $20.8 ^{+1.8}_{-2.2}$ (22.0)& $14.5 ^{+1.1}_{-1.3}$ (25.4)& $2.9 \pm 0.2$ (13.1) & $11.6 ^{+1.0}_{-1.2}$ (12.3)  \\[5pt]
			18\textasteriskcentered &  $5.9 \pm 0.8$ (8.6)& $1.6 \pm 0.2$ (4.1) & $4.3 \pm 0.7$ (4.5)& $3.3 ^{+0.5}_{-0.4}$ (4.8)& $0.9 \pm 0.1$ (2.3) & $2.4 \pm 0.4$ (2.5)  \\[5pt]
			19\textasteriskcentered &  $5.0 ^{+0.9}_{-1.3}$ (8.2)& $0.9 ^{+0.1}_{-0.2}$ (3.9) & $4.1 ^{+0.9}_{-1.2}$ (4.3)& $2.8 ^{+0.5}_{-0.8}$ (4.6)& $0.5 \pm 0.1$ (2.2) & $2.3 ^{+0.5}_{-0.7}$ (2.4)  \\[5pt]
			20\textasteriskcentered &  $14.3 ^{+1.6}_{-3.7}$ (28.2)& $1.0 ^{+0.1}_{-0.3}$ (12.8) & $13.3 ^{+1.9}_{-3.6}$ (15.5)& $8.0 ^{+0.9}_{-2.1}$ (15.8)& $0.6 \pm 0.1$ (7.2) & $7.4 ^{+1.1}_{-2.0}$ (8.7)  \\[5pt]
			21\textasteriskcentered & $1.8_{-0.6}^{+0.5}$ (2.6) & $1.1_{-0.5}^{+0.2}$ (1.9) & $0.7_{-0.4}^{+0.5}$ (0.7) & $1.1_{-0.4}^{+0.3}$ (1.6) & $0.7_{-0.3}^{+0.1}$ (1.1) & $0.4_{-0.2}^{+0.3}$ (0.4) \\[5pt] 
			22 &  $1.4 ^{+0.4}_{-0.5}$ (2.0)& $0.5 ^{+0.1}_{-0.2}$ (1.1) & $0.9 \pm 0.4$ (1.0)& $0.8 ^{+0.2}_{-0.3}$ (1.1)& $0.3 \pm 0.1$ (0.6) & $0.5 \pm 0.2$ (0.5)  \\[5pt]		
			23$\dagger$ & $0.3_{-0.3}^{+0.4}$ (0.3) & $0.2_{-0.2}^{+0.4}$ (0.2) & $0.1_{-0.1}^{+0.2}$ (0.1) & $0.2\pm 0.2$ (0.2) & $0.1_{-0.1}^{+0.2}$ (0.1) & $0.05_{-0.05}^{+0.14}$ (0.05) \\[5pt]
			24\textasteriskcentered &  $14.0 \pm 1.3$ (21.9)& $4.3 \pm 0.3$ (11.9) & $9.7 \pm 1.3$ (10.0)& $7.9 \pm 0.7$ (12.3)& $2.4 \pm 0.2$ (6.7) & $5.5 \pm 0.7$ (5.6)  \\[5pt]
			25 &  $ 1.7 ^{+11.7}_{-1.5}$ (63.6)& $ 0.3 ^{+1.5}_{-0.3}$ (62.0) & $ 1.2 ^{+11.4}_{-1.1}$ (1.7)& $ 0.9 ^{+6.5}_{-0.9}$ (35.6)& $ 0.2 ^{+0.8}_{-0.1}$ (34.7) & $ 0.7 ^{+6.4}_{-0.6}$ (0.9)  \\[5pt]			
			26 & $0.3_{-0.3}^{+3.5}$ (1.1) & $0.1_{-0.1}^{+0.7}$ (0.9) & $0.2_{-0.2}^{+5.0}$ (0.2) & $0.2_{-0.2}^{+2.1}$ (0.6) & $0.1_{-0.1}^{+0.4}$ (0.5) & $0.1_{-0.1}^{+3.0}$ (0.1) \\[5pt]
			27 &  $1.2 \pm 0.4$ (1.9)& $0.6 ^{+0.1}_{-0.2}$ (1.2) & $0.6 \pm 0.3$ (0.6)& $0.7 \pm 0.2$ (1.0)& $0.3 \pm 0.1$ (0.7) & $0.3 ^{+0.2}_{-0.1}$ (0.4)  \\[5pt]			
			28 &  $ 0.3 ^{+13.1}_{-0.3}$ (1.5)& $ 0.1 ^{+0.7}_{-0.1}$ (1.5) & $ 0.1 ^{+6.8}_{-0.1}$ (0.1)& $ 0.2 ^{+7.3}_{-0.2}$ (0.9)& $ 0.1 ^{+0.4}_{-0.1}$ (0.8) & $ 0.1 ^{+3.8}_{-0.1}$ (0.0)  \\[5pt]
			29\textasteriskcentered &  $12.2 \pm 1.3$ (15.1)& $3.9 ^{+0.4}_{-0.3}$ (6.8) & $8.2 \pm 1.1$ (8.3)& $6.8 \pm 0.7$ (8.5)& $2.2 \pm 0.2$ (3.8) & $4.6 \pm 0.6$ (4.7)  \\[5pt]
			30 &  $ 0.4 ^{+1.5}_{-0.4}$ (2.3)& $ 0.2 ^{+0.6}_{-0.2}$ (2.1) & $ 0.2 ^{+1.5}_{-0.1}$ (0.1)& $ 0.2 ^{+0.9}_{-0.2}$ (1.3)& $ 0.1 ^{+0.3}_{-0.1}$ (1.2) & $ 0.1 ^{+0.9}_{-0.1}$ (0.1)  \\[5pt]
			31 &  $ 0.2 ^{+8.2}_{-0.2}$ (0.3)& $ 0.1 \pm 0.1$ (0.3) & $ 0.1 ^{+0.2}_{-0.1}$ (0.1)& $ 0.1 ^{+4.6}_{-0.1}$ (0.2)& $ 0.05 ^{+0.05}_{-0.05}$ (0.1) & $ 0.05 ^{+0.1}_{-0.05}$ (0.05)  \\[5pt]	
			\hline
			\end{tabular} 	
			\label{tab.luminosities3310}
            \smallskip
{\az{            Column 1: The source ID, ULXs are indicated with \textasteriskcentered and diffuse emission clumps with  $\dagger$; columns 2, 3, and 4: fluxes in the broad, soft, and hard bands respectively; columns 5, 6, and 7: luminosities in the broad, soft, and hard bands respectively. The unabsorbed fluxes and luminosities are given in parenthesis.}}
			\end{minipage}
			\end{table*}
							
		\begin{table*}
			\centering                             
			\begin{minipage}{150mm}
				\caption{NGC\,3310 observed and absorption-corrected fluxes and luminosities (OBSID 19891).}
				\hskip-2.0cm\begin{tabular}[!htbp]{@{}lllllll@{}}
					\hline   
					Src& $\mathrm{fx^{obs}(f_x^{corr})}$ & $\mathrm{fx^{obs}(f_x^{corr})}$  & $\mathrm{fx^{obs}(f_x^{corr})}$  &  $\mathrm{L_x^{obs}(L_x^{corr})}$ & $\mathrm{L_x^{obs}(L_x^{corr})}$   & $\mathrm{L_x^{obs}(L_x^{corr})}$  \\
					ID & ($\mathrm{0.3-10.0\ keV}$) &($\mathrm{0.3-2.0\ keV}$) & ($\mathrm{2.0-10.0\ keV}$)& ($\mathrm{0.3-10.0\ keV}$) &($\mathrm{0.3-2.0\ keV}$) & ($\mathrm{2.0-10.0\ keV}$)\\
					& $\mathrm{10^{-14}\ erg\ s^{-1}\ cm^{-2}}$  &  $\mathrm{10^{-14}\ erg\ s^{-1}\ cm^{-2}}$&  $\mathrm{10^{-14}\ erg\ s^{-1}\ cm^{-2}}$& $\mathrm{10^{39}\ erg\ s^{-1}}$     & $\mathrm{10^{39}\ erg\ s^{-1}}$  & $\mathrm{10^{39}\ erg\ s^{-1}}$    \\
					(1) & (2) & (3) & (4) & (5) & (6) & (7)\\
					\hline		
			1 &  $ 0.7 ^{+9.7}_{-0.6}$ (2.6)& $ 0.2 ^{+1.0}_{-0.2}$ (2.4) & $ 0.3 ^{+9.7}_{-0.3}$ (0.2)& $ 0.4 ^{+5.4}_{-0.3}$ (1.5)& $ 0.1 ^{+0.6}_{-0.1}$ (1.3) & $ 0.1 ^{+5.4}_{-0.1}$ (0.1)  \\[5pt]
			2 &  $ 0.4 ^{+6.3}_{-0.4}$ (4.2)& $ 0.2 ^{+1.2}_{-0.2}$ (3.9) & $ 0.2 ^{+5.6}_{-0.2}$ (0.2)& $ 0.2 ^{+3.5}_{-0.2}$ (2.3)& $ 0.1 ^{+0.7}_{-0.1}$ (2.2) & $ 0.1 ^{+3.1}_{-0.1}$ (0.1)  \\[5pt]
			3 &  $ 0.4 \pm 0.3$ (0.5)& $ 0.1 ^{+0.1}_{-0.1}$ (0.3) & $ 0.2 \pm 0.1$ (0.2)& $ 0.2 \pm 0.2$ (0.3)& $ 0.05 ^{+0.05}_{-0.05}$ (0.1) & $ 0.1 \pm 0.1$ (0.1)  \\[5pt]
			4\textasteriskcentered &  $13.4 ^{+1.7}_{-1.6}$ (19.3)& $7.4 ^{+1.0}_{-0.7}$ (13.2) & $6.0 \pm 0.1$ (6.1)& $7.5 \pm 0.9$ (10.8)& $4.2 ^{+0.5}_{-0.4}$ (7.4) & $3.4 ^{+0.6}_{-0.5}$ (3.4)  \\[5pt]
			5$\dagger$ &  $ 0.4 \pm 0.3$ (0.5)& $ 0.2 ^{+0.2}_{-0.1}$ (0.4) & $ 0.2 \pm 0.1$ (0.2)& $ 0.2 ^{+0.2}_{-0.1}$ (0.3)& $ 0.1 \pm 0.1$ (0.2) & $ 0.1 \pm 0.1$ (0.1)  \\[5pt]
			6$\dagger$ &-&-&-&-&-&-\\
			7\textasteriskcentered &  $2.5 ^{+0.7}_{-0.6}$ (4.2)& $0.8 \pm 0.2$ (2.5) & $1.7 \pm 0.5$ (1.8)& $1.4 ^{+0.4}_{-0.3}$ (2.4)& $0.4 \pm 0.1$ (1.4) & $0.9 \pm 0.3$ (1.0)  \\[5pt]
			8\textasteriskcentered &  $2.6 ^{+0.6}_{-0.7}$ (8.1)& $0.6 ^{+0.1}_{-0.2}$ (6.0) & $1.9 \pm 0.5$ (2.1)& $1.4 ^{+0.3}_{-0.4}$ (4.6)& $0.4 \pm 0.1$ (3.4) & $1.1 \pm 0.3$ (1.2)  \\[5pt]
			9$\dagger$ &  $ 0.3 ^{+0.4}_{-0.3}$ (1.3)& $ 0.1 ^{+0.3}_{-0.1}$ (1.1) & $ 0.2 ^{+0.2}_{-0.1}$ (0.3)& $ 0.2 ^{+0.2}_{-0.1}$ (0.7)& $ 0.1 ^{+0.2}_{-0.1}$ (0.6) & $ 0.1 \pm 0.1$ (0.1)  \\[5pt]
			10\textasteriskcentered &  $6.7 ^{+1.4}_{-1.4}$ (9.2)& $0.7 \pm 0.1$ (2.8) & $6.0 \pm 1.3$ (6.4)& $3.8 \pm 0.8$ (5.2)& $0.4 \pm 0.1$ (1.6) & $3.4 \pm 0.7$ (3.6)  \\[5pt]
			11\textasteriskcentered &  $3.2 ^{+1.3}_{-1.0}$ (3.9)& $0.8 \pm 0.2$ (1.4) & $2.4 ^{+1.1}_{-0.9}$ (2.5)& $1.8 ^{+0.7}_{-0.6}$ (2.2)& $0.4 \pm 0.1$ (0.8) & $1.4 ^{+0.6}_{-0.5}$ (1.4)  \\[5pt]
			12 &  $ 1.0 ^{+0.6}_{-0.5}$ (2.3)& $ 0.4 ^{+0.5}_{-0.2}$ (1.7) & $ 0.6 \pm 0.3$ (0.6)& $ 0.6 ^{+0.4}_{-0.3}$ (1.3)& $ 0.2 ^{+0.3}_{-0.1}$ (0.9) & $ 0.3 \pm 0.2$ (0.4)  \\[5pt]
			13 &  $ 0.9 ^{+0.3}_{-0.2}$ (0.7)& $ 0.2 ^{+0.2}_{-0.1}$ (0.6) & $ 0.1 \pm 0.1$ (0.1)& $ 0.5 ^{+0.2}_{-0.1}$ (0.4)& $ 0.1 \pm 0.1$ (0.3) & $ 0.1 ^{+0.0}_{-0.0}$ (0.1)  \\[5pt]
			14 &  $ 0.3 ^{+0.4}_{-0.2}$ (2.5)& $ 0.1 ^{+0.2}_{-0.1}$ (2.2) & $ 0.2 ^{+0.3}_{-0.2}$ (0.3)& $ 0.2 ^{+0.2}_{-0.1}$ (1.4)& $ 0.1 \pm 0.1$ (1.2) & $ 0.1 \pm 0.1$ (0.2)  \\[5pt]
			15 &  $ 1.2 ^{+1.1}_{-0.8}$ (12.3)& $ 0.2 ^{+0.8}_{-0.2}$ (11.0) & $ 0.9 ^{+0.6}_{-0.5}$ (1.3)& $ 0.7 ^{+0.6}_{-0.4}$ (6.9)& $ 0.1 ^{+0.4}_{-0.1}$ (6.2) & $ 0.5 \pm 0.3$ (0.7)  \\[5pt]
			16\footnote{Nucleus of the galaxy.} &  $11.5 ^{+1.8}_{-1.6}$ (14.9)& $1.4 \pm 0.2$ (4.4) & $10.1 ^{+1.6}_{-1.5}$ (10.5)& $6.4 ^{+1.0}_{-0.9}$ (8.3)& $0.8 \pm 0.1$ (2.5) & $5.6 ^{+0.9}_{-0.8}$ (5.9)  \\[5pt]
			17\textasteriskcentered &  $11.7 ^{+1.2}_{-1.4}$ (20.5)& $2.3 \pm 0.2$ (10.5) & $9.4 ^{+1.1}_{-1.2}$ (9.9)& $6.6 ^{+0.7}_{-0.8}$ (11.5)& $1.3 \pm 0.1$ (5.9) & $5.2 ^{+0.6}_{-0.7}$ (5.6)  \\[5pt]
			18\textasteriskcentered &  $13.9 \pm 1.8$ (20.1)& $3.7 ^{+0.5}_{-0.4}$ (9.6) & $10.2 ^{+1.5}_{-1.6}$ (10.5)& $7.8 \pm 0.1$ (11.2)& $2.1 ^{+0.3}_{-0.2}$ (5.4) & $5.7 \pm 0.9$ (5.9)  \\[5pt]
			19\textasteriskcentered &  $3.4 ^{+1.2}_{-1.1}$ (5.7)& $0.6 \pm 0.2$ (2.7) & $2.8 ^{+1.1}_{-1.0}$ (3.0)& $1.9 ^{+0.7}_{-0.6}$ (3.2)& $0.4 \pm 0.1$ (1.5) & $1.6 ^{+0.6}_{-0.5}$ (1.7)  \\[5pt]
			20\textasteriskcentered &  $7.1 ^{+1.5}_{-2.1}$ (14.0)& $0.5 \pm 0.1$ (6.4) & $6.6 ^{+1.6}_{-2.0}$ (7.7)& $4.0 ^{+0.8}_{-1.2}$ (7.9)& $0.3 \pm 0.1$ (3.6) & $3.7 ^{+0.9}_{-1.1}$ (4.3)  \\[5pt]
			21\textasteriskcentered &  $ 23.8 ^{+2.6}_{-2.9}$ (34.3) & $ 5.3 \pm 0.6$ (15.2) & $ 18.5 ^{+2.5}_{-2.8}$ (19.1) & $ 13.3 ^{+1.5}_{-1.6}$ (19.2)& $ 3.0 \pm 0.3$ (8.5)  & $ 10.4 ^{+1.4}_{-1.6}$ (10.7)  \\[5pt]				
			22 &  $1.3 \pm 0.5$ (1.9)& $0.4 ^{+0.1}_{-0.2}$ (1.0) & $0.9 \pm 0.4$ (0.9)& $0.7 \pm 0.3$ (1.1)& $0.2 \pm 0.1$ (0.6) & $0.5 \pm 0.2$ (0.5)  \\[5pt]
			23$\dagger$ &-&-&-&-&-&-\\
			24\textasteriskcentered &  $11.0 ^{+1.4}_{-1.3}$ (17.2)& $3.4 ^{+0.3}_{-0.4}$ (9.3) & $7.6 ^{+1.2}_{-1.0}$ (7.9)& $6.2 ^{+0.8}_{-0.7}$ (9.6)& $1.9 \pm 0.2$ (5.2) & $4.3 ^{+0.7}_{-0.6}$ (4.4)  \\[5pt]
			25 &  $ 0.6 ^{+0.6}_{-0.4}$ (31.4)& $ 0.1 ^{+0.2}_{-0.1}$ (30.6) & $ 0.5 ^{+0.4}_{-0.3}$ (0.8)& $ 0.3 ^{+0.3}_{-0.2}$ (17.6)& $ 0.1 ^{+0.1}_{-0.0}$ (17.1) & $ 0.3 \pm 0.2$ (0.5)  \\[5pt]
			26 &  $ 0.2 ^{+1.3}_{-0.2}$ (0.4)& $ 0.1 \pm 0.1$ (0.2) & $ 0.1 ^{+1.0}_{-0.1}$ (0.1)& $ 0.1 ^{+0.7}_{-0.1}$ (0.2)& $ 0.05 ^{+0.1}_{-0.05}$ (0.1) & $ 0.05 ^{+0.5}_{-0.05}$ (0.1)  \\[5pt]		
			27 &  $1.4 \pm 0.5$ (2.3)& $0.7 \pm 0.2$ (1.5) & $0.8 ^{+0.4}_{-0.3}$ (0.8)& $0.8 \pm 0.3$ (1.3)& $0.4 \pm 0.1$ (0.8) & $0.4 \pm 0.2$ (0.4)  \\[5pt]
			28 &  $ 0.2 ^{+0.2}_{-0.1}$ (1.9)& $ 0.1 ^{+0.4}_{-0.2}$ (1.8) & $ 0.1 \pm 0.1$ (0.1)& $ 0.1 \pm 0.1$ (1.1)& $ 0.05 ^{+0.2}_{-0.1}$ (1.0) & $ 0.05 ^{+0.05}_{-0.05}$ (0.05)  \\[5pt]
			29\textasteriskcentered &  $15.2 ^{+1.9}_{-1.7}$ (18.9)& $4.9 ^{+0.6}_{-0.5}$ (8.5) & $10.3 ^{+1.6}_{-1.5}$ (10.4)& $8.5 ^{+1.1}_{-0.9}$ (10.6)& $2.8 \pm 0.3$ (4.7) & $5.8 ^{+0.9}_{-0.8}$ (5.8)  \\[5pt]		
			30 &  $ 0.3 ^{+0.4}_{-0.2}$ (2.2)& $ 0.2 ^{+0.3}_{-0.1}$ (2.1) & $ 0.1 \pm 0.1$ (0.1)& $ 0.2 ^{+0.2}_{-0.1}$ (1.3)& $ 0.1 ^{+0.2}_{-0.1}$ (1.2) & $ 0.1 ^{+0.05}_{-0.05}$ (0.1)  \\[5pt]
			31 &  $ 0.2 ^{+0.2}_{-0.1}$ (1.0)& $ 0.2 ^{+0.2}_{-0.1}$ (0.9) & $ 0.1 \pm 0.1$ (0.1) & $ 0.1 \pm 0.1$ (0.6)& $ 0.1 \pm 0.1$ (0.5) & $ 0.1 ^{+0.05}_{-0.05}$ (0.1)  \\[5pt]					
					\hline
						\end{tabular} 	
						\label{tab.luminosities3310_19891}
                        \smallskip
{\az{            Column 1: The source ID, ULXs are indicated with \textasteriskcentered and diffuse emission clumps with  $\dagger$; columns 2, 3, and 4: fluxes in the broad, soft, and hard bands respectively; columns 5, 6, and 7: luminosities in the broad, soft, and hard bands respectively. The unabsorbed fluxes and luminosities are given in parenthesis.}}
						\end{minipage}
						\end{table*}		
\begin{table*}
	\centering                             
	\begin{minipage}{140mm}
		\caption{NGC\,2276 observed and absorption-corrected fluxes and luminosities for OBSID 4968 (top) and for OBSID 15648 (bottom).}
		\hskip-2.5cm
		\setlength{\tabcolsep}{4pt}
		\begin{tabular}[!htbp]{@{}lllllll@{}}
			\hline   
			Src& $\mathrm{fx^{obs}(f_x^{corr})}$ & $\mathrm{fx^{obs}(f_x^{corr})}$  & $\mathrm{fx^{obs}(f_x^{corr})}$  &  $\mathrm{L_x^{obs}(L_x^{corr})}$ & $\mathrm{L_x^{obs}(L_x^{corr})}$   & $\mathrm{L_x^{obs}(L_x^{corr})}$  \\
			ID & ($\mathrm{0.3-10.0\ keV}$) &($\mathrm{0.3-2.0\ keV}$) & ($\mathrm{2.0-10.0\ keV}$)& ($\mathrm{0.3-10.0\ keV}$) &($\mathrm{0.3-2.0\ keV}$) & ($\mathrm{2.0-10.0\ keV}$)\\
			& $\mathrm{10^{-14}\ erg\ s^{-1}\ cm^{-2}}$  &  $\mathrm{10^{-14}\ erg\ s^{-1}\ cm^{-2}}$&  $\mathrm{10^{-14}\ erg\ s^{-1}\ cm^{-2}}$& $\mathrm{10^{39}\ erg\ s^{-1}}$     & $\mathrm{10^{39}\ erg\ s^{-1}}$  & $\mathrm{10^{39}\ erg\ s^{-1}}$    \\
			(1) & (2) & (3) & (4) & (5) & (6) & (7)\\
			\hline
			\hline
			&&& OBSID 4968&&&\\
			\hline
	    	1\textasteriskcentered & $0.5^{+135.7}_{-0.5}$  (58.4) & $0.3^{+7.3}_{-0.3}$  (38.4) & $0.2^{+27.9}_{-0.2}$  (20.0) & $1.0^{+270.3}_{-1.0}$  (116.3) & $0.6^{+14.5}_{-0.6}$  (76.5) & $0.4^{+55.6}_{-0.4}$  (39.8)\\[2pt]
		    2\textasteriskcentered & $8.5_{-1.2}^{+1.0}$ (17.8) & $2.1\pm 0.2$ (10.7) & $6.4\pm 1.2$ (6.8) & $16.9_{-2.4}^{+2.0}$ (35.5) & $4.2\pm 0.4$ (21.3) & $12.7\pm 2.4$ (13.5) \\[2pt]
			3\textasteriskcentered& $1.1_{-0.6}^{+1.4}$ (1.1) & $0.2\pm 0.1$ (0.2) & $0.9_{-0.6}^{+1.4}$ (0.9) & $2.2_{-1.2}^{+2.8}$ (2.2) & $0.4\pm 0.2$ (0.4) & $1.8_{-1.2}^{+2.8}$ (1.8) \\[2pt]
			4	&$0.06\pm 0.07$ (0.08)&$0.019\pm 0.02$ (0.04)&$0.04\pm 0.04$ (0.04)&$0.15\pm 0.16$ (0.19)& $0.05\pm 0.05$ (0.07)& $0.1\pm 0.1$ (0.1)	\\
			5\textasteriskcentered & $0.6^{+2.9}_{-0.5}$  (1.9) & $0.2^{+0.8}_{-0.2}$  (1.7) & $0.2^{+2.3}_{-0.2}$  (0.2) & $1.2^{+5.8}_{-1.0}$  (3.8) & $0.6^{+1.6}_{-0.4}$  (3.4) & $0.6^{+4.6}_{-0.4}$  (0.4)\\[2pt]
			6\textasteriskcentered & $3.6\pm 0.7$ (4.8) & $1.1\pm 0.2$ (2.1) & $2.5\pm 0.7$ (2.6) & $7.2\pm 1.4$ (9.6) & $2.2\pm 0.4$ (4.2) & $5.0\pm 1.4$ (5.2) \\[2pt]
			7\textasteriskcentered & $3.6_{-1.3}^{+0.7}$ (6.2) & $0.4_{-0.1}^{+0.6}$ (2.6) & $3.2_{-1.1}^{+0.7}$ (3.6) & $7.2_{-2.6}^{+1.4}$ (12.3) & $0.8_{-0.1}^{+0.5}$ (5.2) & $6.4_{-1.8}^{+1.4}$ (7.2) \\[2pt]
			8\textasteriskcentered & $5.8_{-0.6}^{+0.5}$ (12.9) & $1.7\pm 0.1$ (8.2) & $4.1\pm 0.5$ (4.7) & $11.6_{-1.2}^{+1.0}$ (25.7) & $3.4\pm 0.2$ (16.3) & $8.2\pm 1.0$ (9.4) \\[2pt]
			9\footnote{Nucleus of the galaxy.} & $0.2^{+1.9}_{-0.2}$  (11.7) & $0.15^{+0.8}_{-0.1}$  (11.6) & $0.05^{+0.7}_{-0.05}$  (0.05) & $0.4^{+3.8}_{-0.4}$  (23.3) & $0.3^{+1.6}_{-0.2}$  (23.1) & $0.1^{+1.4}_{-0.1}$  (0.1)\\[2pt]
			10  & $0.2_{-0.1}^{+0.2}$ (0.2) & $0.2_{-0.1}^{+0.2}$ (0.2) & $0.04_{-0.03}^{+0.17}$ (0.04) & $0.5_{-0.2}^{+0.4}$ (0.4) & $0.4_{-0.2}^{+0.4}$ (0.4) & $0.1_{-0.1}^{+0.3}$ (0.1) \\[2pt]
			11 &$0.13\pm 0.06$ (0.17)&$0.04\pm 0.02$ (0.08)&$0.09\pm 0.05$ (0.01)&$0.31\pm 0.16$ (0.5)&$0.09\pm 0.04$ (0.2)&$0.22\pm 0.10$ (0.22)\\[2pt]
			12 & $0.3\pm 0.2$  (0.3) & $0.1\pm 0.1$  (0.2) & $0.1\pm 0.1$  (0.1) & $0.6^{+0.4}_{-0.3}$  (0.6) & $0.3\pm 0.1$  (0.4) & $0.3\pm 0.2$  (0.2)\\[2pt]
			13\textasteriskcentered & $0.8^{+4.4}_{-0.7}$  (15.5) & $0.5^{+3.4}_{-0.5}$  (15.4) & $0.1^{+1.3}_{-0.1}$  (0.1) & $1.6^{+8.8}_{-1.4}$  (30.9) & $1.2^{+6.8}_{-1.0}$  (30.7) & $0.4^{+2.6}_{-0.2}$  (0.2)\\[2pt]
			14 & $0.2^{+0.4}_{-0.1}$  (0.5) & $0.1^{+0.2}_{-0.1}$  (0.3) & $0.2^{+4.7}_{-0.2}$  (0.2) & $0.4^{+0.8}_{-0.2}$  (1.0) & $0.1^{+0.4}_{-0.2}$  (0.6) & $0.3^{+9.4}_{-0.4}$  (0.4)\\[2pt]
			15& $0.2_{-0.2}^{+42.1}$ (6.3) & $0.14_{-0.14}^{+2.74}$ (6.31) & $0.03_{-0.001}^{+11.7}$ (0.03) & $0.4_{-0.4}^{+83.9}$ (12.5) & $0.3_{-0.3}^{+5.5}$ (12.6) & $0.1_{-0.1}^{+23.3}$ (0.1) \\[2pt]
			16\textasteriskcentered  & $0.8^{+2.1}_{-0.6}$  (2.2) & $0.3^{+0.6}_{-0.3}$  (1.7) & $0.5^{+2.0}_{-0.4}$  (0.5) & $1.6^{+4.2}_{-1.2}$  (4.4) & $0.6^{+1.2}_{-0.6}$  (3.4) & $1.0^{+4.0}_{-0.8}$  (1.0)\\[2pt]
			17\textasteriskcentered& $2.3_{-1.9}^{+6.7}$ (2.2) & $0.05_{-0.03}^{+0.04}$ (0.05) & $2.3_{-1.9}^{+8.4}$ (2.3) & $4.6_{-3.8}^{+13.3}$ (4.6) & $0.1\pm 0.1$ (0.1) & $4.5_{-3.8}^{+16.7}$ (4.6) \\[2pt]
			18\textasteriskcentered & $0.5\pm 0.2$  (0.7) & $0.2\pm 0.1$  (0.3) & $0.4\pm 0.2$  (0.4) & $1.0\pm 0.4$  (1.4) & $0.4\pm 0.2$  (0.6) & $0.6\pm 0.4$  (0.8)\\[2pt]
			19 & $0.3^{+5.5}_{-0.3}$  (0.4) & $0.1^{+0.2}_{-0.1}$  (0.2) & $0.3^{+5.1}_{-0.3}$  (0.3) & $0.6^{+11.0}_{-0.6}$  (0.8) & $0.2^{+0.4}_{-0.2}$  (0.4) & $0.4^{+10.2}_{-0.6}$  (0.6)\\[2pt]		
			\hline		
			\hline
			&&& OBSID 15648&&&\\
			\hline
			1 & $0.19\pm 0.10$ (62.7) & $0.18\pm 0.09$ (62.69) & $0.01\pm 0.01$ (0.01) & $0.47\pm 0.23$ (150.31) & $0.44\pm 0.22$ (150.28) & $0.03\pm 0.01$ (0.03) \\[2pt]
			2\textasteriskcentered & $4.4\pm 1.0$ (9.1) & $1.1\pm 0.2$ (5.3) & $3.2\pm 0.8$ (3.5) & $8.2\pm 2.0$ (18.1) & $2.2\pm 0.4$ (10.6) & $7.0\pm 1.6$ (7.0) \\[2pt]
			3 & $0.4\pm 0.3$ (0.4) & $0.1\pm 0.1$ (0.1) & $0.3\pm 0.2$ (0.3) & $0.8\pm 0.6$ (0.8) & $0.2\pm 0.2$ (0.2) & $0.6\pm 0.4$ (0.6) \\[2pt]
			4 & $1.0\pm 0.3$ (1.4) & $0.3\pm 0.1$ (0.6) & $0.7\pm 0.2$ (0.8) & $2.2\pm 0.6$ (2.8) & $0.6\pm 0.2$ (1.2) & $1.6\pm 0.4$ (1.6) \\[2pt]
			5 & $0.3\pm 0.1$ (1.2) & $0.2\pm 0.1$ (1.1) & $0.1\pm 0.1$ (0.2) & $0.8\pm 0.2$ (2.4) & $0.4\pm 0.4$ (2.2) & $0.4\pm 0.2$ (0.4) \\[2pt]
			6\textasteriskcentered & $4.0\pm 0.5$ (10.6) & $1.2\pm 0.2$ (2.3) & $2.8\pm 0.9$ (2.8) & $8.0\pm 1.0$ (10.2) & $2.4\pm 0.6$ (4.6) & $5.6\pm 1.8$ (5.6) \\[2pt]
			7\textasteriskcentered& $11.1\pm 0.4$ (13.8) & $1.2\pm 0.3$ (8.7) & $10.7\pm 0.3$ (11.8) & $23.9\pm 0.8$ (57.5) & $2.4\pm 0.4$ (32.3) & $21.5\pm 0.6$ (23.5) \\[2pt]
			8\textasteriskcentered& $6.3\pm 0.6$ (14.9) & $1.9\pm 0.2$ (8.7) & $4.4\pm 0.4$ (5.0) & $12.5\pm 1.2$ (29.7) & $3.8\pm 0.2$ (17.3) & $8.8\pm 0.8$ (10.0) \\[2pt]
			9 & $0.26\pm 0.13$ (6.68) & $0.2\pm 0.1$ (6.61) & $0.06\pm 0.03$ (0.07) & $0.63\pm 0.3$ (16.01) & $0.48\pm 0.23$ (15.84) & $0.15\pm 0.07$ (0.17) \\[2pt]
			10 & $0.084\pm 0.073$ (562.8) & $0.082\pm 0.071$ (562.8) & $0.002\pm 0.001$ (0.002) & $0.201\pm 0.174$ (1349.1) & $0.197\pm 0.171$ (1349.1) & $0.004\pm 0.004$ (0.006) \\[2pt]
			11 & $0.6\pm 0.2$ (0.8) & $0.2\pm 0.1$ (0.3) & $0.4\pm 0.2$ (0.4) & $1.2\pm 0.4$ (1.6) & $0.4\pm 0.2$ (0.6) & $0.8\pm 0.4$ (1.0) \\[2pt]
			12\textasteriskcentered & $2.6^{+4.2}_{-1.9}$  (3.8) & $1.2^{+1.0}_{-0.6}$  (2.4) & $1.4^{+2.4}_{-0.9}$  (1.5) & $5.2^{+8.4}_{-3.8}$  (7.6) & $2.4^{+1.6}_{-1.2}$  (4.8) & $2.8^{+4.8}_{-1.8}$  (3.0)\\[2pt]
			13\textasteriskcentered  & $0.9^{+1.0}_{-0.5}$  (24.1) & $0.7^{+0.8}_{-0.4}$  (23.9) & $0.1\pm 0.1$  (0.1) & $1.6^{+2.0}_{-1.0}$  (48.0) & $1.4^{+1.6}_{-0.8}$  (47.6) & $0.2\pm 0.2$  (0.2)\\[2pt]
			14 & $0.2\pm 0.2$ (0.4) & $0.1\pm 0.1$ (0.3) & $0.1\pm 0.1$ (0.1) & $0.4\pm 0.4$ (0.8) & $0.2\pm 0.2$ (0.6) & $0.2\pm 0.2$ (0.2) \\[2pt]		
			15 & $0.04\pm 0.06$ (2.64) & $0.03\pm 0.05$ (2.63) & $0.01\pm 0.01$ (0.01) & $0.1\pm 0.15$ (6.32) & $0.08\pm 0.12$ (6.3) & $0.02\pm 0.03$ (0.02) \\[2pt]
			16\textasteriskcentered & $0.8^{+0.6}_{-0.5}$  (2.2) & $0.3^{+0.5}_{-0.2}$  (1.7) & $0.5\pm 0.2$  (0.5) & $1.6^{+1.2}_{-1.0}$  (4.4) & $0.6^{+1.0}_{-0.4}$  (3.4) & $1.0^{+0.5}_{-0.4}$  (1.0)\\[2pt]			
			17 & $0.04\pm 0.11$ (0.05) & $0.01\pm 0.03$ (0.02) & $0.03\pm 0.08$ (0.03) & $0.1\pm 0.28$ (0.12) & $0.03\pm 0.08$ (0.05) & $0.07\pm 0.19$ (0.07) \\[2pt]
			18\textasteriskcentered & $2.3^{+2.5}_{-1.2}$  (2.9) & $0.7^{+0.4}_{-0.3}$  (1.2) & $1.6^{+2.3}_{-0.9}$  (1.7) & $4.6^{+5.0}_{-2.4}$  (5.8) & $1.4\pm 0.8$  (2.4) & $3.2^{+4.6}_{-1.8}$  (3.4)\\[2pt]
			19 & $0.4\pm 0.3$ (0.5) & $0.1\pm 0.1$ (0.2) & $0.3\pm 0.2$ (0.3) & $0.8\pm 0.6$ (1.0) & $0.2\pm 0.2$ (0.4) & $0.6\pm 0.6$ (0.6) \\[2pt]
			\hline			
		\end{tabular} 	
		\label{tab.luminosities2276}
        \smallskip
        {\az{            Column 1: The source ID, ULXs are indicated with \textasteriskcentered and diffuse emission clumps with  $\dagger$; columns 2, 3, and 4: fluxes in the broad, soft, and hard bands respectively; columns 5, 6, and 7: luminosities in the broad, soft, and hard bands respectively. The unabsorbed fluxes and luminosities are given in parenthesis.}}
	\end{minipage}
\end{table*}

We note here that the number of counts, flux, and luminosity of the nuclear region (Src 9) in NGC\,2276 reported in Tables \ref{tab.propertiesngc2276} and \ref{tab.luminosities2276}, correspond to a region consistent with a point-like source. We used this small region in order to extract a conservative estimate of a nuclear X-ray source (e.g. an XRB or an AGN). We have also measured the flux from the full extent of the nuclear region using an elliptical aperture with major and minor radius of 7.5 and 6.3 arcsecs (consistent with the nuclear region in infrared images; Spitzer 8 microns) respectively, which correspond to a physical scale of about 1.5 kpc for the major axis. We calculated the net counts for both OBSIDs as well as for the co-added observation. We found that the extended nuclear region has $113.0\pm 13.8$, $67.2\pm 10.2$, and $183.2\pm 16.8$ broad-band (0.5-7.0 keV)  net counts for OBSID 4968, OBSID 15648 and the co-added observation respectively. We then extracted the spectrum of the longer observation. The best-fit model ($\chi^2/dof=1.92/3$; model in XSPEC: \texttt{phabs(po+apec)}) was an absorbed power-law component plus a thermal plasma component with best-fit parameters $\mathrm{N_H=(0.24\pm 0.36)\times 10^{22}\ cm^{-2}}$, $\Gamma=3.62\pm 2.50$, and $\mathrm{kT=1.10\pm 0.34\ keV}$.
We followed the same procedure as mentioned in the beginning of this section, to calculate the absorbed fluxes and luminosities of both OBSIDs from their corresponding counts.  The 0.3-10.0 keV band absorbed fluxes and luminosities are $(1.09\pm 1.00)\times \mathrm{10^{-14}\ erg\ s^{-1}\ cm^{-2}}$ and $(2.5\pm 2.3)\times \mathrm{10^{39}\ erg\ s^{-1}}$ for OBSID 4968 and $(1.66\pm 1.55\times \mathrm{10^{-14}\ erg\ s^{-1}\ cm^{-2}})$  and $(3.8\pm 3.4)\times \mathrm{10^{39}\ erg\ s^{-1}}$ for the OBSID 15648.

\subsection{Variability}\label{variability}

\az{We searched for inter-observation variability of the sources detected in \ngca\ and \ngcb\, using the CIAO \textit{glvary} tool. This tool is based on the  Gregory-Loredo variability detection method \citep{gl} which compares the photon-arrival times with a uniform distribution. Since the dither pattern of \chandra\ may result in a false variability signal because of pixel-to-pixel effective area variations we first ran the \textit{dither\_region} tool which generates  the time-dependent source area accounting for bad pixels, chip gaps, node boundaries etc. The variability of a source is encoded in the variability index (varindex) which takes into account the variability probability and the odds ratio for the photon arrival times to be non-uniformly distributed. Sources with a  variability index (varindex) larger than 6 are considered variable, with larger values of the   varindex indicate higher degree of variability. }

For NGC\,3310 and OBSID 2939, the nucleus (Src 16) showed evidence of variability with \textit{varindex=9} indicating that it is definitely a variable source. Inspecting its lightcurve (Fig. \ref{fig.variable_src16}) one can see a gradual increase in the observed counts for about 2/3 of the exposure time and a decrease at the last 1/3. This translates to a luminosity change of a factor of 1.7 starting with $ (1.65\pm 0.08)\times \mathrm{10^{40}\ erg\ s^{-1}}$ at the lowest to $ (2.84\pm 0.08)\times \mathrm{10^{40}\ erg\ s^{-1}}$ at the highest count rate, assuming the best-fit spectrum from the total exposure.
However analysing data from the shorter observation (OBSID 19891), we found that Src 16 is not variable at this observation with varindex=0 but corresponds to a luminosity of $6.4\pm 0.1\times \mathrm{10^{39}\ erg\ s^{-1}}$, which is about 2.5 times lower than the average luminosity of OBSID 2939.

Comparing also the luminosities of the two OBSIDs (Tables \ref{tab.luminosities3310} and \ref{tab.luminosities3310_19891}) we find that 8 out of 31 sources show long-term variability based on the broad-band luminosities of the two observations. Two sources are brighter (Sources 18 and 21) in the shorter exposure and 6 sources are brighter in the longer observation (Sources 7, 8, 10, 16, 17, and 20). All of these variable sources have luminosities above the ULX limit in both observations. Sources 16 (nucleus) and 21 show the most extreme variability of all by changing the broad-band luminosity 4 and 13 times respectively.

\begin{figure}	
	\resizebox{\hsize}{!}{\includegraphics[scale=1]{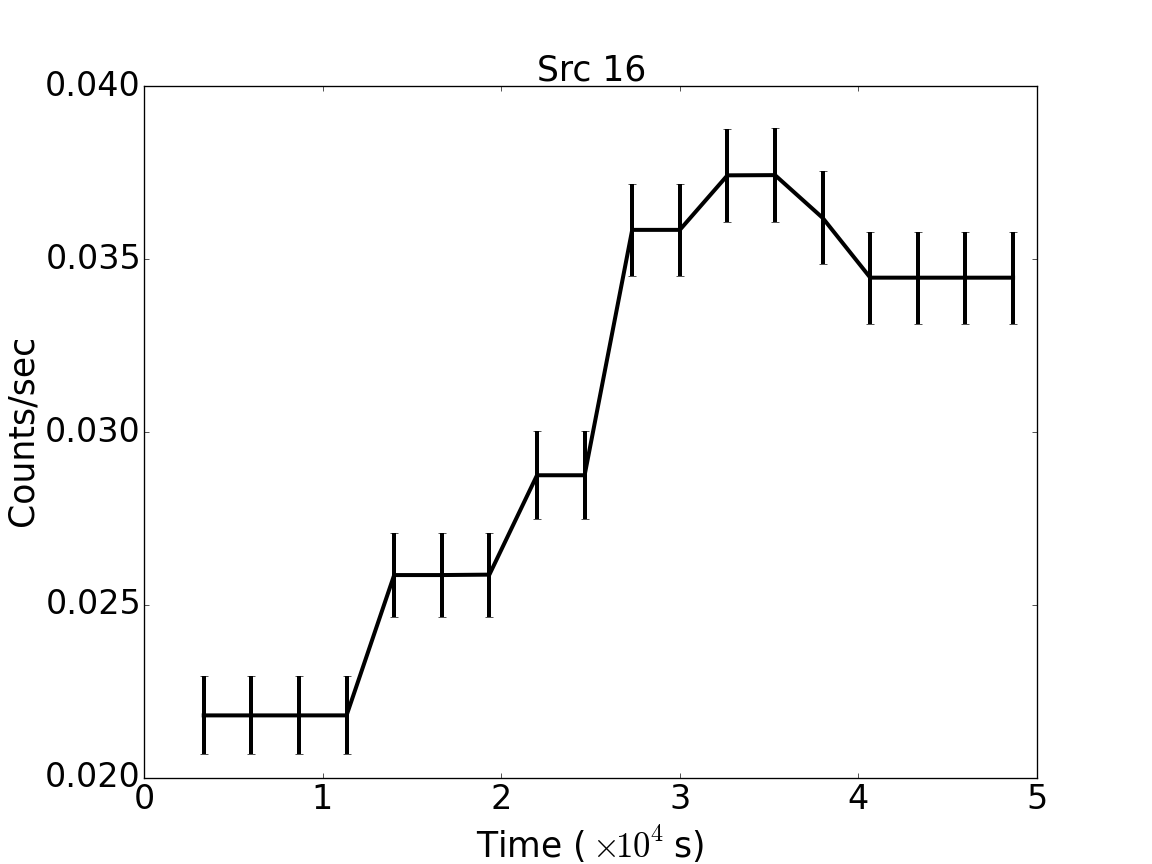}}
		\caption{
{\ka{        Lightcurve of the nucleus of NGC\,3310 (Src 16), extracted from OBSID 2939 divided in bins of $\sim$ 2.67\,ks. We  see a clear rise of the source intensity during the observation.
}}
}
	\label{fig.variable_src16}
\end{figure}

For NGC\,2276, none of the sources observed during the OBSID 4968 showed any evidence of short-term variability. In OBSID 15648 sources 1 and 2 have non-zero variability indices and are probably not variable (\textit{varindex=2}) and considered not variable (\textit{varindex=1}) respectively. However Src 3 is likely to be variable (\textit{varindex=4}) and its lightcurve (Fig. \ref{fig.variable_ngc2276}) shows characteristics of a fast rise and exponential decay (FRED) flare with an amplitude of 0.0013 counts/s. This corresponds to a luminosity change of $ (4.5\pm 1.0\times \mathrm{10^{39}\ erg\ s^{-1}})$.

We also searched for long-term variability in NGC\,2276 by comparing the luminosities {\az{between}} the two OBSIDs (Table \ref{tab.luminosities2276}). We find that 7 out of 19 sources show long-term variability based on the broad-band luminosities of the two observations. Five sources are brighter (Sources 4, 7, 11, 12, and 18) in the shorter exposure  and 2 sources are brighter in the longer observation (Sources 2 and 17). Out of the 7 variable sources, four (Sources 2, 7, 17, and 18) have luminosities above the ULX limit in the longer exposure whereas one (Src 12) has a 
luminosity in excess of $\mathrm{10^{39}ergs^{-1}}$ in the shorter exposure.

\begin{figure}	 
	\resizebox{\hsize}{!}{\includegraphics[scale=1]{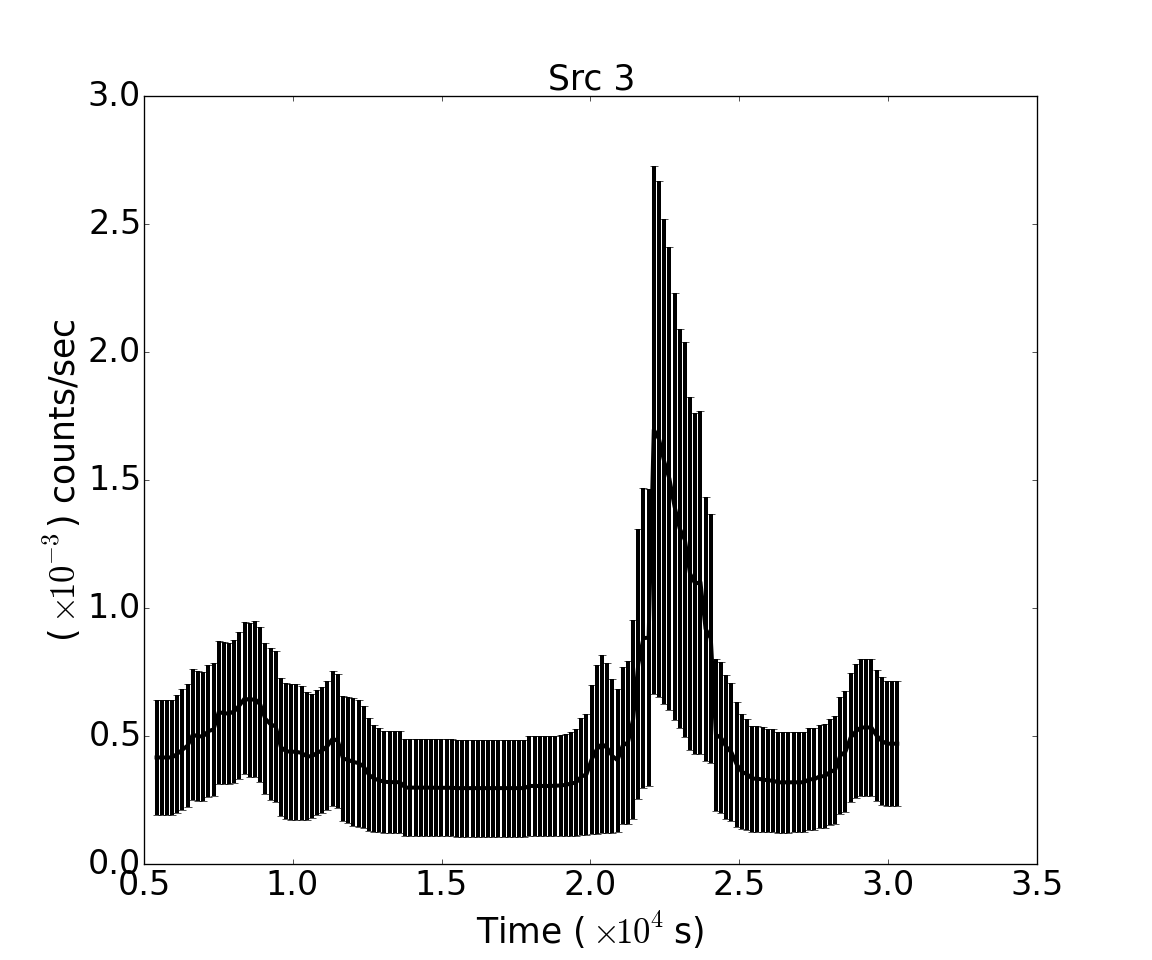}}
	\caption{
 {\ka{   Lightcurve of source 3 in NGC\,2276, extracted from OBSID 15648  in bins of $\sim$ 174\,s. We clearly see  a flare with a duration of $\sim500$\,sec. 
 }}
 }
	\label{fig.variable_ngc2276}
\end{figure}

\subsection{Integrated and extended X-ray emission of NGC\,3310}\label{integratedxrayemissionfthegalaxy}

In the following section we present the analysis on the integrated and extended emission of NGC\,3310. We do not perform the same analysis for NGC\,2276 since its was analysed previously in detail \citep{rasmussen06,wolter11,wolter15}. 
Therefore we extracted the integrated spectrum of NGC\,3310 included in the outline of the galaxy from each observation. We determined the background spectrum from a source-free area outside the galaxy using the \textit{specextract} tool.
We fitted simultaneously the spectra from the two separate observations (OBSID 2939 and OBSID 19891) with all the model parameters apart from the normalizations tied together.
We obtained the best-fit ($\mathrm{\chi_{\nu}^2/dof=626.04608}$) with a model consisting of two thermal plasma components and a power-law component (\texttt{phabs(powerlaw+apec+apec)}) seen through a common absorber. The best-fit parameters are $\Gamma=1.57\pm 0.06$, $\mathrm{N_{H}=(0.17\pm 0.03)\times 10^{22}\ cm^{-2}}$, $\mathrm{kT_1=0.22\pm 0.02\ keV}$ and $\mathrm{kT_2=0.77_{-0.04}^{+0.05}\ keV}$ (Table \ref{tab.specparam_galaxy3310}).
The total absorbed and unabsorbed luminosities based on this joint fit in the broad ($\mathrm{0.3-10.0\ keV}$), soft ($\mathrm{0.3-2.0\ keV}$), and hard ($\mathrm{2.0-10.0\ keV}$) bands are shown in Table \ref{tab.binulxslum}.

\begin{figure}
	\resizebox{\hsize}{!}{\includegraphics[scale=1]{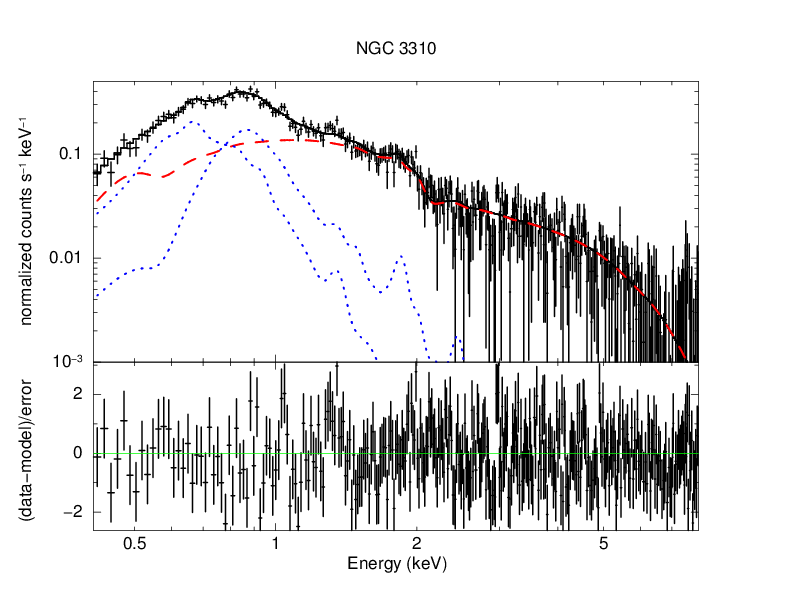}}
	\caption{(Top panel) The integrated X-ray spectrum of NGC\,3310 (OBSID 2939), along with the best-fit folded model (Table \ref{tab.specparam_galaxy3310}) consisting of: an absorbed power-law  (dashed red line), and two APEC components (dotted blue lines). (Bottom panel) The fit residuals in terms of sigma with error bars of size 1$\mathrm{\sigma}$.}		
	\label{fig.spectotallx}
\end{figure}

\subsubsection{Luminosity of X-ray binaries and ULXs}\label{luminosityofxraybinariesandulxs}

In order to determine the luminosity of the XRB populations, we measured the integrated emission of both resolved and unresolved XRBs in NGC\,3310 by extracting the integrated spectrum of the galaxy and measuring the flux of only the power-law component (best-fit results in Table \ref{tab.specparam_galaxy3310}). 
Since NGC\,3310 may host an AGN, we also measured the luminosity of the extra-nuclear XRB population by extracting the spectrum of the galaxy, but this time excising the nucleus. We fitted these spectra from each observation using the same model as for the total galaxy spectrum (\texttt{phabs(powerlaw+apec+apec))} resulting in a good fit ($\mathrm{\chi^2/dof=620.01/582}$). The best-fit parameters are listed in Table \ref{tab.specparam_galaxy3310}, and the total and XRB (power-law component) luminosity in Table \ref{tab.binulxslum}.

We calculated the total flux of the ULXs by extracting a spectrum from a region consisting of all their apertures and using a background spectrum from a source-free area outside the outline of the galaxy. However due to variability (namely 14 ULXs for OBSID 2939 and 12 ULXs for OBSID 19891), we used different apertures for each observation  and fitted the spectra for each observation separately
A model consisting of an absorbed power-law and a thermal-plasma model gave good fit for both observations ($\mathrm{\chi^2=214.67/185}$ and $\mathrm{\chi^2=124.67/132}$). The best-fit parameters for this fit are presented in Table \ref{tab.specparam_galaxy3310} and the corresponding absorbed and unabsorbed luminosities for the overall spectrum in Table \ref{tab.binulxslum}.

We also tried a model consisting of an absorbed thermal-plasma and a disk black-body model which has been successfully used to model the spectra of ULXs \citep{gladstone,rana}. This resulted in unrealistic values for the component parameters of the shorter observation (OBSID 19891) but gave a good fit for the longer observation (OBSID 2939; $\mathrm{\chi^2= 216.54/185}$) with best-fit parameters $\mathrm{N_{H}=(0.08\pm 0.03)\times 10^{22}\ cm^{-2}}$, $\mathrm{kT=0.42_{-0.19}^{+1.58}\ keV}$ and $\mathrm{T_{in}=1.58_{-0.08}^{+0.07}\ keV}$ which resulted in a total observed luminosity of $\mathrm{L(0.3-10\ keV)=6.2\pm 0.5\times 10^{40} erg\  s^{-1}}$.

\begin{table*}
	\centering
	\begin{minipage}{140mm}
		\caption{Spectral fitting parameters of integrated spectrum of galaxy, binaries, ULXs, and diffuse emission of NGC\,3310}
		\hskip-2.5cm
		\setlength{\tabcolsep}{4pt}		
		\begin{tabular}{@{\extracolsep{1pt}}lllllllll@{}}
			\hline 			
			& \multicolumn{4}{l}{Power-law} & \multicolumn{3}{l}{Thermal plasma} & \\[5pt]
			\cline{2-5}
			\cline{6-8}
			&&&&&&&&\\[-5pt]
			{\az{Region}} 
& $\mathrm{N_H}$ & $\Gamma$ & Norm & Norm & kT & Norm  & Norm& \\
			& $10^{22}\ cm^{-2}$ & & (OBSID 2939) & (OBSID 19891) & keV  & (OBSID 2939)  & (OBSID 19891)&     $\mathrm{\chi^2/dof}$ \\ 
			(1)&(2)&(3)&(4)&(5)&(6)&(7)&(8)&(9)\\
			\hline
			\hline
			Total galaxy & $0.17\pm 0.03$ & $1.57\pm 0.06$& $38.11_{-2.36}^{+2.88}$ & $34.10_{-2.37}^{+2.86}$& $0.22\pm 0.02$ &$40.5_{-11.0}^{+30.7}$ & $40.3_{-12.4}^{+32.8}$& 626.04/608\\[5pt]
			& & &&& $0.77_{-0.04}^{+0.05}$&$12.0\pm 2.5$&$10.5\pm 2.5$&\\[10pt]	
			Total galaxy-  & $0.17\pm 0.04$ & $1.65\pm 0.07$ & $33.89_{-2.26}^{+2.45}$ & $34.42_{-2.56}^{+2.79}$&$0.22\pm 0.02$ &$40.8_{-11.0}^{+29.7}$&$39.5_{-12.3}^{+31.6}$ & 620.01/582 \\ [5pt]
			(no nucleus)&&&&&  $0.78\pm 0.05$ &$12.2\pm 3.0$&$10.0\pm 3.0$&\\[10pt]		
			ULXs (OBSID 2939) &$0.27_{-0.04}^{+0.05}$&$1.72\pm{0.07}$&$25.5_{-2.04}^{+2.38}$&-&$0.18\pm{0.03}$&$9.90_{-5.85}^{+10.4}$&-&214.67/185\\	[5pt]	
			ULXs (OBSID 19891) &$0.49_{-0.15}^{+0.22}$&$2.08\pm{0.15}$&$34.7_{-6.20}^{+9.12}$&-&$0.28\pm{0.04}$&$14.72_{-14.18}^{+93.76}$&-& 124.73/132\\	[5pt]							
			Diffuse emission (total) & $0.15\pm 0.04$& $1.82_{-0.22}^{+0.24}$& $10.1_{-1.5}^{+1.8}$ &$10.1_{-1.5}^{+1.8}$& $0.78\pm 0.04$ & $11.0_{-2.1}^{+3.4}$&$11.0_{-2.1}^{+3.4}$& 495.13/460\\[5pt]	
			&&&&&$0.22\pm 0.02$&$31.8_{-10.4}^{+21.1}$&$31.8_{-10.4}^{+21.1}$&\\[5pt]
			Diffuse emission (north) & $0.71_{-0.08}^{+0.07}$& $1.92_{-0.30}^{+0.31}$& $1.5_{-0.3}^{+0.4}$ &$1.5_{-0.3}^{+0.4}$ & $0.19_{-0.01}^{+0.02}$ & $116.7_{-60.2}^{+97.5}$&$116.7_{-60.2}^{+97.5}$& 64.76/44\\[5pt]	
			Diffuse emission (south) & $0.50_{-0.21}^{+0.09}$& $2.66_{-0.58}^{+0.67}$& $1.2_{-0.5}^{+0.7}$ &$1.2_{-0.5}^{+0.7}$& $0.19\pm 0.03$ &$46.33_{-39.20}^{+50.86}$&$46.33_{-39.20}^{+50.86}$& 33.51/34\\[5pt]	
			Diffuse emission (ring) & $0.16\pm 0.04$& $2.09_{-0.15}^{+0.16}$& $5.3_{-0.7}^{+0.9}$ &$5.3_{-0.7}^{+0.9}$ &$0.23\pm 0.02$ &$13.1_{-4.30}^{+7.05}$&$13.1_{-4.30}^{+7.05}$& 201.94/154\\[5pt]	
			&&&&&$0.77_{-0.04}^{+0.05}$&$6.30_{-1.13}^{+1.55}$&$6.30_{-1.13}^{+1.55}$&\\[5pt]
			\hline
		\end{tabular} 	
		\label{tab.specparam_galaxy3310}
        \smallskip
{\ka{        Column 1: the regions used to extract the spectra. For the ULX population, we present spectral fit results for each OBSID separately as described in  Section \ref{luminosityofxraybinariesandulxs}. The regions of diffuse emission total, north, south, and ring correspond to the diffuse emission of the entire galaxy, the north, south spiral arm and ring respectively (after removing all resolved sources); column 2: \ion{H}{i} column density along the line of sight; columns 3, 4, and 5: the power-law photon index and the normalisation for each observation (in units of $\mathrm{10^{-5}\,photons\,keV^{-1}\,cm^{-2}\,s^{-1}}$ at 1 keV); columns 6, 7, and 8 : the temperature of the thermal plasma component (APEC) and the normalisation of each observation (expressed as $\frac{10^{-19}}{4\pi D^2}\int n_e n_H dV$ where $n_e$ and $n_H$ are the electron and hydrogen densities integrated over the volume V of the emitted region and D is the distance to the source in cm); column 9: the \chisq\ and the degrees of freedom for each spectral fit.
}}
	\end{minipage}
\end{table*}

\begin{table*}
	\centering
	\begin{minipage}{120mm}
		\caption{NGC\,3310 luminosities of the integrated spectrum of the galaxy, X-ray binaries, ULXs, and the diffuse emission}
		\setlength{\tabcolsep}{10pt}
		\begin{tabular}{@{\extracolsep{-2pt}}lcccc@{}}
			\hline    			
			Region & OBSID
&  $\mathrm{L_x^{obs}(L_x^{corr})}$ & $\mathrm{L_x^{obs}(L_x^{corr})}$   & $\mathrm{L_x^{obs}(L_x^{corr})}$   \\
			&& ($\mathrm{0.3-10.0\ keV}$) &($\mathrm{0.3-2.0\ keV}$) & ($\mathrm{2.0-10.0\ keV}$)  \\
			&&  $\mathrm{10^{40}\ erg\ s^{-1}}$     & $\mathrm{10^{40}\ erg\ s^{-1}}$  & $\mathrm{10^{40}\ erg\ s^{-1}}$  \\
			(1)&(2)&(3)&(4) & (5)\\
			\hline  	
			\hline 
			 \multirow{2}{*}{Total galaxy}& 2939 & $17.2 \pm 0.7\ (24.9)$ & $6.0\pm 0.2 \ (13.5)$ & $11.2\pm 0.7 \ (11.4)$ \\
			  &  19891 & $15.6 \pm 0.7\ (22.7)$ & $5.6\pm 0.3 \ (12.5)$ & $10.0\pm 0.6 \ (10.2)$ \\ 
			\multirow{2}{*}{Binaries} & 2939 & 11.6$\pm 0.7$ ($14.4$) & 3.0$\pm 0.2$ (5.7) &  8.6$\pm 0.7$ ($8.6$) \\ 
			 & 19891 & 11.8$\pm 0.7$ ($14.6$) & 3.1$\pm 0.3$ (5.8) &  8.7$\pm 0.6$ ($8.9$) \\
			\multirow{2}{*}{ULXs} & 2939 &7.3 $\pm 0.3$ ($10.6$)&1.8 $\pm 0.1$ ($5.1$)&5.4 $\pm 0.3$ ($5.6$)\\
			 & 19891& 6.2 $\pm 0.5$ ($13.0$)&1.8 $\pm 0.2$ ($8.3$)&4.4 $\pm 0.3$ ($4.7$)\\
			\multirow{2}{*}{Diffuse emission (total)} & 2939 & 4.8$\pm 0.6$ ($4.9$) & 3.3$\pm 0.1$ ($4.0$)& 0.8$\pm 0.5$ ($0.8$) \\
			 & 19891 & 5.1$\pm 0.6$ ($6.3$) & 2.9$\pm 0.2$ ($4.2$)& 0.8$\pm 0.6$ (0.8) \\	
			\hline	
		\end{tabular} 	
		\label{tab.binulxslum}
\smallskip
{\az{ 
Column 1: The region IDs (see Table \ref{tab.specparam_galaxy3310}); column 2: the OBSID; columns 3, 4, and 5: the broad, soft and hard-band luminosity respectively. The unabsorbed luminosities are given in the parenthesis. The component indicated as ``total galaxy" includes emission from the entire galaxy whereas the ``Binaries" component include the emission from the power-law component of the galaxy  without its nuclear region (see row 2; Table \ref{tab.specparam_galaxy3310}). ``ULXs" corresponds to the power-law component of the integrated spectrum of the ULXs described in Section \ref{luminosityofxraybinariesandulxs}. The diffuse emission (total) corresponds to the diffuse emission of the entire galaxy (after excising all resolved sources, and correcting for the corresponding area).}}
	\end{minipage}
\end{table*}

\subsubsection{Luminosity of the diffuse emission}\label{luminosityofthediffuseemission}

We measure the total diffuse emission of NGC\,3310 and the diffuse emission of three morphologically distinct areas (ring, north and south spiral arm; Fig.\ref{fig.annuli}), defined using the 8$\mu$m IRAC \textit{Spitzer} infrared image, in the broad ($\mathrm{0.3-10.0\ keV}$) , soft ($\mathrm{0.3-2.0\ keV}$), and hard ($\mathrm{2.0-10.0\ keV}$) bands. 
In order to measure the luminosity of the diffuse X-ray emission of NGC\,3310, we created ``swiss cheese'' images for each observation and in each of the three bands by removing the regions corresponding to the 27 detected X-ray sources.
To obtain a picture of its spectral parameters we extracted a spectrum from the entire region of the galaxy (also excluding the discrete X-ray sources) and a background spectrum from a source-free area outside the outline of the galaxy using the \textit{specextract} tool. We did the same for the north spiral arm, the ring, and the south spiral arm areas (Fig. \ref{fig.annuli}). The spectra from the two observations were fitted simultaneously with an absorbed power-law model (to account for unresolved X-ray binaries) plus two thermal plasma model components for the total and the ring diffuse emission. For the spiral arms the spectra were fitted with an absorbed power-law model plus a thermal plasma model component. The spectral parameters for the two observations were tied together (including the model normalisation) and a constant that was free to float was introduced in order to account for variations of the overall intensity between the two observations (e.g due to calibration deferences). Since the normalisation of the power-law component of the spectral fits to the integrated ULX spectra, do not show any significant variation (Table \ref{tab.specparam_galaxy3310}) we opted to tie together the two normalisations for the fits of the two observations. The best-fit parameters for these models and their corresponding errors are shown in Table \ref{tab.specparam_galaxy3310}.

Since the spectrum of the diffuse emission of the galaxy does not account for the diffuse emission that lies within the excised source regions we cannot directly compute the flux and consequently the luminosity of the diffuse emission of NGC\,3310. For this reason, following the same approach as in \citet{konna16}, we created an image of the diffuse emission by interpolating the pixel values in the source regions based on the intensity in annular regions surrounding them, using the \textit{dmfilth} tool. These annular regions were the same as those used to measure the background in the spectral analysis. Using the best-fit spectral model of the diffuse emission of the galaxy from each of the two observations, and making the implicit assumption that the spectrum in the source regions is on average the same as in the rest of the galaxy, we calculated the flux and the corresponding luminosity of the diffuse emission in the soft and hard bands by rescaling the model-predicted fluxes by the ratio of the counts in the interpolated image and the swiss-cheese image in each band. The absorbed, as well as the corrected luminosities for OBSIDs 2939 and 19891 are reported in Table \ref{tab.binulxslum}.
 
For NGC\,2276 we do not perform any further analysis on the diffuse emission of the galaxy since it was studied in great detail by \citet{rasmussen06}. In summary, they find that the interstellar medium is compressed at the western edge as the galaxy moves supersonically through the IGM ($\sim \mathrm{850\,km\, s^{-1}}$). The detailed temperatures and luminosities of the diffuse emission can be seen in Table 2 of \citet{rasmussen06}.

\subsection{Sub-galactic scaling relations}\label{subgalactic}

In this section we study the link between the X-ray emission of the X-ray binary populations of NGC\,3310 and NGC\,2276 with their star formation rates (SFR) and stellar masses at sub-galactic scales. Furthermore we examine the applicability of the galaxy-wide scaling relations in these smaller scales.

\subsubsection{NGC\,3310}\label{sub3310}

In order to examine whether sub-galactic regions of the galaxy follow the galaxy-wide scaling relations or not, we used as reference the linear relation $\mathrm{L^{XRBS}_{0.5-8.0keV}(ergs^{-1})=2.61\times 10^{39} SFR (M_{\odot} yr^{-1})}$ of \citet{mineo12a} between the integrated luminosity of HMXBs and SFR. 
We defined morphologically three sub-galactic regions (ring, north and south spiral arms; Section \ref{luminosityofthediffuseemission}; Fig. \ref{fig.annuli}). 
For these three regions as well as for the entire galaxy we calculated the XRB  X-ray luminosity by simultaneously fitting the X-ray spectra of both observations, accounting only for the power-law component of each spectrum which is representative of the X-ray binaries. The total luminosity of the X-ray binaries is reported in Table \ref{tab.binulxslum} for each observation. We then calculated the SFR for each of these regions from the \textit{Spitzer} MIPS $\mathrm{24\mu m}$ image using the calibration from \citet{rieke09}. 
In Fig. \ref{fig.mineo} we report the average luminosities of these two observations, calculated in the $\mathrm{0.5-8.0\ keV}$ band reported in \citet{mineo12a} for the X-ray luminosity versus SFR correlation. This correlation is shown in Fig. \ref{fig.mineo} as the reference solid line. 
We interestingly notice (Fig. \ref{fig.mineo}; top panel) that the total galaxy and its south spiral arm follow the scaling relation but the regions containing the ULXs (north spiral arm and ring) show an excess of the HMXB X-ray luminosity (Fig.\ref{fig.mineo}).

In order to examine the scaling relations in sub-galactic scales regardless of particular morphological features, we calculated the SFR of 11 semi-annuli  on the north and 11 on the south around the nucleus of the galaxy with a thickness of about 10 arcsec. The annuli are shown also in Fig.\ref{fig.annuli}. Moreover, since the regions we study may contain a non-negligible contribution of LMXBs, we then compared with the scaling relation of \citet{lehmer10} which correlates the hard X-ray luminosity (2.0-10 keV) of XRBs per SFR to the sSFR. In this comparison we used the 2.0-10.0\,keV luminosity of the power-law component from the spectral fits in each of the four regions (total galaxy, ring, north and south spiral arms). 
Due to the small number of counts in each annular region, it was not possible to fit the individual spectra. Instead, we calculated the X-ray luminosity in each annulus based on the counts of the co-added image in the 2-10 keV band (the effective area and the background  between the two observations do not change significantly), using as background region an annulus around the galaxy.
To convert the count rate to flux we used the exposure weighted sum of the conversion factors from the two observations, resulting from the power-law component in the nucleus-free integrated spectrum of the galaxy. Alternatively weighting by the number of counts gives essentially identical results (less than 1\%). The SFR of the annular regions was computed again from the \textit{Spitzer} MIPS $\mathrm{24\mu m}$ image using the calibration from \citet{rieke09}.
The stellar mass of the sub-galactic regions was computed using IRAC images at $\mathrm{3.6\mu m}$ and the calibration relation from \citet{zhu10}.
In Fig. \ref{fig.lehmer} we plot the X-ray luminosity per SFR of each of those regions against the sSFR. The annuli are enumerated from  1-11 starting from the inner annulus to the outer annulus and we show them (Fig. \ref{fig.annuli}) with black stars for the north part and red circles for the south part. Some annuli are not shown because they have too few counts to be considered significant (north annuli: 8, 9, 10, and 11). 

We find (Fig. \ref{fig.lehmer}; top panel) that there is an excess in the Lx/SFR-sSFR scaling relation for the north spiral arm and the ring of the galaxy  which host the ULXs, whereas the entire galaxy and the south spiral arm follow the \citet{lehmer10} relation. Furthermore, we observe  that there is indeed an excess of hard luminosity per SFR for the inner annuli which contain the ULXs. The outer annuli of the galaxy unless containing a ULX fall bellow the scaling relation line but within the errors are consistent with the correlation. Overall in spite of the fact that the entire galaxy follows the scaling relations, we observe an excess in the hard X-ray luminosity of the XRBs for the regions containing ULXs.

\subsubsection{NGC\,2276}

For NGC\,2276 we followed the same analysis procedure for measuring the X-ray luminosity, SFR, and stellar mass as for NGC\,3310 (see Section \ref{sub3310}). We divided the galaxy in two regions (west and east), in order to investigate whether the ram pressure on the west side of the galaxy enhanced its SFR in comparison to the east side. We also divide the galaxy in 8 semi-annuli with thickness of about 20 arcsec in order to include enough counts to allow measurement of the X-ray luminosity. The different regions of the galaxy and the 8 semi-annuli are shown in Fig.\ref{fig.annuli}. We calculated the SFR and we found that there is marginal difference between the two sides of the galaxy (Table \ref{tab.sfr}).

Plotting the integrated luminosity of the galaxy and the luminosity of the two sides on the galaxy-wide scaling relation of \citet{mineo12a} we see that there is an excess of the HMXBs X-ray luminosity at the west side of the galaxy which contains the brightest ULXs, (bottom plot; Fig. \ref{fig.mineo}) {as well as for the entire galaxy.
In Fig. \ref{fig.lehmer} we plot the X-ray luminosity per SFR of the entire galaxy, the two sides as well as the annular regions against the sSFR. The annuli are enumerated from 1 to 4 starting from the inner annulus to the outer annulus. The black stars correspond to the west part and red circles to the east part of the galaxy. Some annuli are not shown because the number of their counts is too low to be considered significant (west: 1; east: 4).  We observe again, as in NGC\,3310, an excess in the hard X-ray luminosity of the XRBs for the regions containing ULXs (especially the bright ones) but this time also an excess in the hard X-ray luminosity of the entire galaxy although of smaller scale. However the east side of the galaxy follows the standard scaling relation of \citet{mineo12a}.

\begin{figure}
	\resizebox{\hsize}{!}{\includegraphics[scale=1]{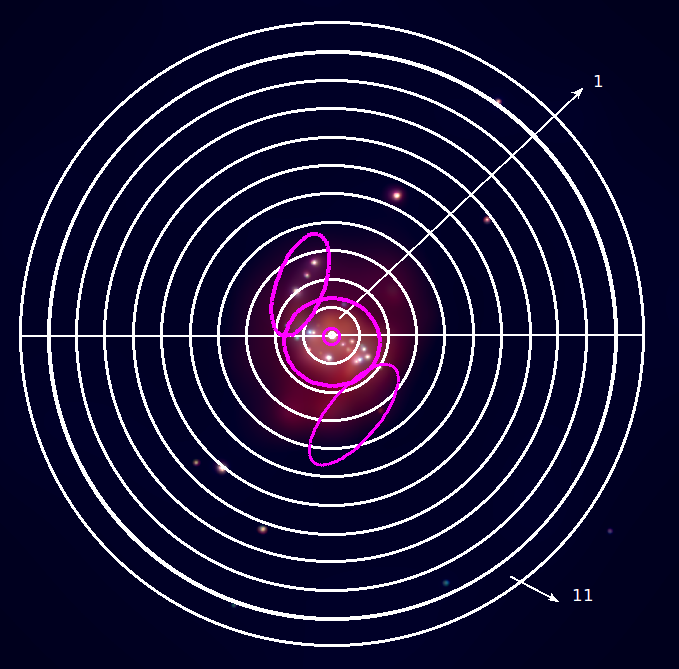}}	\resizebox{\hsize}{!}{\includegraphics[scale=1]{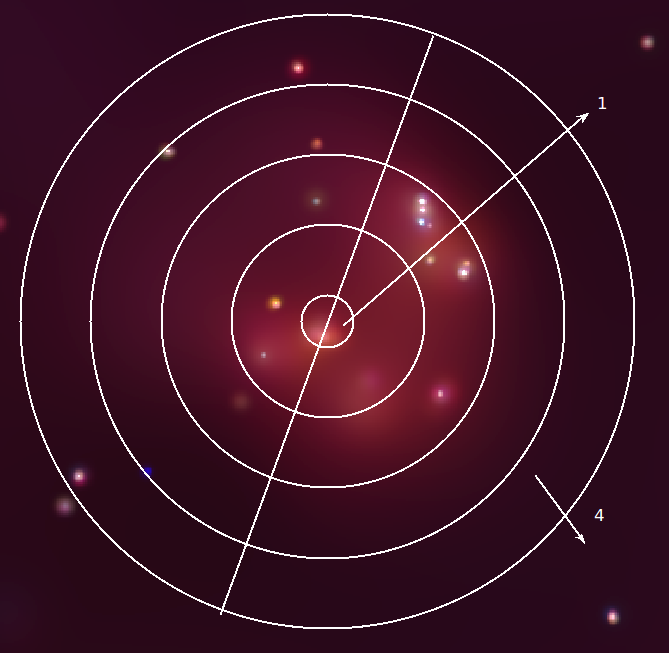}}
	\caption{
 {\az{   Top: X-ray colour image of NGC\,3310. Overlaid are the three regions used to measure the X-ray luminosity of the X-ray binary populations in the ring, north, and south spiral arm (magenta colours). We also show the 11 semi-annuli used to measure the X-ray binary X-ray emission in different regions of  \ngca. Bottom:  X-ray colour image of NGC\,2276. 
 The diagonal line shows the separation between the west and east side of the galaxy. The annuli show the regions used to measure the X-ray emission of X-ray binaries in \ngcb in sub-galactic scales.   The orientation of the images is top-north and left-east.
 }}
 }		
	\label{fig.annuli}
\end{figure}

\begin{table}	
	\caption{Star formation rates of NGC\,2276 and NGC\,3310}	
		\begin{tabular}{@{}cc@{}}		
			\hline   
		Region & SFR (24$\mu$m)   \\
		&M$_{\odot}$yr$^{-1}$\\
		\hline
		NGC\,3310&44.7\\
		ring &4.7\\
		north&1.1\\
		south&1.0\\
		\hline
		NGC\,2276 west&3.6\\
		NGC\,2276 east&2.5\\
		\hline
		\end{tabular} 
		\label{tab.sfr}	
\end{table}

\begin{figure}
	\resizebox{\hsize}{!}{\includegraphics[scale=1]{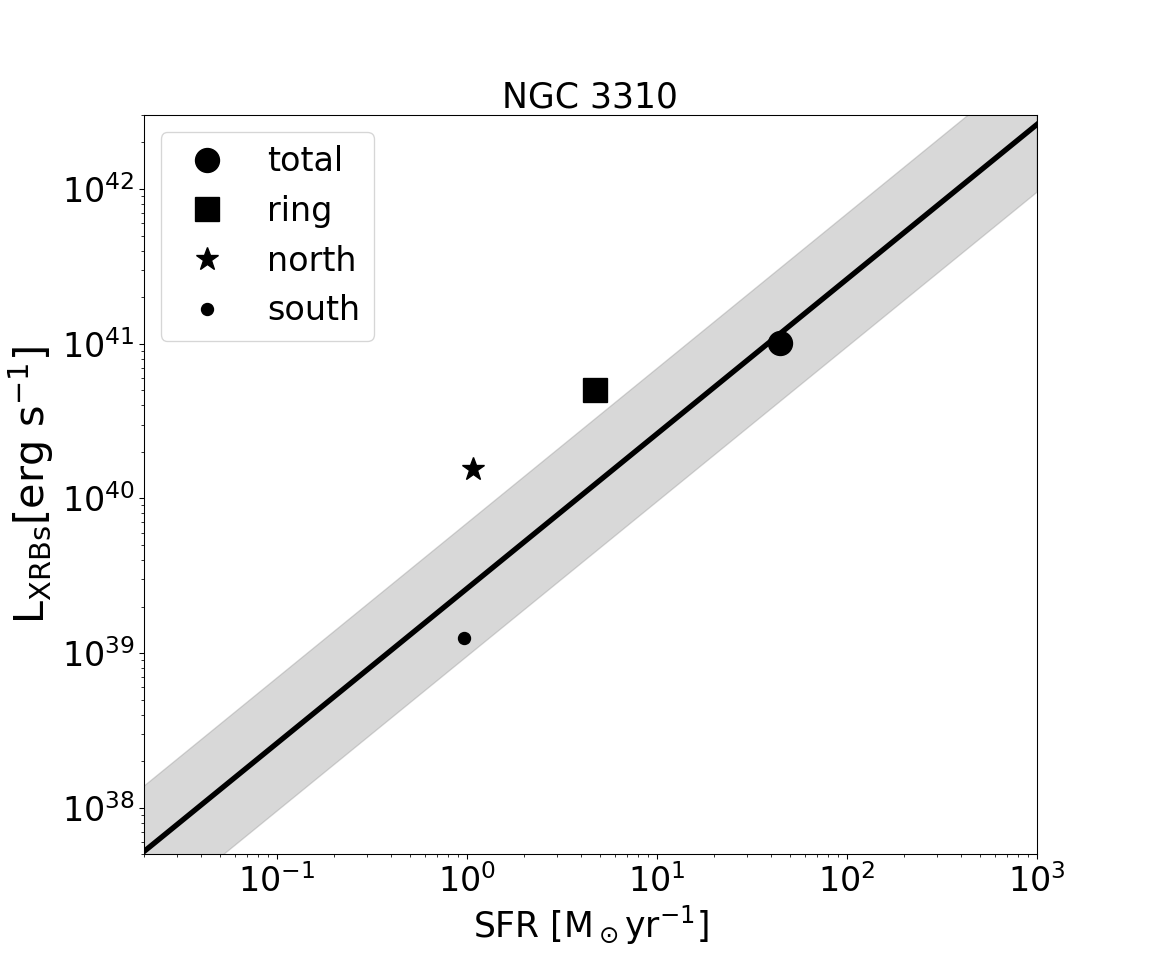}}
	\resizebox{\hsize}{!}{\includegraphics[scale=1]{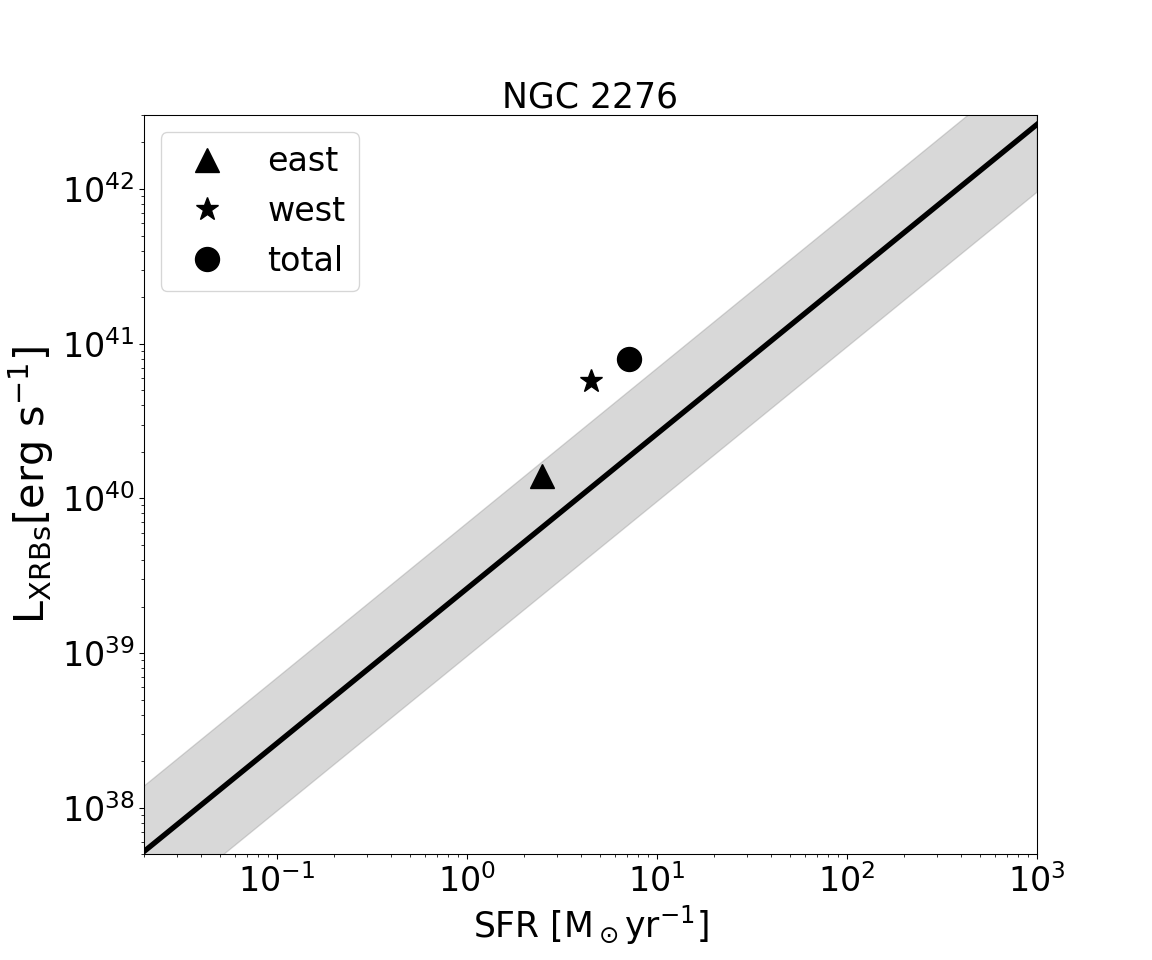}}	
	\caption{
 {\az{   Luminosity of HMXBs (0.5-8.0\,keV) versus SFR in different regions of NGC 3310 (top) and NGC 2276 (bottom).
		For NGC\,3310 the emission from the ring, the north spiral  arm and the south spiral arm are shown with the square, the star, and  the dot respectively. For NGC\,2276 the emission from the west and east side are depicted with the star and the triangle respectively. The black circle in both cases corresponds to the total emission of the galaxy. The black line and shaded area indicate the best-fit correlation of \citet{mineo12a} and its $1\sigma$ scatter.
        }}
        }		
	\label{fig.mineo}
\end{figure}

\begin{figure}
	\resizebox{\hsize}{!}{\includegraphics[scale=1]{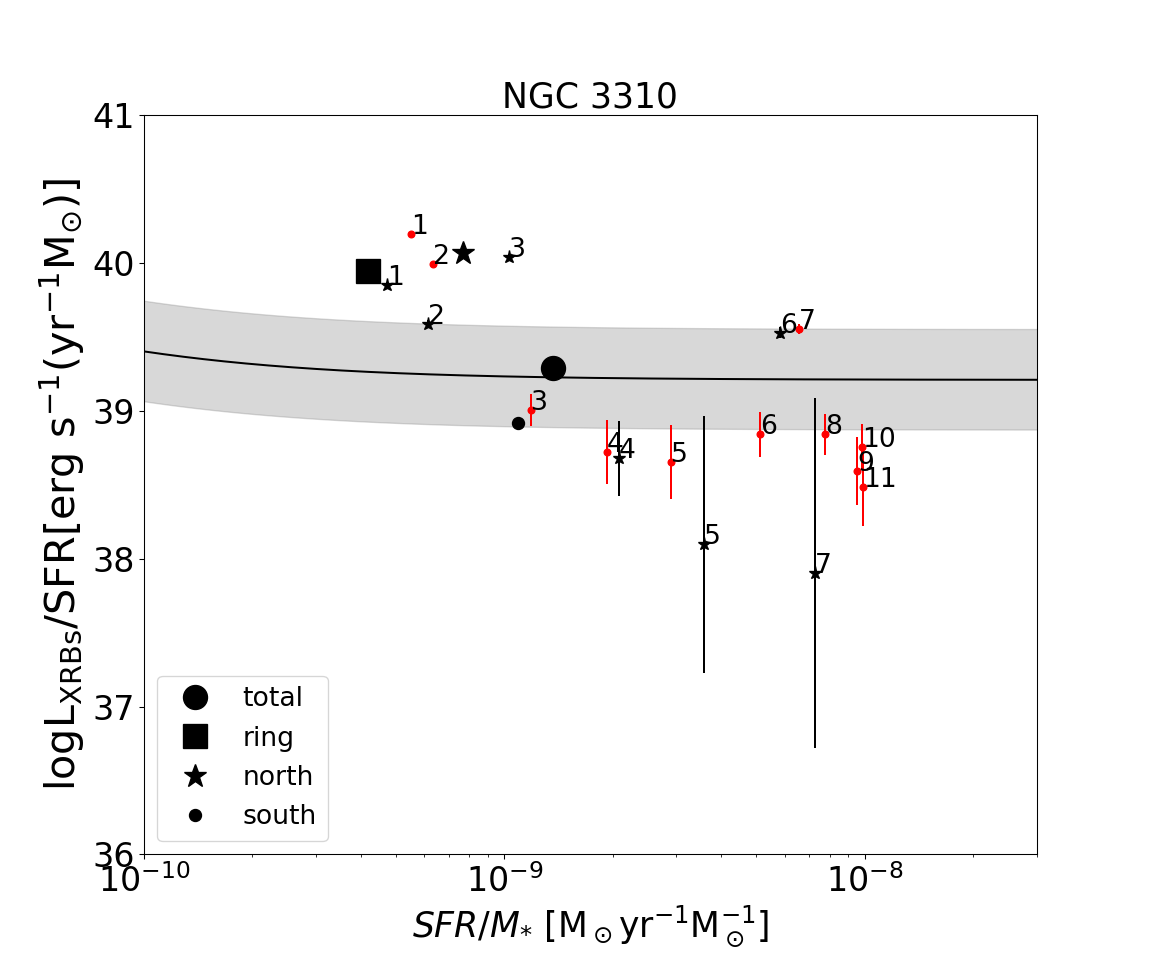}}
	\resizebox{\hsize}{!}{\includegraphics[scale=1]{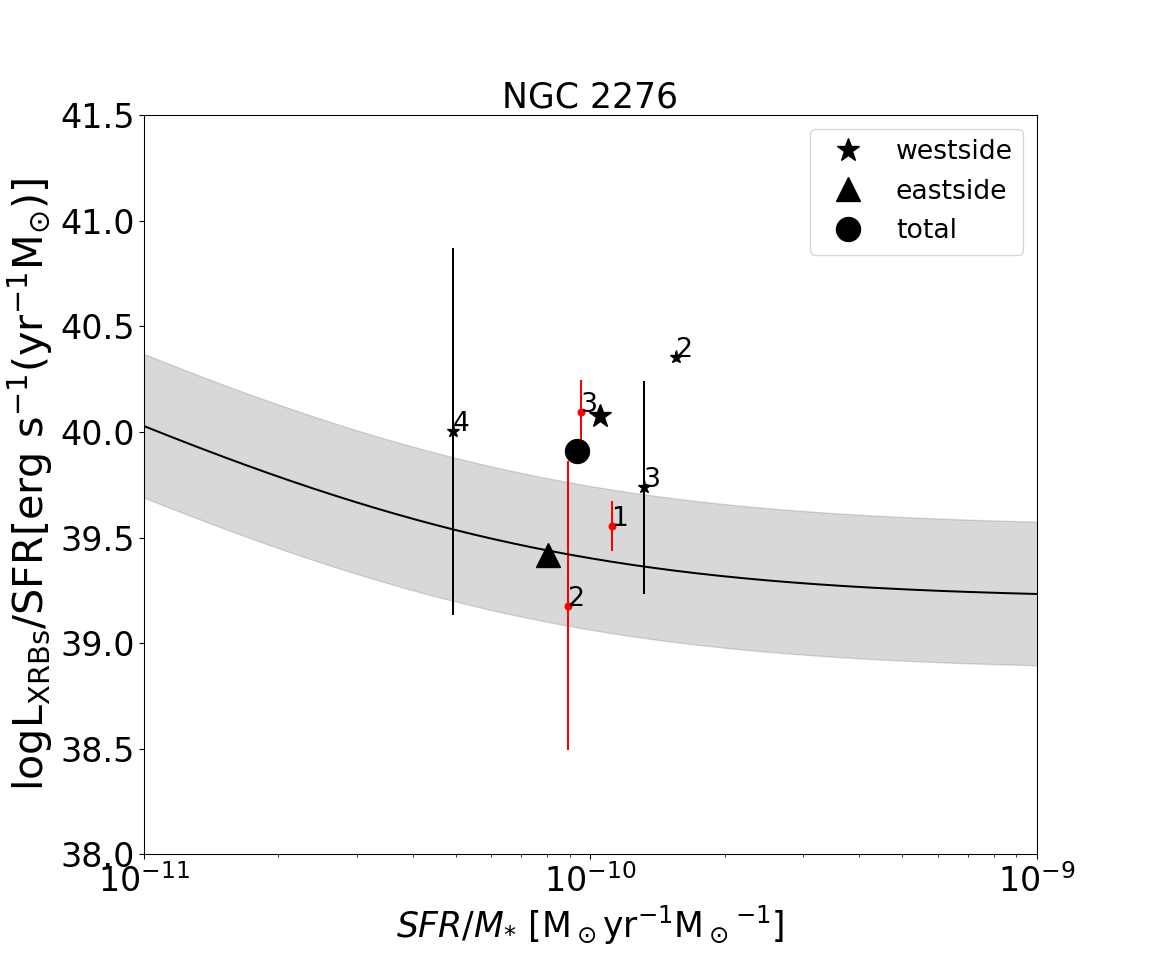}}
	\caption{
 {\az{   
    Luminosity of XRBs (2-10 keV) per SFR versus sSFR for different regions in NGC\,3310 (top) and NGC\,2276 (bottom).
		For NGC\,3310 the emission from the ring, the north spiral  arm and the south spiral arm are shown with the square, the star and the dot respectively. The enumerated small black stars indicate the north semi-annuli while the enumerated red dots indicate the south semi-annuli. For NGC\,2276 the emission form the west and east side are depicted with the star and the triangle respectively. The enumerated small black stars indicate the west semi-annuli while the enumerated red dots indicated the east semi-annuli. The black circle in both cases corresponds to the total  emission of the galaxy. The black line and shaded area indicate the best-fit correlation of \citet{lehmer10} and its $1\sigma$ scatter.
        }}
        }		
	\label{fig.lehmer}
\end{figure}

\section{DISCUSSION}\label{discussion}

In the previous sections we presented the results from the analysis of the discrete sources (photometry, spectral, timing analysis), the spectral properties of the integrated galactic emission as well as the applicability of galaxy-wide scaling relations at sub-galactic scales. In this section we examine the contribution of the XRBs, ULXs, and the diffuse emission to the integrated luminosity of the NGC\,3310. We also examine whether the variations of the sub-galactic regions on the galaxy-wide scaling relations are a result of stochastic effects or underlying factors such as age or metallicity the stellar populations.

\subsection{Diffuse emission of NGC\,3310}\label{diffuseemmision}

Based on the results reported in Table \ref{tab.binulxslum} we see that the contribution of the diffuse emission observed (i.e. absorbed) to the absorbed total luminosity of the galaxy is: 30\% in the broad band (0.3-10.0\,keV), 57\% in the soft band (0.3-2.0\,keV), and 7\% in the hard band (2.0-10.0\,keV). 

The soft X-ray luminosity of the galaxy is dominated by the diffuse emission, indicating thermal gas which is mostly concentrated on the ring of the galaxy and is characterized by a low ($\mathrm{kT\sim 0.20\,keV}$) and a high ($\mathrm{kT\sim 0.70\,keV}$) temperature component. These temperatures (Table \ref{tab.specparam_galaxy3310}) are consistent within the errors with results found in other star-forming galaxies \citep[e.g. Antennae, M101,Arp299][]{fabbiano,baldii,kuntz,mineo12b,konna16}.

We also notice that the hard diffuse emission of the ring and the north spiral arm contribute all ($\sim$80\% and $\sim$20\%) of the diffuse hard emission of the galaxy. Since the vast majority of the sources lay on these areas, it is expected that they would also dominate the diffuse hard X-ray emission of the galaxy. Moreover, the ring seems to be the main source of soft diffuse emission since it is responsible for half the diffuse emission of the galaxy.

\subsection{Nature of the X-ray sources}\label{natureofthexraysources}

From the spectral analysis results for NGC\,3310 (Table \ref{tab.binulxslum}) we find that the resolved and unresolved XRBs (power-law component) account for 70\% broad-band ($\mathrm{0.3-10\ keV}$), 50\%  soft-band ($\mathrm{0.3-2.0\ keV}$), and 77\% hard-band ($\mathrm{2.0-10\ keV}$,) absorbed luminosity of the galaxy. Their corresponding contribution to the absorption-corrected luminosity is 61\% in the broad band, 45\% in the soft band and 78\% in the hard band. As expected from other star-forming galaxies \citep[e.g.][]{lira, fabbianorev} the XRBs dominate in the hard X-ray emission of the galaxy.

In NGC\,2276 we observe that the ratio between ULX and lower luminosity sources is larger in the west (shocked) region of the galaxy. Interestingly, the total number of XRBs is marginally larger on the east side of the galaxy. More specifically, on the  west side we find 8 sources out of which 7 are ULXs, whereas on the east side we find 10 sources out of which 4 are ULXs. In Fig.~\ref{fig.hists} we plot the cumulative distribution of the luminosity of the XRBs in the two sides of the galaxy. We notice that the X-ray sources on the west side are about five times more luminous than those on the east side of the galaxy and appear to have a flatter distribution of luminosities. This behaviour does not change when we adopt the distance of 32.9\,Mpc. We discuss in following sections what could be the cause of this difference in the total luminosity between the two sides of the galaxy.
\begin{figure}
	\resizebox{\hsize}{!}{\includegraphics[scale=1]{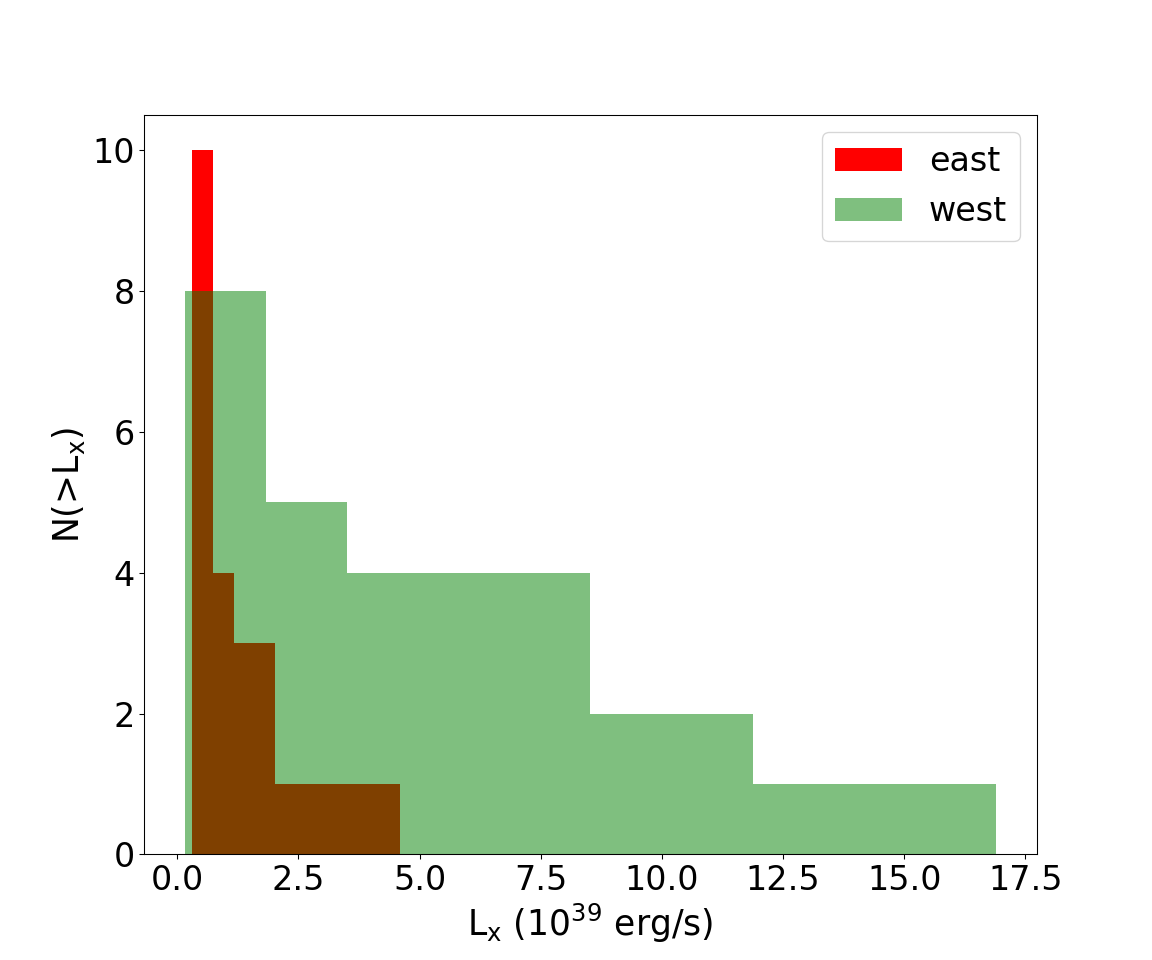}}
	\caption{Cumulative distribution of the
	luminosity of the X-ray sources in the two sides
	of NGC 2276, {\az{ showing an excess of luminous sources in the west side.}}
    }
	\label{fig.hists}
\end{figure}

\subsection{What is the cause for the excess?}\label{s:simul}

As described previously (Section \ref{subgalactic}), sub-galactic regions of NGC\,3310 and NGC\,2276 containing ULXs, and especially the brightest ones in the case of NGC\,2276, are located above the galaxy-wide scaling relations of the hard X-ray luminosity versus the SFR and the stellar mass. The important question is whether this  is a statistical sampling effect or if it has a physical origin.

In order to evaluate the significance of this excess we simulated the luminosity we would expect in the different regions of the two galaxies based on the galaxy-wide scaling relations, and we calculated the probability to get the observed luminosity by chance. 

In more detail, we first calculated the expected number of LMXBs and HMXBs in each region, based on the normalisation of the XLFs with stellar mass \citep{gilfanov04} and SFR \citep{mineo12a} respectively. We assumed two limiting luminosities ($\mathrm{L_{min}=10^{36}erg\ s^{-1}}$ and $\mathrm{L_{min}=10^{37} erg\ s^{-1}}$) for integrating the number of XRBs, in the XLF of \citet{mineo12a} and limiting luminosity of $\mathrm{L_{min}=2\times 10^{37} erg\ s^{-1}}$ for the XLF of \citet{gilfanov04}. {\az{This is supported by the finding of \citep{,mineo14} that in NGC\,2206/IC\,2163 the global scaling relations between the X-ray luminosity of ULXs and SFR also hold in local scales.}}

Then, we obtained 500 samples of LMXBs and HMXBs from a Poisson distribution, with mean equal to the expected number of HMXBs and LMXBs, in each sub-galactic region based on the aforementioned scaling relations and their local SFR and stellar mass (derivation described in Section \ref{subgalactic}). For each of these number of HMXBs and LMXBs in each region we obtained 500 samples of luminosities drawn from their corresponding XLF, for each population. The total luminosity for each region was calculated by summing the luminosities of the individual sources in the region. This resulted in two distributions of 250,000 total luminosities for each region, one for LMXBs and one for HMXBs. We then added the two distributions for each region to get the distribution of total XRB luminosities in each region. This way we accounted for fluctuations on the number of sources in each region as well as stochastic effects on their luminosity.

\begin{figure}
	\resizebox{\hsize}{!}{\includegraphics[scale=1]{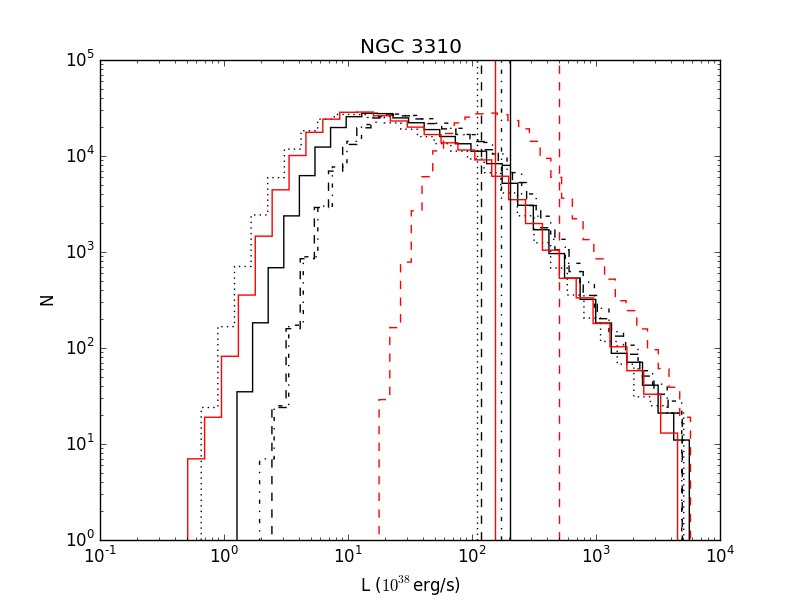}}
	\resizebox{\hsize}{!}{\includegraphics[scale=1]{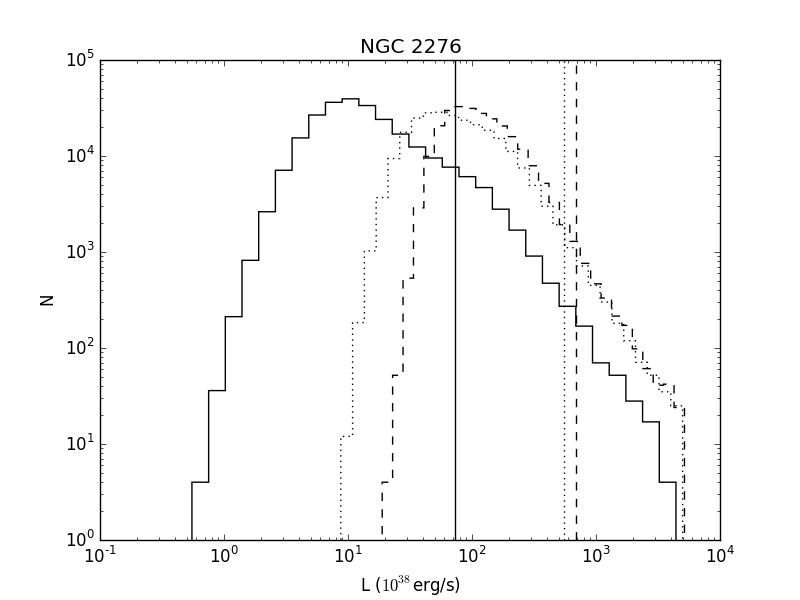}}
	\caption{
    {\az{Distributions of the expected luminosity in different regions of \ngca\ (top) and \ngcb\ (bottom) drawn from the scaling relations of X-ray binaries with SFR and stellar mass (see Section \ref{s:simul}.   }}
    Each histogram corresponds to a different region which show an excess in the luminosity of the X-ray binaries they host. {\ka{The vertical lines correspond to the observed luminosities of each region.}}  For NGC\,3310: dashed: 1st north semi-annulus, dotted: 3rd north semi-annulus, solid: 1st south semi-annulus, dash-dot: 2nd south semi-annulus, red dashed: ring, red solid: north spiral arm. For NGC\,2276: dashed: west side, dotted: 2nd west semi-annulus, solid: 3rd east semi-annulus.}	
	\label{fig.simulations}
\end{figure}

Fig. \ref{fig.simulations} shows the histograms of simulated total luminosities. Each histogram corresponds to a different region which shows an excess in the XRB luminosity for NGC\,3310 and NGC\,2276. The probability to get the observed excess by chance is given by the tail of the luminosity histograms for each region.
We find that for every region showing an excess in the scaling relations the probability of measuring such a high value only due to statistical fluctuation is between 1\% and 7\% for the various regions of NGC\,2276 and between 3\% and 13\% for the various regions of NGC\,3310. 
This test indicates that the excess of X-ray luminosity for the XRBs we observe in the scaling relations could have a physical origin.

\subsubsection{What is the physical origin of this excess?}
An obvious answer could be that differences in the SFR are responsible for the excess in the luminosity of the XRB population. However we argue that this is not the case, since all relations presented in this paper are normalised by the SFR. In NGC\,2276 in particular we measure more or less the same SFR between the two sides of the galaxy.

\textbf{A metallicity effect?}

Metallicity could be a factor causing the observed excess in the hard X-ray luminosity. Theoretical models suggest that lower metallicities are associated with higher X-ray luminosities for a given stellar population \citep{fragos13}. Recent studies have shown that low-metalicity regions could result in higher numbers of HMXBs and in particular ULXs \citep{linden10,prestwich13,brorby,douna15}, although with significant scatter. 

NGC\,3310 is a low metallicity galaxy \citep[young star clusters peak at Z=0.4Z$_\odot$;][]{degrijs03a,degrijs03b} which could explain the large number of ULXs observed. However, the areas containing the ULXs do not seem to be of much lower metallicity. According to \citet{degrijs03b}, who studied the star clusters in NGC\,3310, the distribution of metallicities in the ring of the galaxy (where the majority of ULXs reside) and outside the ring do not show any differences. Additionally, \citet{miralles14a} found a rather flat gaseous abundance gradient for about a hundred HII regions located on the disk and the spiral arms.

Measurements of the diffuse X-ray gas on the main body of NGC\,2276 show that it is of low metallicity  \citep[$\sim 0.06-0.11  Z/Z_\odot$,][]{rasmussen06}, with no metallicity differences between the two sides of the galaxy, though we could not find any reliable metallicity measurements of the stellar populations. However, since it has been found that the metallicity gradient of galaxies \citep[e.g.][]{maragkoudakis18} has a radial dependence and we are studying two {\az{ symmetric sides of the galaxy with respect to its center}} we would not expect significant differences of the metallicity between the two sides of the galaxy.

\textbf{An age effect?}

Another factor explaining the observed excess in the hard X-ray luminosity of the different regions of the galaxy could be the age variations of the stellar populations. Theoretical work \citep{linden10,fragos13} supports that the luminosity of the HMXBs peaks at younger HMXB populations and that the younger X-ray binaries populations result in more luminous sources. There is also increasing observational evidence for measurable dependence of the number and/or the X-ray luminosity of XRBs as function of their age \citep[e.g.][]{antoniou16,lehmer17,antoniou18}. {\az{Similarly, \citep{mineo14} tentatively attributed a dependence of the number of ULXs in NGC\,2206/IC\,2163 on the FIR to UV luminosity ratio on variations of the local star-formation timescales (although FIR to UV luminosity variations could also result from dust extinction).}} 

In the case of  NGC\,3310  hundreds of star clusters \citep[HST;][]{elmegreen02,degrijs03a} have been found. According to \citet{degrijs03b}, young clusters {\az{(ages peaking at $\sim30$\,Myr)}} reside predominately at the ring and northern spiral arm where the  majority of ULXs are located.

The difference in the total luminosity of the XRBs between two sides of NGC\,2276 (the west side {\az{is $\sim5$}} times more luminous)  could be explained by a younger XRB population on the west side of the galaxy. Younger stellar population in this side of the galaxy are expected from the compression-induced star-forming activity. In fact, {\ka{ H$\alpha$ and FUV images (see Fig. \ref{fig.ha}) show that the west side of the galaxy is brighter than the east side. The FUV emission is produced by stars up to $\sim$ 100\,Myr old while the Ha emission is powered by stars up to 10\,Myr old \citep{ken12}. The stark contrast of the Ha intensity between the west and the east side of the galaxy strongly indicates that the  west side is dominated by young (up to $\sim$ 10Myr) stellar populations.}

\begin{figure}
    \resizebox{\hsize}{!}{\includegraphics[scale=1]{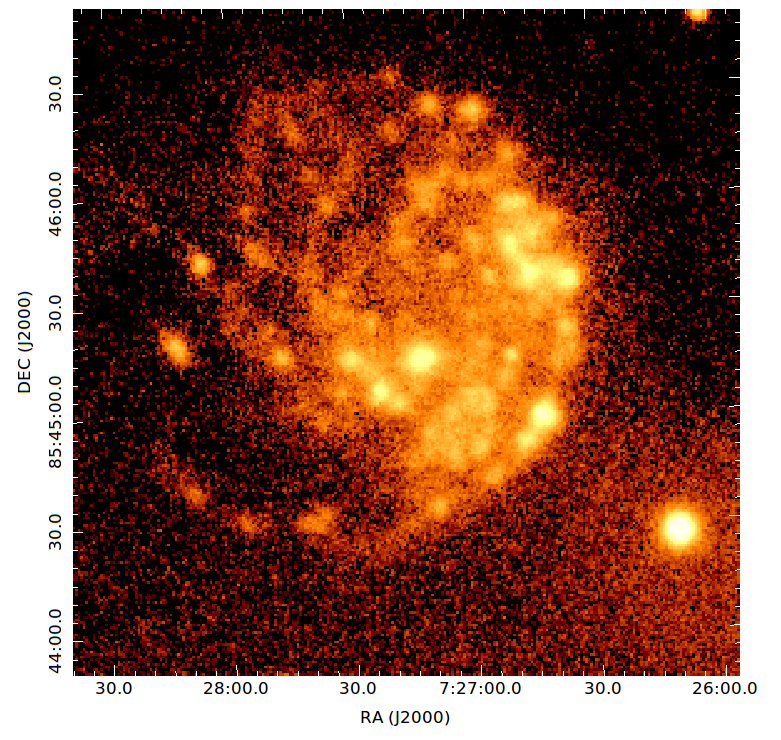}}
     \resizebox{\hsize}{!}{\includegraphics[scale=1]{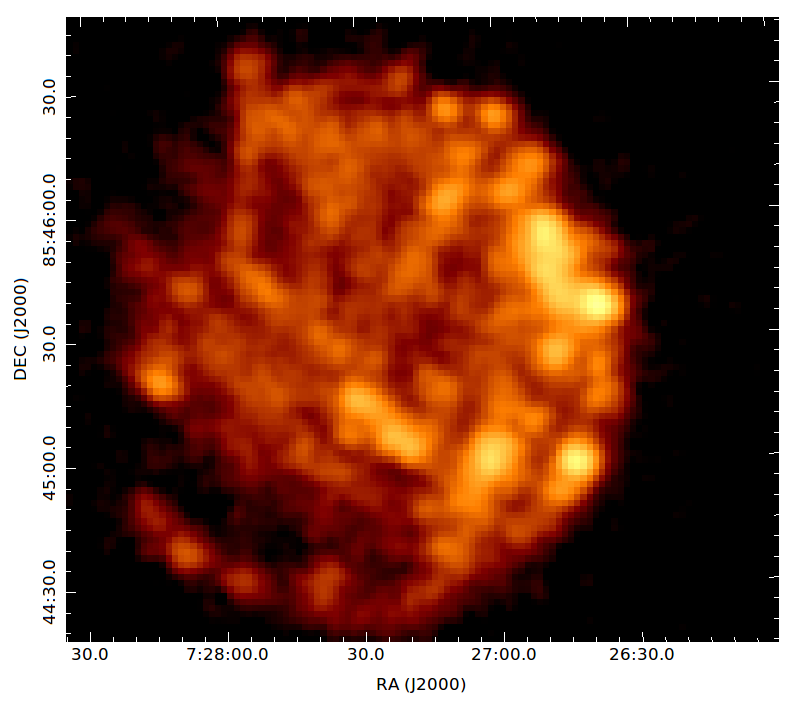}}
	\caption{ \az{H$\alpha$ (top) and GALEX FUV (bottom) images of NGC\,2276 showing intense star-formation on the compression front in the west side of the galaxy. The presence of strong H$\alpha$ and FUV emission suggests the presence of young ($\sim10$\,Myr) stellar populations. (The H$\alpha$ observations are taken at the Observatoire de Haute Provence(OHP), France)}}
	\label{fig.ha}
\end{figure}

\textbf{Implication for scaling relations}

{\az{ Based on the above arguments we favor the age-dependence of the ULX populations as the driving factor for the excess we see in the sub-galactic X-ray luminosity scaling relations with respect to the galaxy-wide scaling relations. Although metallicity may have an effect which needs to be explored more systematically, the relatively shallow metallicity gradients typically seen within star-forming galaxies \citep{maragkoudakis18,moustakas10} support the notion that the dominant factor of scatter in sub-galactic X-ray luminosity scaling relations is stellar population age variations. For the younger populations ($<100$\,Myr) such age variations may have a dramatic effect since their X-ray output may change by more than one order of magnitude per unit stellar mass for age differences as small as $\sim10-20$\,Myr \cite{fragos13}. Interestingly, the regions we are probing have SFR and sSFR similar to those observed in local dwarf galaxies. This has the important implication that star-formation history variations  (in addition to stochastic sampling of the X-ray luminosity function) could play a possibly important role in producing the scatter we observe in the galaxy-wide X-ray luminosity - SFR scaling relations in the low SFR regime.      }}

\section{SUMMARY}\label{summary}

In this work we have analysed ACIS-S \textit{Chandra} observations for the galaxies NGC\,3310 and NGC\,2276. 
For NGC\,3310 we find 27 X-ray discrete sources (SNR$>$3.0) down to 1.0$\times \mathrm{10^{38}\ erg\ s^{-1}}$. Fourteen of those sources are ULXs located on the ring and north spiral arm reaching 1.5$\times \mathrm{10^{40}\ erg\ s^{-1}}$. The majority of sources are fitted well with an absorbed power-law model ($\mathrm{N_{H}}$ greater than Galactic; $\Gamma\sim$1.7-2.0; typical for XRBs). We also find that the nucleus of the galaxy is variable but there is no sign of an AGN. 
The contribution of XRBs is more than 80\% to the hard (2.0-10.0 keV) galaxy luminosity. The diffuse emission component (kT=0.2 keV and 0.8 keV) is 60\% of the soft (0.3-10.0 keV) emission of the whole galaxy.

For NGC\,2276 we find 19 X-ray discrete sources (SNR>3.0) down to $\mathrm{L(0.3-10.0 keV)}=1.0\times \mathrm{10^{38}\ erg\ s^{-1}}$. Eleven of those sources are ULXs. Five of them are located on the east and six on the west side of the galaxy. We also find that the total luminosity of the XRBs on the west side of the galaxy is five times larger than the luminosity of XRBs in the east side and almost every ULX on the west side is brighter than those in the east side.

Moreover we find that the ULX-hosting areas of both galaxies are located above the $\mathrm{L_{X}}$-SFR and $\mathrm{L_{X}}$/SFR-sSFR scaling relations. This indicates that sub-galactic regions follow the galaxy-wide scaling relations but with much larger scatter resulting from the age (and possibly metallicity) of the local stellar populations in agreement with recent theoretical and observational results. This indicates age differences could be the origin of the scatter we observe in the low SFR regime in the $\mathrm{L_{X}}$-SFR scaling relations.

\section*{Acknowledgements}
K. A., A. Z., and K. K. acknowledge funding from the European Research Council under the European Union's Seventh Framework Programme (FP/2007-2013)/ERC Grant Agreement n. 617001. This project has received funding from the European Union's Horizon 2020 research and innovation programme under the Marie Sklodowska-Curie RISE action, grant agreement No 691164 (ASTROSTAT). We also made use of the NASA's Astrophysics Data System and observations made with the NASA/ESA Hubble Space Telescope, and obtained from the Hubble Legacy Archive, which is a collaboration between the Space Telescope Science Institute (STScI/NASA), the Space Telescope European Coordinating Facility (ST-ECF/ESA) and the Canadian Astronomy Data Centre (CADC/NRC/CSA).

\bsp	
\label{lastpage}
\end{document}